%% file: techreport.tex
\documentclass[preprint]{sig-alternate-05-2015}

\usepackage{acronym}
\usepackage[backend=bibtex, sortcites=true, firstinits=true]{biblatex}
\usepackage{booktabs}
\usepackage{comment}
\usepackage{etoolbox}
\usepackage{graphicx}
\usepackage[pdfborder={0 0 0}]{hyperref}
\usepackage{listings}
\usepackage{placeins}
\usepackage{subcaption}
\usepackage{xspace}

\makeatletter
\patchcmd{\maketitle}{\@copyrightspace}{}{}{}
\makeatother

\conferenceinfo{}{}

\input{acronyms}
\bibliography{bibliography,traff}

\usepackage{color}

\usepackage{listings}
\newcommand{\klsm}{$k$-LSM\xspace}
\newcommand{\mars}{\texttt{mars}\xspace}

\newcommand{\kpqurl}{\url{https://github.com/klsmpq/klsm}}

\makeatletter
\renewcommand*\AC@acs[1]{%
    \expandafter\AC@get\csname fn@#1\endcsname\@firstoftwo{#1}}
\makeatother

\begin{document}

\title{Benchmarking Concurrent Priority Queues: \\
Performance of k-LSM and Related Data Structures}
\subtitle{[Brief Announcement]}

\numberofauthors{3}
\author{
\alignauthor
Jakob Gruber\\
\affaddr{TU Wien, Austria}\\
\email{gruber@par.tuwien.ac.at}
\alignauthor
Jesper Larsson Tr\"aff\\
\affaddr{TU Wien, Austria}\\
\email{traff@par.tuwien.ac.at}
\alignauthor
Martin Wimmer\\
\affaddr{Google}\\
\email{wimmerm@google.com}
}

\setkeys{Gin}{width=\columnwidth}

\maketitle



\keywords{Priority Queues, Concurrency, Relaxation, Benchmarking}

\begin{refsection}

\begin{abstract} 
A number of concurrent, relaxed priority queues have recently been
proposed and implemented. Results are commonly reported for a
\emph{throughput benchmark} that uses a uniform distribution of keys
drawn from a large integer range, and mostly for single systems.  We
have conducted more extensive benchmarking of three recent, relaxed
priority queues on four different types of systems with different key
ranges and distributions. While we can show superior throughput and
scalability for our own $k$-LSM priority queue for the uniform key
distribution, the picture changes drastically for other distributions,
both with respect to achieved throughput and relative merit of the
priority queues. The throughput benchmark alone is thus not sufficient
to characterize the performance of concurrent priority queues. Our
benchmark code and $k$-LSM priority queue are publicly available to
foster future comparison.  
\end{abstract}

\section{Concurrent Priority Queues}

Due to the increasing number of processors in modern computer systems,
there is significant interest in concurrent data structures with
scalable performance that goes beyond a few dozen
processor-cores. However, data structures (e.g., priority queues) with
strict sequential semantics often present (inherent) bottlenecks
(e.g., the \texttt{delete\_min} operation) to scalability, which
motivates weaker correctness conditions or data structures with
relaxed semantics (e.g., one of the smallest $k$ items for some $k$
allowed to be deleted). Applications can often accomodate such
relaxations, and in many such cases (discrete event simulation,
shortest path algorithms, branch-and-bound), the priority queue is a
key data structure.

Many lock-free designs have been based on
Skip\-lists~\cite{shavitlotan00,sundelltsigas05,lindenjonsson13}.  In
contrast, the recently proposed, relaxed $k$-LSM priority
que\-ue~\cite{traff15:klsm} is based on a deterministic \ac{LSM}, and
combines an efficient thread-local variant for scalability with a
shared, relaxed variant for semantic guarantees. The $k$-LSM priority
queue is lock-free, linearizable, and provides configurable guarantees
of \texttt{delete\_min} returning one of the $kP$ smallest items,
where $k$ is a configuration parameter and $P$ the number of cores
(threads).  The SprayList~\cite{alistarhkopinskylishavit15} uses a
lock-free Skiplist, and allows \texttt{delete\_min} to remove a random
element from the $O(P \log^3 P)$ items at the head of the
list. MultiQueues~\cite{rihanisandersdementiev15} randomly spread both
insertions and deletions over $cP$ local priority queues, each
protected by a lock, with tuning parameter $c$, but gives no obvious
guarantees on the order of deleted elements.

\section{A Configurable Benchmark}

Priority queue performance is often measured by counting the number of
\texttt{insert} and \texttt{delete\_min} operations that can be
performed in a given amount of time, i.e., the \emph{throughput},
which would ideally increase linearly with the number of threads. Like
recent
studies~\cite{alistarhkopinskylishavit15,lindenjonsson13,shavitlotan00,sundelltsigas05,traff15:klsm},
we also measure throughput, but additionally we experiment with
different \emph{workloads}: (a) \emph{uniform}, where each thread
performs 50\% insertions and 50\% deletions, randomly chosen, (b)
\emph{split}, where half the threads perform only insertions, and the
other half only deletions; and (integer) \emph{key distributions}: (a)
\emph{uniform}, where keys are drawn uniformly at random from the
range of 32-bit, 16-bit, or 8-bit integers, and (b) \emph{ascending
(descending)}, where keys are drawn from a 10-bit integer range which
is shifted up\-wards (down\-wards) at each operation (plus/minus one).

Queues are prefilled with $10^6$ elements with keys taken from the
chosen distribution. This benchmark provides more scope for
investigating locality (split workload), distribution and range
sensitivity. The benchmark could be parameterized
further~\cite{Gramoli15} to provide for wider synthetic workloads,
e.g., sorting as in~\cite{LarkinSenTarjan14}.  To some extent, our
ascending/descending distributions correspond to the \emph{hold model}
advocated in~\cite{Jones86}.

For relaxed priority queues, it is as important to characterize the
deviation from strict priority queue behavior, also for verifying
whether claimed relaxation bounds hold. We have implemented a rank
error benchmark as in~\cite{rihanisandersdementiev15}, where the rank
of an item is its position within the priority queue as it is deleted.

\section{Experimental Results}

We have benchmarked variants of the \klsm priority
que\-ue~\cite{traff15:klsm} with different relaxation settings
(\texttt{klsm128}, \texttt{klsm256}, \texttt{klsm4096})
against a Skiplist based queue~\cite{lindenjonsson13}
(\texttt{linden}), the MultiQueue~\cite{rihanisandersdementiev15}
(\texttt{multiq}) and the
SprayList~\cite{alistarhkopinskylishavit15} (\texttt{spray}).  
As a baseline we have used a sequential heap protected by a lock
(\texttt{globallock}). The benchmarks ran on four different
machines, but we give only results from an 80-core Intel Xeon E7-8850 2 GHz 
system (\texttt{mars}) here (without hyperthreading);
see the appendix for full details and results on all machines.
Each benchmark is executed 30 times, and we report on the mean values and
confidence intervals. 
Our benchmark code can be found at \kpqurl.

\begin{figure}
\begin{center}
\includegraphics{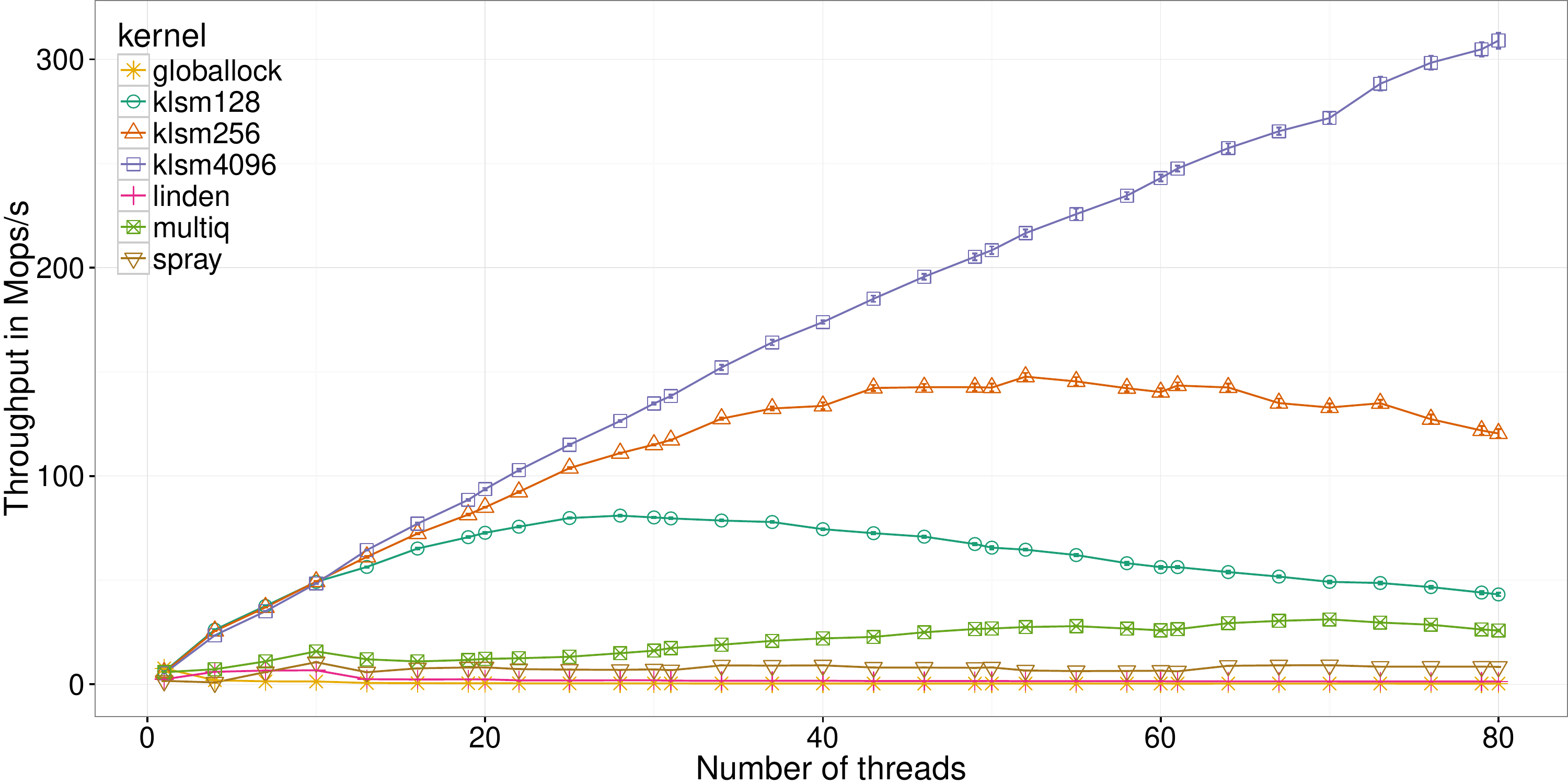}
\end{center}
\caption{\texttt{mars}: Uniform workload, uniform keys.}
\label{mars:uniform80}
\end{figure}

\begin{figure}
\begin{center}
\includegraphics{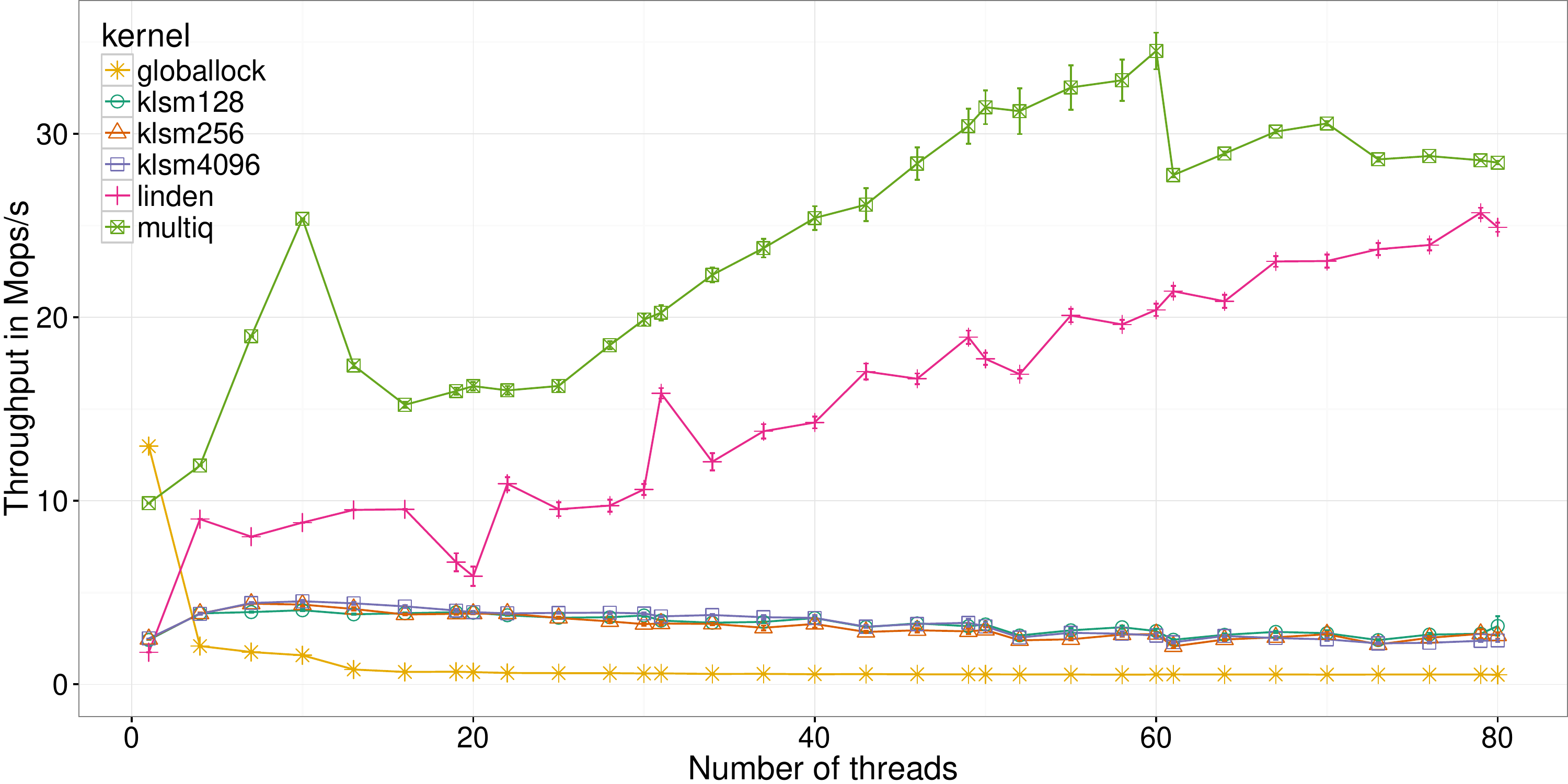}
\end{center}
\caption{\texttt{mars}: Split workload, ascending keys.}
\label{mars:split80}
\end{figure}

\begin{figure}
\begin{center}
\includegraphics{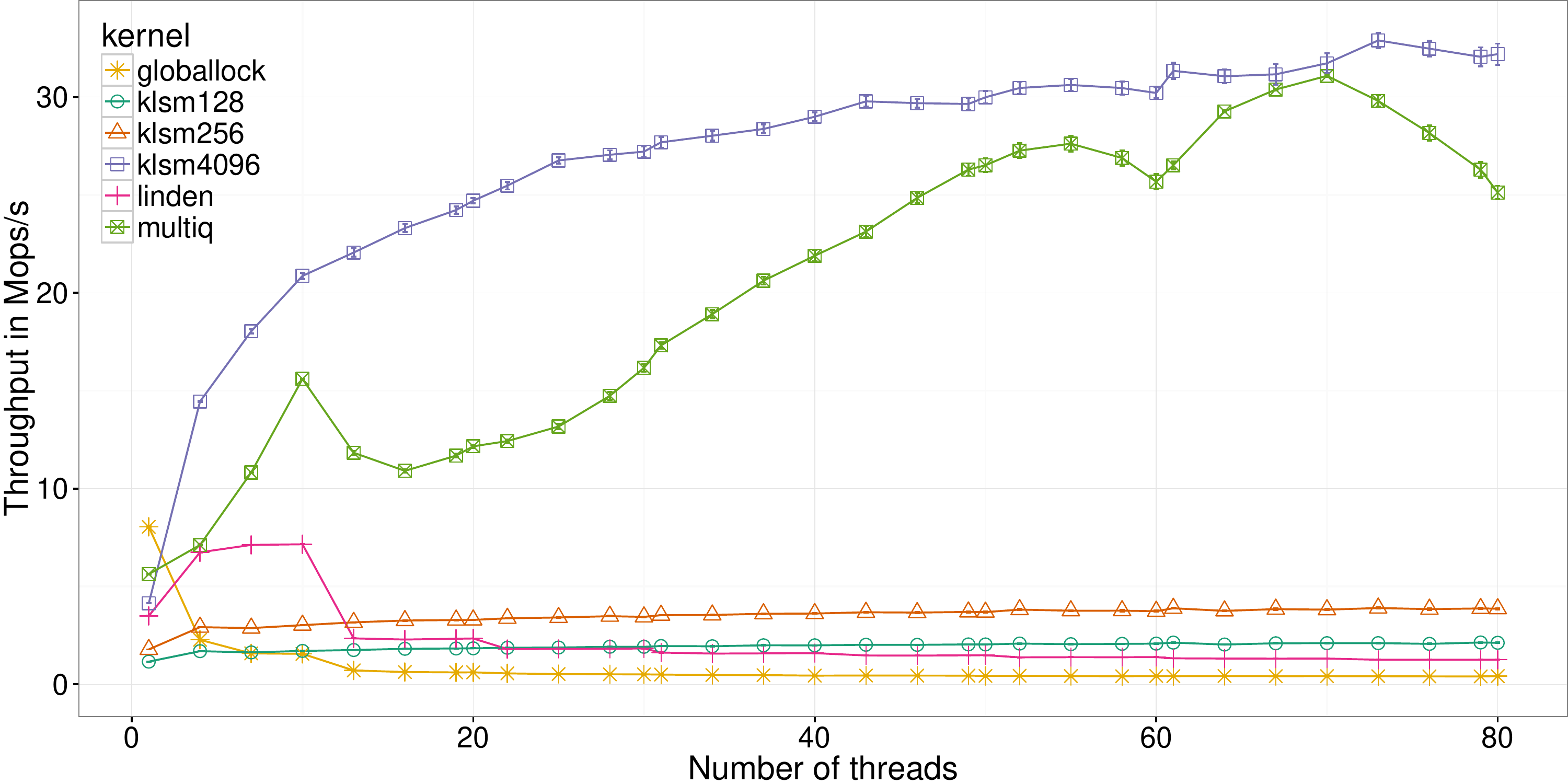}
\end{center}
\caption{\texttt{mars}: Uniform workload, 8-bit restricted keys.}
\label{mars:8bit80}
\end{figure}

Figure~\ref{mars:uniform80} compares the seven priority queue variants
under \emph{uniform workload}, \emph{uniform keys}. The \klsm variant
with $k=4096$ exhibits superior scalability and throughput of more
than 300 \ac{MOps/s}, and vastly outperforms the other priority
queues. Changing to a \emph{split workload} and \emph{ascending keys},
this picture changes dramatically, as shown in
Figure~\ref{mars:split80} where the throughput drops by a factor of
10. Here \texttt{multiq} performs best, also in terms of scalability,
surprisingly closely followed by \texttt{linden}. Restricting the key
range likewise dramatically reduces the throughput, but the \klsm
performs better in this case, (Figure~\ref{mars:8bit80}). In the
latter two benchmark configurations, the SprayList code was not stable
and it was not possible to gather results. Similar behavior and
sensitivity can be observed for the other three
machines. Hyperthreading only in rare cases leads to a throughput
increase. Overall, \texttt{multiq} delivers the most consistent
performance.

The rank error results in Table~\ref{tab:marsrankerror80} for the
uniform workload, uniform key situation show that 
\texttt{delete\_min} for all queues return keys that are not far from
the minimum, much better than the the worst-case analyses predict.


\begin{table}
\small
\begin{center}
\begin{tabular}{lrrrrrr}
  \toprule
& \multicolumn{2}{c}{20 threads} & \multicolumn{2}{c}{40 threads} & \multicolumn{2}{c}{80 threads} \\ \cmidrule(r){2-3}\cmidrule(r){4-5}\cmidrule(r){6-7}  & Mean & St.D. & Mean & St.D. & Mean & St.D. \\ 
  \midrule
klsm128 & 32 & 29 & 57 & 48 & 298 & 288 \\ 
  klsm256 & 42 & 42 & 68 & 57 & 635 & 464 \\ 
  klsm4096 & 422 & 729 & 1124 & 1287 & 13469 & 13980 \\ 
  multiq & 1163 & 3607 & 2296 & 7881 & 3753 & 12856 \\ 
   \bottomrule
\end{tabular}\end{center}
\caption{\texttt{mars}: Rank error, uniform workload and keys.}
\label{tab:marsrankerror80}
\end{table}

\vspace{-2mm}
\begin{sloppypar}
\printbibliography
\end{sloppypar}
\end{refsection}

\begin{refsection}
\FloatBarrier
\clearpage
\appendix

This appendix contains the full set of experimental results and
details on the experimental setup. We briefly recapitulate the main
three priority queue implementations, and describe the benchmarks in
more detail. The remainder are our current results from
the four available machines.

\section{Concurrent, relaxed priority queues}

The priority queues considered here support two operations on key-value pairs,
namely:
\begin{itemize}
\item
\texttt{insert}, which inserts a key-value pair into the priority queue, and
\item
\texttt{delete\_min}, which removes a key-value pair with a smallest key and
copies the corresponding value into a given location.
\end{itemize}
Concurrent priority queues so far neither support operations on specific
key-value pairs, like for instance \texttt{decrease\_key} as needed for 
asymptotically efficient single-source shortest path algorithms, nor
operations on the queues as a whole like \texttt{meld}.

A \emph{linearizable}, \emph{strict} priority queue imposes a real
time order on priority queue operations, and in such an order each
\texttt{delete\_min} must return a least key-value pair. Relaxed
consistency conditions like \emph{quasi-linearizability} allow some
non-determinism by accepting runs that are some bounded distance away
from a strict, linear order~\cite{afek2010quasi} as correct. Recently,
an even weaker conditions called \emph{local linearizability} was
proposed~\cite{Haas15local}, but this may be too weak to be useful for
priority queues.  On the other hand, relaxed priority queue semantics
relax the sequential priority queue semantics to allow that one of the
$k$ smallest keys is returned, for some value $k$, at the
\texttt{delete\_min} operation~\cite{traff15:klsm,Wimmer14:diss}. Such
relaxed priority queues are considered here. Presumably, the better
and the more precisely $k$ can be controlled, the better for the
applications.

\section{The $k$-LSM priority queue}

The \klsm~\cite{traff15:klsm,Gruber16:master} is a lock-free,
linearizable, relaxed priority queue consisting of a global component
called the \ac{SLSM}, and a thread-local component called the
\ac{DLSM}. As their names imply, both the \ac{SLSM} and \ac{DLSM} are
based on the \ac{LSM} (Log-Structured Merge-Tree)~\cite{o1996log} data
structure. The \ac{LSM} was first introduced to the database community
in \citeyear{o1996log} and later reinvented independently by Wimmer
driven by the requirements of relaxed, concurrent priority
queues~\cite{Wimmer14:diss}.  Both the \ac{SLSM} and the \ac{DLSM} may
be used as standalone priority queues, but have complementary
advantages and disadvantages which can be balanced against each other
by their composition.

The \ac{LSM} consists of a logarithmic number of sorted arrays (called
blocks) storing key-value containers (items). Blocks have capacities
$C = 2^i$ and capacities within the \ac{LSM} are distinct. A block
with capacity $C$ must contain more than $\frac{C}{2}$ and at most $C$
items.  Insertions initially add a new singleton block to the
\ac{LSM}, and then merge blocks with identical capacities until all
block capacities within the \ac{LSM} are once again distinct.
Deletions simply return the smallest of all blocks' minimal item. It
is easy to see that both insertions and deletions can be supported in
$O(\log n)$ operations where $n$ is the number of items in the
\ac{LSM}.

\begin{sloppypar}
The \ac{DLSM} is a distributed data structure containing a single thread-local
\ac{LSM} per thread. Operations on the \ac{DLSM} are essentially embarassingly
parallel, since inter-thread communication occurs only when a deletion finds
the local \ac{LSM} empty, and then attempts to copy another thread's items.
Items returned by \texttt{delete\_min} are guaranteed to be minimal on the
current thread.
\end{sloppypar}

The \ac{SLSM} consists of a single global, centralized \ac{LSM},
together with a corresponding range of items called the pivot
range. The \ac{SLSM}'s pivot range depicts a subset of the $k + 1$
smallest items (where $k$ is the relaxation parameter). Deletions
randomly choose an item from this range, and thus are allowed to skip
at most $k$ items.

Finally, the \klsm itself is a very simple data structure: it contains
a \ac{DLSM}, limited to a maximum capacity of $k$ per thread; and a
\ac{SLSM} with a pivot range containing at most $k + 1$ of its
smallest items. Items are initially inserted into the local
\ac{DLSM}. When its capacity overflows, its largest block is
batch-inserted into the \ac{SLSM}. Deletions simply peek at both the
\ac{DLSM} and \ac{SLSM}, and return the smaller item. Since deletions
from the \ac{DLSM} skip at most $k(P-1)$ items (where $P$ is the
number of threads) and deletions from the \ac{SLSM} skip at most $k$
items, \klsm deletions skip a maximum of $kP$ items in total.

We implemented the \klsm using the C++11 memory model.  A memory model
determines the order in which changes to memory locations by one thread
become visible to other threads; for instance, usage of the the
\lstinline|std::atomic| type together with its \lstinline|load()| and
\lstinline|store()| operations ensures portable multithreaded behavior
across different architectures. It is possible to vary the strictness
of provided guarantees between sequential consistency (on the strict
end) and relaxed behavior (guaranteeing only atomicity).

In our implementation, we extensively use the previously mentioned
\lstinline|std::atomic| type together with its \lstinline|load|,
\lstinline|store|, \lstinline|fetch_add|, and
\lstinline|compare_exchange_strong| operations. When possible, we
explicitly use relaxed memory ordering as it is the potentially most
efficient (and weakest) of all memory ordering types, requiring only
atomicity.

Our implementation, consisting of a standalone \klsm as well as our
parameterizable benchmark, is publicly available at \kpqurl, and
desribed in detail in~\cite{Gruber16:master}.

\section{Algorithms}

Our benchmarks compare the following algorithms:

\textbf{Globallock} (\texttt{globallock}). A simple, standardized
sequential priority queue implementation protected by a global lock is
used to establish a baseline for acceptable performance. We use the
simple priority queue implementation (\lstinline|std::priority_queue|)
provided by the \ac{STL}~\cite{c++11std}.

\textbf{Linden} (\texttt{linden}). The \citeauthor{lindenjonsson13}
priority que\-ue~\cite{lindenjonsson13} is currently one of the most
efficient Skiplist-based designs, improving upon the performance of
previous similar data
structures~\cite{shavitlotan00,sundelltsigas05,herlihy2012art} by up
to a factor of two. It is lock-free and linearizable, but has
strict semantics, i.e., deletions must return the minimal item in some
real-time order.

\textbf{SprayList} (\texttt{spray}). This relaxed priority queue is
based on the lock-free \citeauthor{fraser2004practical}
Skiplist~\cite{fraser2004practical}.  Deletions use a random-walk
method in order to return one of the $O(P \log^3 P)$ smallest items,
where $P$ is the number of threads~\cite{alistarhkopinskylishavit15}.

\textbf{MultiQueue} (\texttt{multiq}). The MultiQueue is a recent
design by
\citeauthor{rihanisandersdementiev15}~\cite{rihanisandersdementiev15}
and consists of $cP$ arbitrary priority queues, where $c$ is a tuning
parameter (set to $4$ in our benchmarks) and $P$ is the number of
threads; our benchmark again uses the simple sequential priority queue
(\lstinline|std::priority_queue|) provided by the
\ac{STL}~\cite{c++11std}, each protected by a lock.  Items are
inserted into a random priority queue, while deletions return the
minimal item of two randomly selected queues. So far, no complete analysis of
its semantic bounds exists.

\textbf{\klsm} (\texttt{klsm128}, \texttt{klsm256},
\texttt{klsm4096}). We evaluate several instantiations of the \klsm
with varying degrees of relaxation, ranging from medium ($k \in \{
128, 256 \}$) to high relaxation ($k = 4096$). Results for low
relaxation ($k = 16$) are not shown since its behavior closely mimics
the \citeauthor{lindenjonsson13} priority queue.

Unfortunately, we were not able to measure every algorithm on each
machine. The \texttt{linden} and \texttt{spray} priority queues
require libraries not present on \texttt{ceres} and
\texttt{pluto}. The SprayList implementation also turned out to be
unstable in our experiments, crashing under most circumstances outside 
the uniform workload, uniform key distribution benchmark.

\section{Other Priority Queues}

The \citeauthor{hunt1996efficient} priority
queue~\cite{hunt1996efficient} is an early concurrent design. It is
based on a Heap structure and attempts to minimize lock contention
between threads by a) adding per-node locks, b) spreading subsequent
insertions through a bit-reversal technique, and c) letting insertions
traverse bottom-up in order to minimize conflicts with top-down
deletions. It has been shown to perform well compared to other efforts
of the time; however, it is easily outperformed by more modern
designs.

\citeauthor{shavitlotan00} were the first to propose the use of
Skiplists for priority queues~\cite{lindenjonsson13}. Their initial
locking implementation~\cite{shavitlotan00} builds on
\citeauthor{pugh1998concurrent}'s concurrent Skiplist
\cite{pugh1998concurrent}, which uses one lock per node per level.
\citeauthor{herlihy2012art}~\cite{herlihy2012art} later described and
implemented a lock-free, quiescently consistent version of this idea
in Java.

\citeauthor{sundelltsigas05} invented the first lock-free concurrent
priority queue in \citeyear{sundell2003fast}~\cite{sundelltsigas05}.
Benchmarks show their queue performing noticeably better than both
locking queues by \citeauthor{shavitlotan00} and
\citeauthor{hunt1996efficient}, and slightly better than a priority
queue consisting of a Skiplist protected by a single global lock.

Mounds~\cite{LiuSpear12,LiuSpear12:mounds} is a recent concurrent
priority queue design based on a tree of sorted
lists. \citeauthor{LiuSpear12} provide two variants of their data
structure; one of them is lock-based, while the other is lock-free and
relies on the \ac{DCAS} instruction, which is not available natively
on most current processors.

One of the latest strict priority queues of interest, called the
\ac{CBPQ}, was presented recently in the dissertation of
\citeauthor{cbpq}~\cite{cbpq}. It is primarily based on two main
ideas: the chunk linked list~\cite{braginsky2011locality} replaces
Skiplists and heaps as the backing data structure, and use of the more
efficient \ac{FAA} instruction is preferred over the \ac{CAS}
instruction.  Benchmarks compare the \ac{CBPQ} against the
\citeauthor{lindenjonsson13} queue and lock-free as well as lock-based
versions of the Mound priority queue~\cite{LiuSpear12} for different
workloads. The \ac{CBPQ} clearly outperforms the other queues in mixed
workloads ($50\%$ insertions, 50\% insertions) and deletion workloads,
and exhibits similar behavior as the \citeauthor{lindenjonsson13}
queue in insertion workloads, where Mounds are dominant.

\section{Machines} \label{sec:machines}

The benchmarks were executed on four machines:

\begin{itemize}
\item \texttt{mars}, an 80-core (8x10 cores) Intel Xeon
      E7-8850 at 2 GHz with 1 TB of RAM main memory, and 32 KB L1, 256 KB L2, 24 MB L3 cache, respectively.
      \texttt{mars} has 2-way hardware hyperthreading.
\item \texttt{saturn}, a 48-core machine with 4 AMD Opteron 6168 processors
      with 12 cores each, clocked at 1.9 GHz, and 125 GB RAM main memory, and 64 KB of L1, 512 KB of L2, and 5 MB of L3 cache, respectively. The AMD processor
does not support hyperthreading.
\item \texttt{ceres}, a 64-core SPARCv9-based machine with 4 processors
      of 16 cores each. Cores are clocked at 3.6 GHz and have 8-way hardware
      hyperthreading. Main memory is 1 TB RAM, and cache is 16 KB L1, 128 KB L2,
      and 8 MB L3 , respectively.
\item \texttt{pluto}, a 61-core Intel Xeon Phi processor clocked at 1.2 GHz 
with 4-way hardware hyperthreading. Main memory is
      15 GB RAM, and cache 32 KB L1, 512 KB L2, respectively.
\end{itemize}

All applications are compiled using \texttt{gcc}, when possible:
version \texttt{5.2.1} on \texttt{mars} and \texttt{saturn}, and
version \texttt{4.8.2} on \texttt{ceres}. We use optimization level of
\texttt{-O3} and enable link-time optimizations using
\texttt{-flto}. Cross-compilation for the Intel Xeon Phi on
\texttt{pluto} is done using Intel's \texttt{icc 14.0.2}. No further
optimizations were performed, in particular vectorization was entirely
delegated to the compiler, which probably leaves the Xeon Phi
\texttt{pluto} at a disadvantage. On the other hand, all
implementations are treated similarly.

\section{Benchmarks}

Our performance benchmarks are (currently) based on \emph{throughput},
i.e., how many operations (insertions and deletions combined) complete
within a certain timeframe. We prefill priority queues with $10^6$
elements prior the benchmark, and then measure throughput for $10$
seconds, finally reporting on the number of operations performed per
second. This metric and a roughly similar setup is used in much recent
work~\cite{alistarhkopinskylishavit15,lindenjonsson13,shavitlotan00,sundelltsigas05,traff15:klsm}. Alternatively,
a number of queue operations could be prescribed, and the time (latency) for
this number and mix of operations measured.

The behavior of our throughput benchmark is controlled by the two parameters
\emph{workload} and \emph{key distribution}. The workload may be
\begin{itemize}
\item 
\emph{uniform}, meaning that
each thread executes a roughly equal amount of insertions and deletions,
\item
\emph{split}, meaning that half the threads insert, while the other half delete, or
\item
\emph{alternating}, in which each thread strictly alternates between insertions and deletions.
\end{itemize}
The key distribution controls how keys are generated for 
inserted key-value pairs, and my may be either
\begin{itemize}
\item 
\emph{uniform} with keys chosen uniformly at random
from some range of integers (we have used 32-, 16-, and 8-bit ranges), or
\item
\emph{ascending} or \emph{descending}, meaning that a uniformly chosen
key from a 10-bit integer range ascends or descends over time by adding or subtracting
the chosen key to the operation number.
\end{itemize}

We would like to supply a parameterized benchmark similar to the Synchrobench
framework
of \citeauthor{gramoli15}~\cite{gramoli15} with the following orthogonal
parameters:
\begin{itemize}
\item
\emph{Key type}: integer, floating point, possibly complex type from some 
ordered set (here, we have experimented with integers only).
\item
\emph{Key base range}, which is the range from which the random component of the
next key is chosen (here, we have used 32-, 16-, 10- and 8-bit ranges).
\item
\emph{Key distribution}, the distribution of keys within their base range
(here, we have used only uniform distributions, but others, e.g., as 
in~\cite{Jones86}, are also possible).
\item
\emph{Key dependency} switch (none, ascending, descending), which determines 
whether the next key for a thread is dependent on the key of the last deleted
element by the thread. A dependent key is formed by adding or subtracting 
the randomly generated base key to the key of the last deleted item (we
have experimented with dependent keys where the next key is formed by adding
to or subtracting from the operation number).
\item
\emph{Operation distribution}: insertions and deletions are chosen randomly with
a prescribed probability of an operation being an insert (we have experimented
with 50\% insertions so that the queues remain in a steady state).
\item
Alternatively, an \emph{operation batch size} can be set to alternate between
batches of insertions and deletions (we have experimented with strictly
alternating insertions and deletions).
\item
\emph{Workload} determines the fraction of threads that perform insertions
and the fraction of threads that perform deletions 
(we have experimented with uniform and split workloads, where in the latter
half the threads perform the insertions and the other half the deletions).
\item
\emph{Prefill} determines the number of items put in the queue before the
time measurement starts; prefilling is done according to the workload and key distribution.
\item
\emph{Throughput/latency} switch, where for throughput a duration (time limit) 
is specified and for latency the total number of operations.
\item
\emph{Repetition count} and other statistic requirements.
\end{itemize}

For instance, giving an operation batch size of one with an insert
following delete with dependent keys under a specific distribution
would correspond to
the \emph{hold model} proposed in~\cite{Jones86} and used in early
studies of concurrent priority
queues~\cite{jones89,ronngrenayani97}. Choosing large batches would
correspond to the sorting benchmark used in~\cite{LarkinSenTarjan14}.

In addition, as in~\cite{rihanisandersdementiev15} we also evaluated
the semantic quality produced by \texttt{multiq} and the \klsm with
several different relaxations by measuring the rank of items (i.e.,
their position within the priority queue) returned by
\texttt{delete\_min}.  The quality benchmark initially records all
inserted and deleted items together with their timestamp in a log;
this log is then used to reconstruct a global, linear sequence of all
operations. A specialized sequential priority queue is then used to
replay this sequence and efficiently determine the rank of all deleted
items. Our quality benchmark is pessimistic, i.e., it may return
artificially inflated ranks when items with duplicate keys are
encountered.

\section{Further Experimental Results} \label{sec:further_results}

Each benchmark is executed 30 times, and we report on the mean values and
confidence intervals. 

\begin{figure*}[ht]
\centering
\begin{minipage}{\columnwidth}
\includegraphics{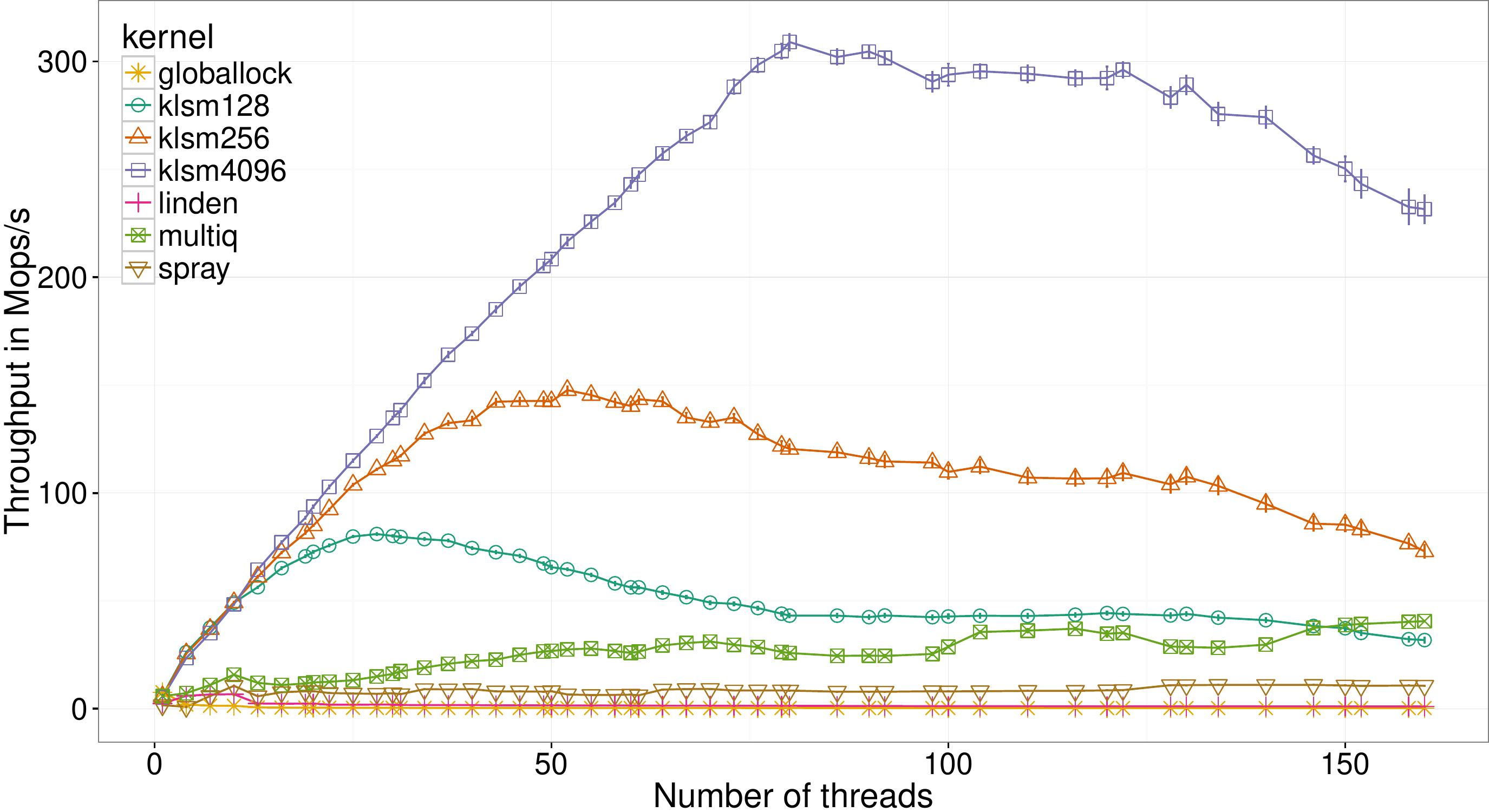}
\subcaption{Uniform workload, uniform keys (32 bits).}
\end{minipage}~%
\begin{minipage}{\columnwidth}
\centering
\includegraphics{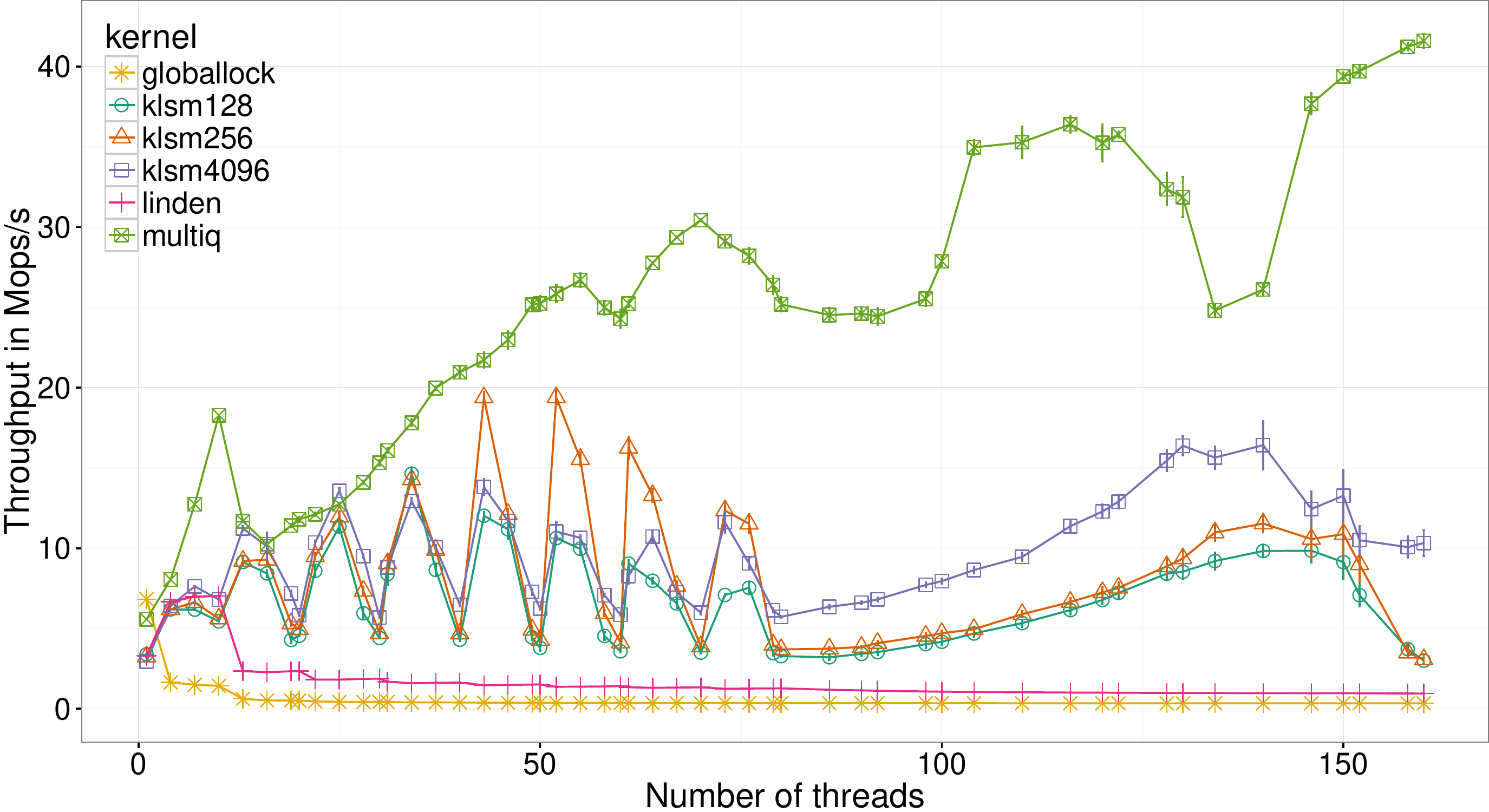}
\subcaption{Uniform workload, ascending keys.}
\label{fig:mars_uni_asc}
\end{minipage}

\begin{minipage}{\columnwidth}
\centering
\includegraphics{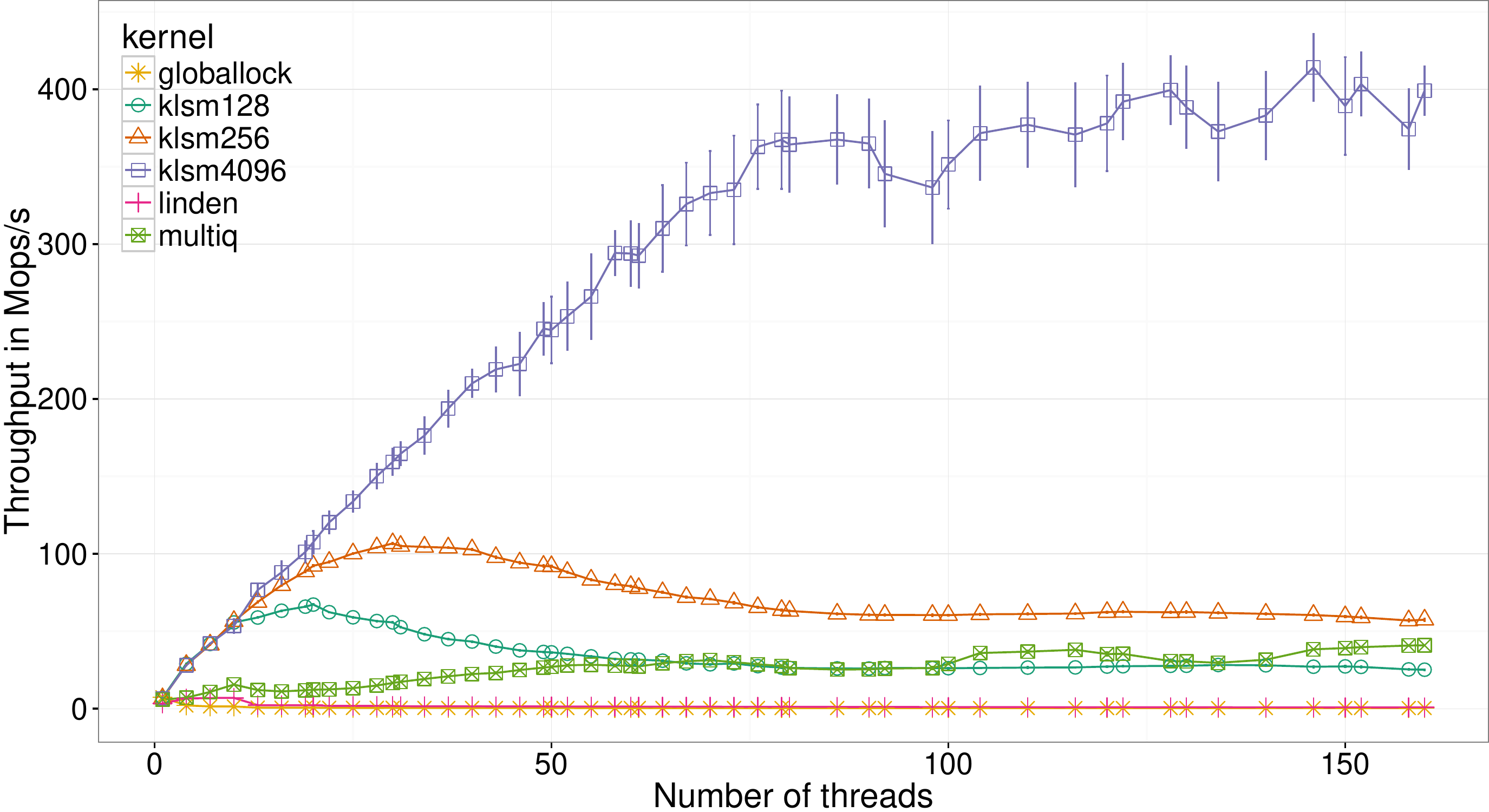}
\subcaption{Uniform workload, descending keys.}
\label{fig:mars_uni_desc}
\end{minipage}~%
\begin{minipage}{\columnwidth}
\centering
\includegraphics{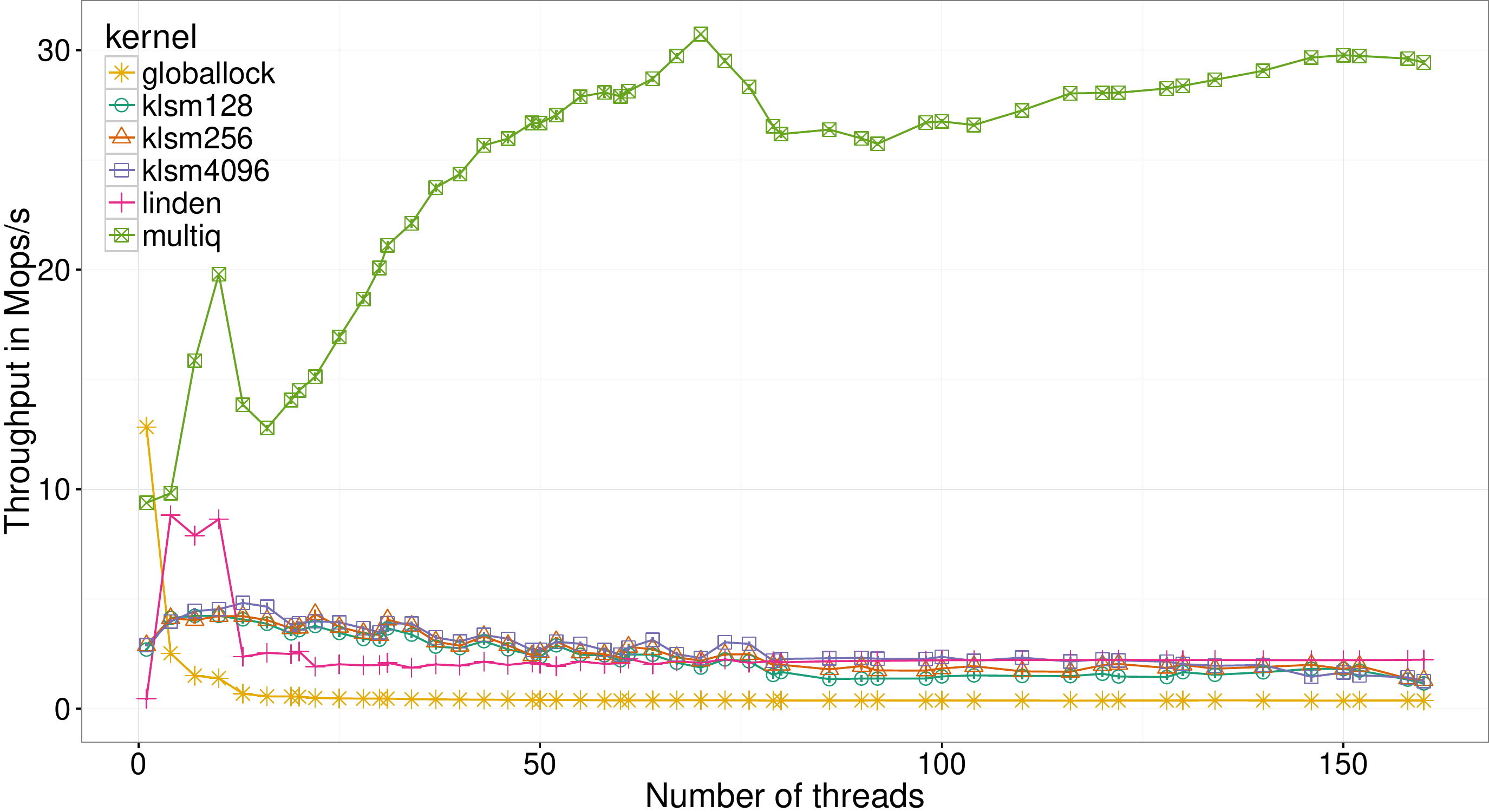}
\subcaption{Split workload, uniform keys (32 bits).}
\label{fig:mars_spl_uni}
\end{minipage}

\begin{minipage}{\columnwidth}
\centering
\includegraphics{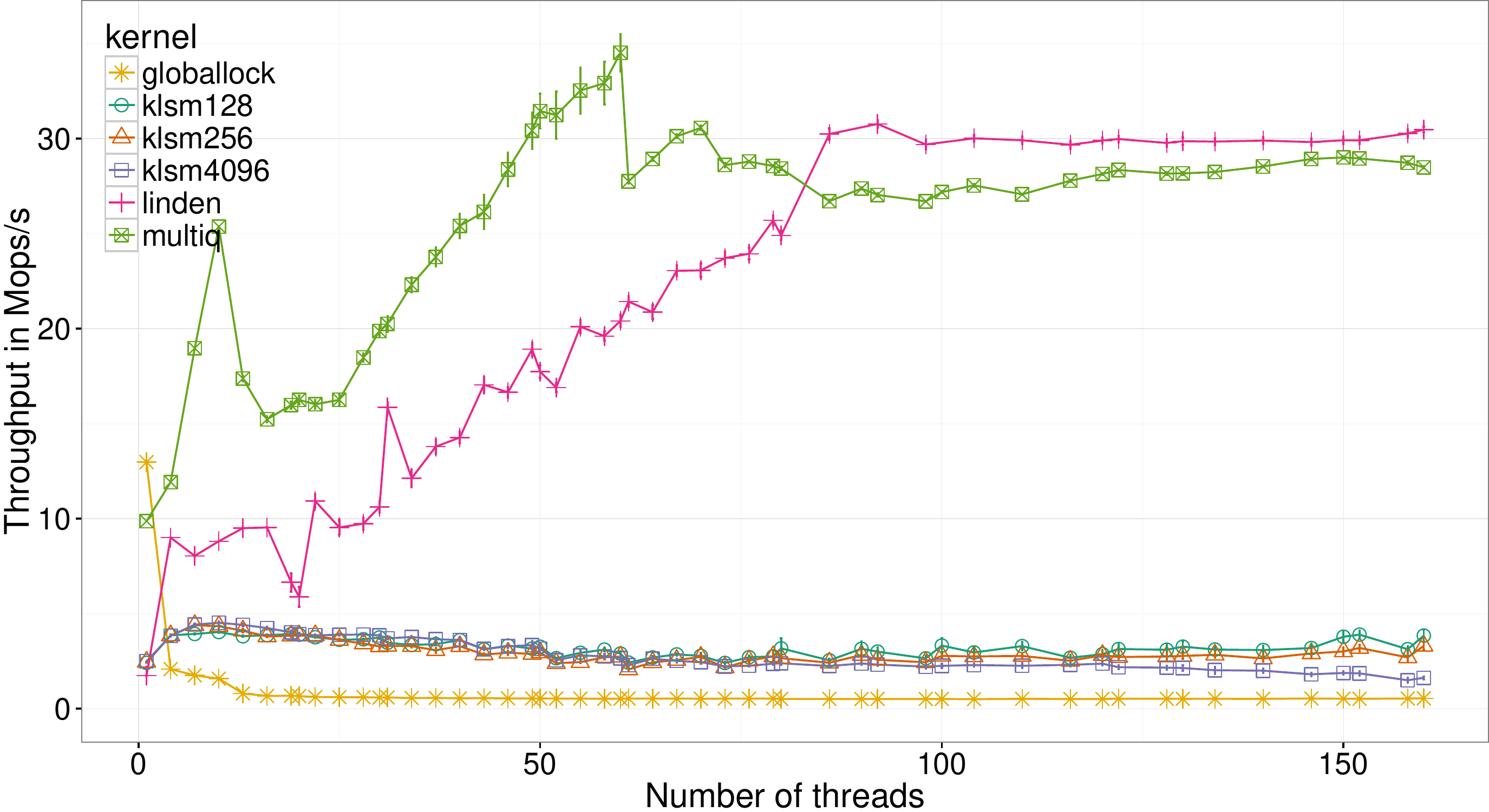}
\subcaption{Split workload, ascending keys.}
\label{fig:mars_spl_asc}
\end{minipage}~%
\begin{minipage}{\columnwidth}
\centering
\includegraphics{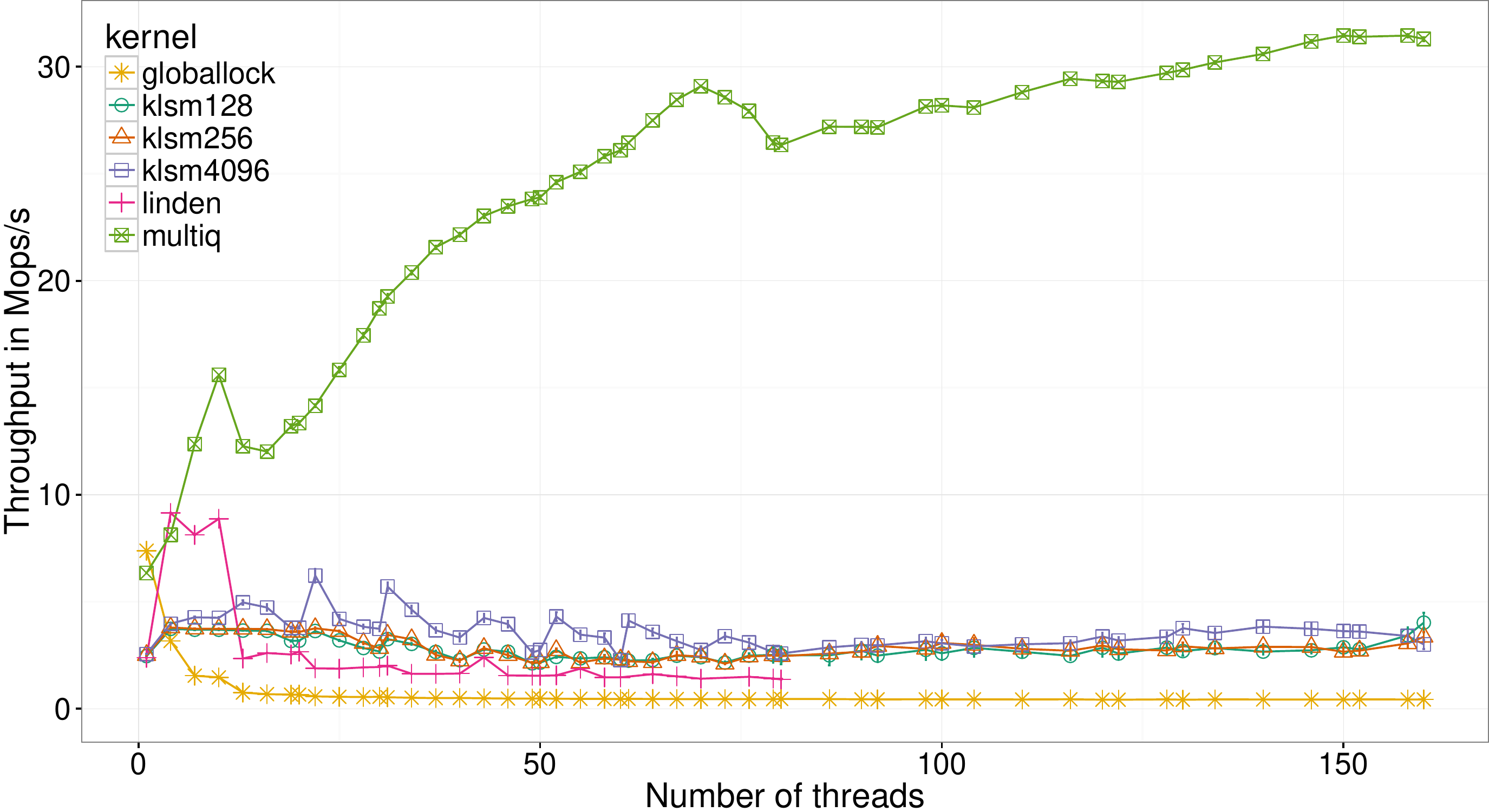}
\subcaption{Split workload, descending keys.}
\label{fig:mars_spl_desc}
\end{minipage}

\begin{minipage}{\columnwidth}
\centering
\includegraphics{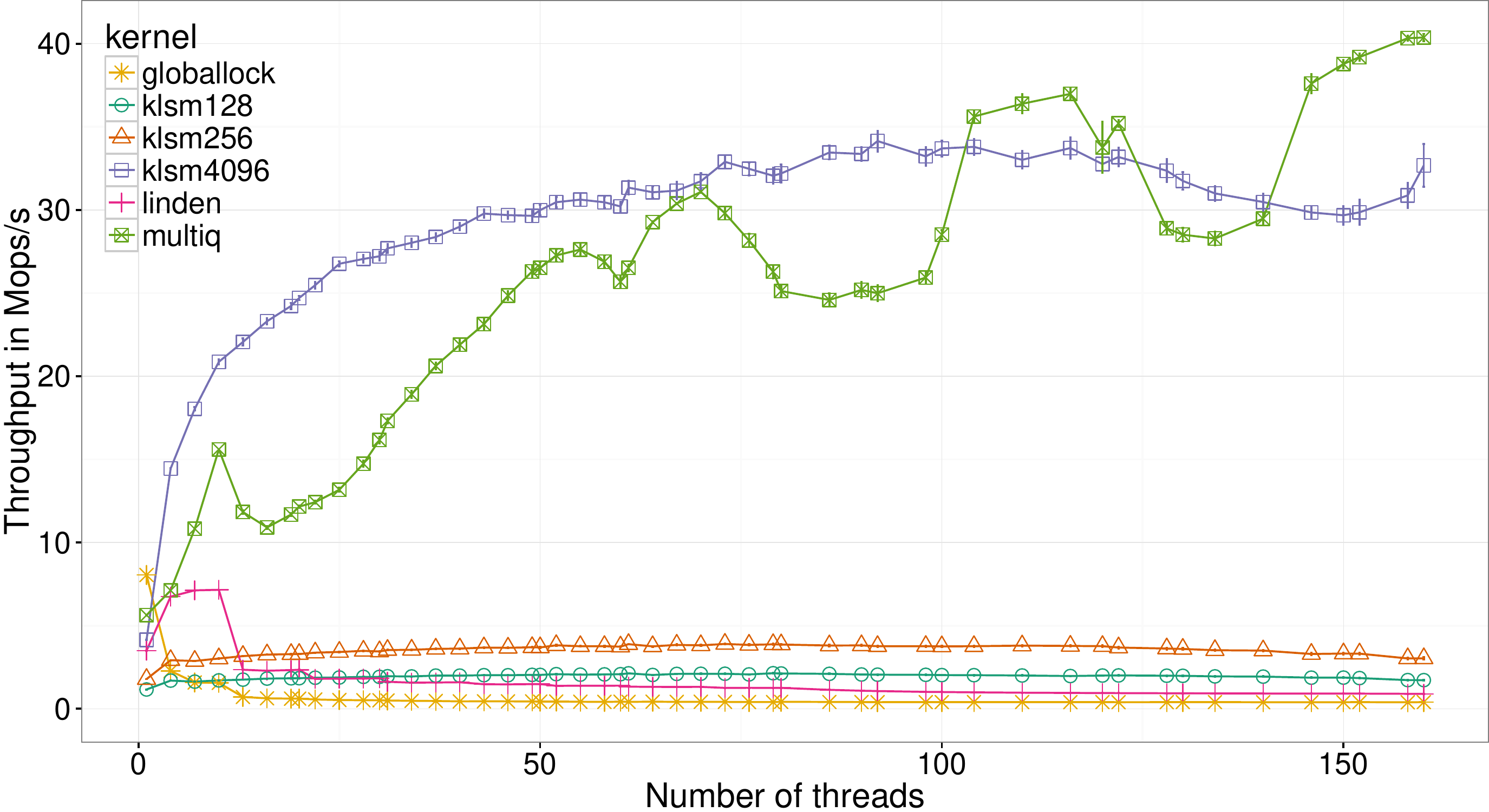}
\subcaption{Uniform workload, uniform keys (8 bits).}
\label{fig:mars_uni_rst8}
\end{minipage}~%
\begin{minipage}{\columnwidth}
\centering
\includegraphics{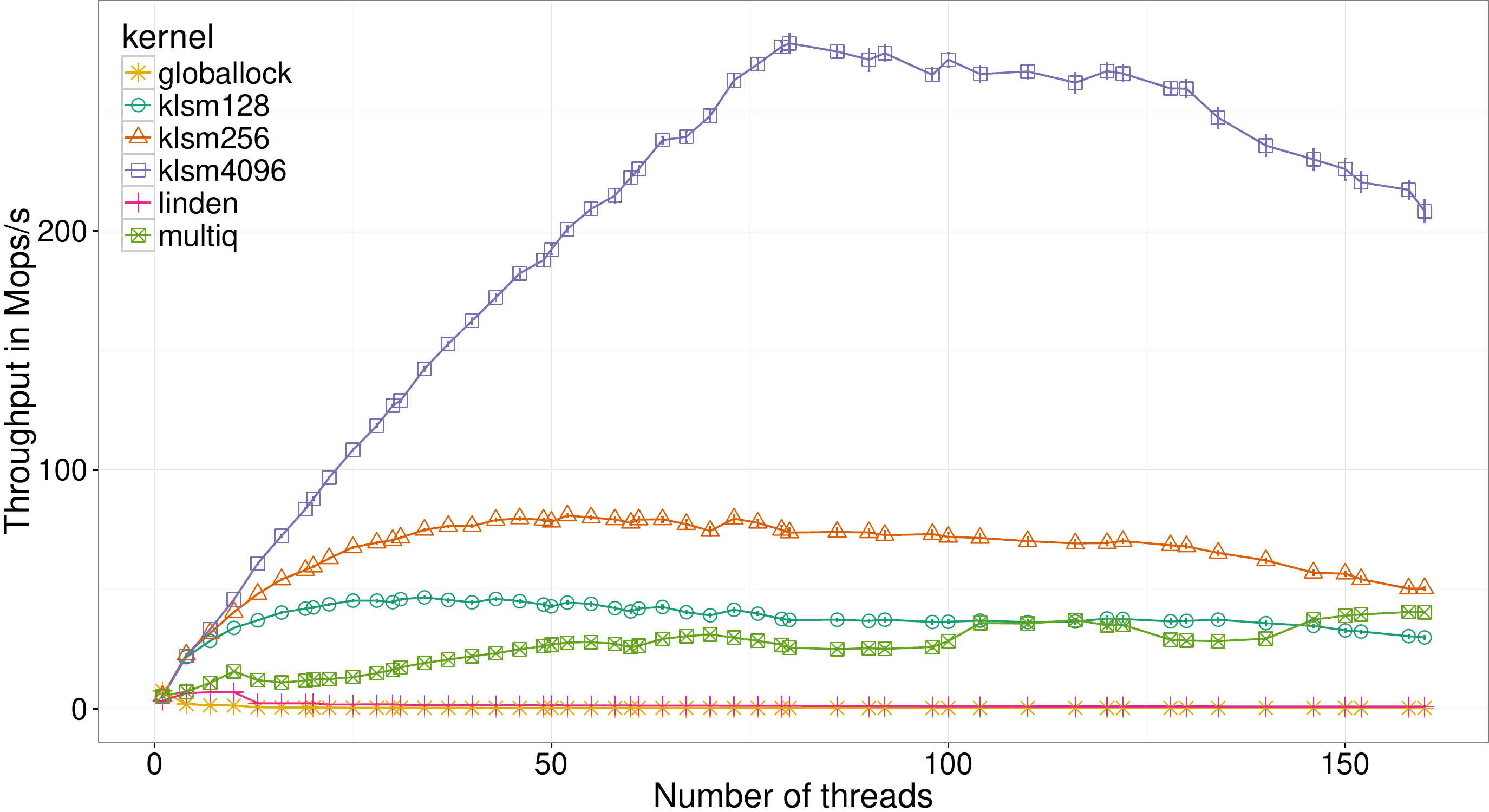}
\subcaption{Uniform workload, uniform keys (16 bits).}
\label{fig:mars_uni_rst16}
\end{minipage}

\par \vspace{\belowdisplayskip} \vspace{\abovedisplayskip}
\caption{Throughput on \texttt{mars}.}
\label{fig:mars}
\end{figure*}

Figure~\ref{fig:mars} shows throughput results for \mars. Up to 80
threads, each thread is pinned to a separate physical core, while
hyperthreading is used at higher thread counts.

Under uniform workload and uniform key distribution, the \klsm has very
high throughput, reaching over 300 \ac{MOps/s} at 80 cores with a
relaxation factor of $k = 4096$.  For the remaining data
structures, \texttt{multiq} shows the best performance, reaching
around 40 \ac{MOps/s} at 160 threads (a decrease over the \klsm by
around a factor 7.5).  The SprayList reaches a maximum of around 11
\ac{MOps/s} at 140 threads, while the \citeauthor{lindenjonsson13}
queue and \texttt{globallock} peak at around 6.7 \ac{MOps/s} (10
threads) and 7.5 \ac{MOps/s} (1 thread), respectively.

Varying the key distribution has a strong influence on the behavior on
the \klsm.  Ascending keys (Figure~\ref{fig:mars_uni_asc}) result in a
significant performance drop for all \klsm variants, all of which
behave similarly: throughput oscillates between around 5 and 15
\ac{MOps/s} until 80 threads, and then slowly increases to a local
maximum at around 140 threads.  On the other hand, descending keys
(Figure~\ref{fig:mars_uni_desc}) cause a performance increase for the
\texttt{klsm4096}, which reaches a new maximum of around 400
\ac{MOps/s}.  Behavior of \texttt{multiq}, \texttt{linden}, and
\texttt{globallock} remain more or less stable in both cases.

Under split workloads (Figures~\ref{fig:mars_spl_uni}
through~\ref{fig:mars_spl_desc}), the \klsm's throughput is very low
and never exceeds the throughput of our sequential baseline
\texttt{globallock} at a single thread. Interestingly, the
\citeauthor{lindenjonsson13} priority queue has drastically improved
scalability when using a combination of split workload and ascending
key distribution.  We assume that this is due to improved
cache-locality: inserting threads access only the tail end of the
list, with deleting threads accessing only the list head.
\lstinline|linden| was unstable at higher thread counts under
split workload and descending keys, and we omit these results.

A key domain restricted to 8-bit integers
(Figure~\ref{fig:mars_uni_rst8}) results in many duplicate key values
within the priority queue. This also causes decreased throughput of
the \klsm: medium relaxations do not appear to scale at all, while the
\texttt{klsm4096} does seem to scale well --- but only to a maximum
throughput of just over 30 \ac{MOps/s}. The larger 16-bit domain of
Figure~\ref{fig:mars_uni_rst16} produces very similar results to the
uniform key benchmark with a 32-bit range.

Hyperthreading is beneficial in only a few cases. For
instance, \texttt{multiq} makes further modest gains beyond 80
threads with uniform workloads (e.g., Figures~\ref{fig:mars_uni_asc}
and~\ref{fig:mars_uni_rst8}). However, in general, most algorithms do
not appear to benefit from hyperthreading.

Table~\ref{tbl:mars} contains our quality results for \mars, showing
both the rank mean and its standard deviation for 20, 40, and 80
threads. In general, the \klsm produces an average quality
significantly better than its theoretic upper bound of a rank of $kP +
1$. For example, the \texttt{klsm128} has an average rank of 32
under the uniform workload, uniform key benchmark at 20
threads (Table~\ref{tbl:mars_uni_uni}), compared to the maximally
allowed rank error of 2561.  Relaxation of \texttt{multiq} appears to
be somewhat comparable to \texttt{klsm4096}, and it seems to
grow linearly with the thread count.  The uniform 8-bit restricted key
benchmark (Table~\ref{tbl:mars_uni_rst8}) has artificially
inflated ranks due to the way our quality benchmark handles key
duplicates.

\begin{figure*}[ht]
\centering
\begin{minipage}{\columnwidth}
\includegraphics{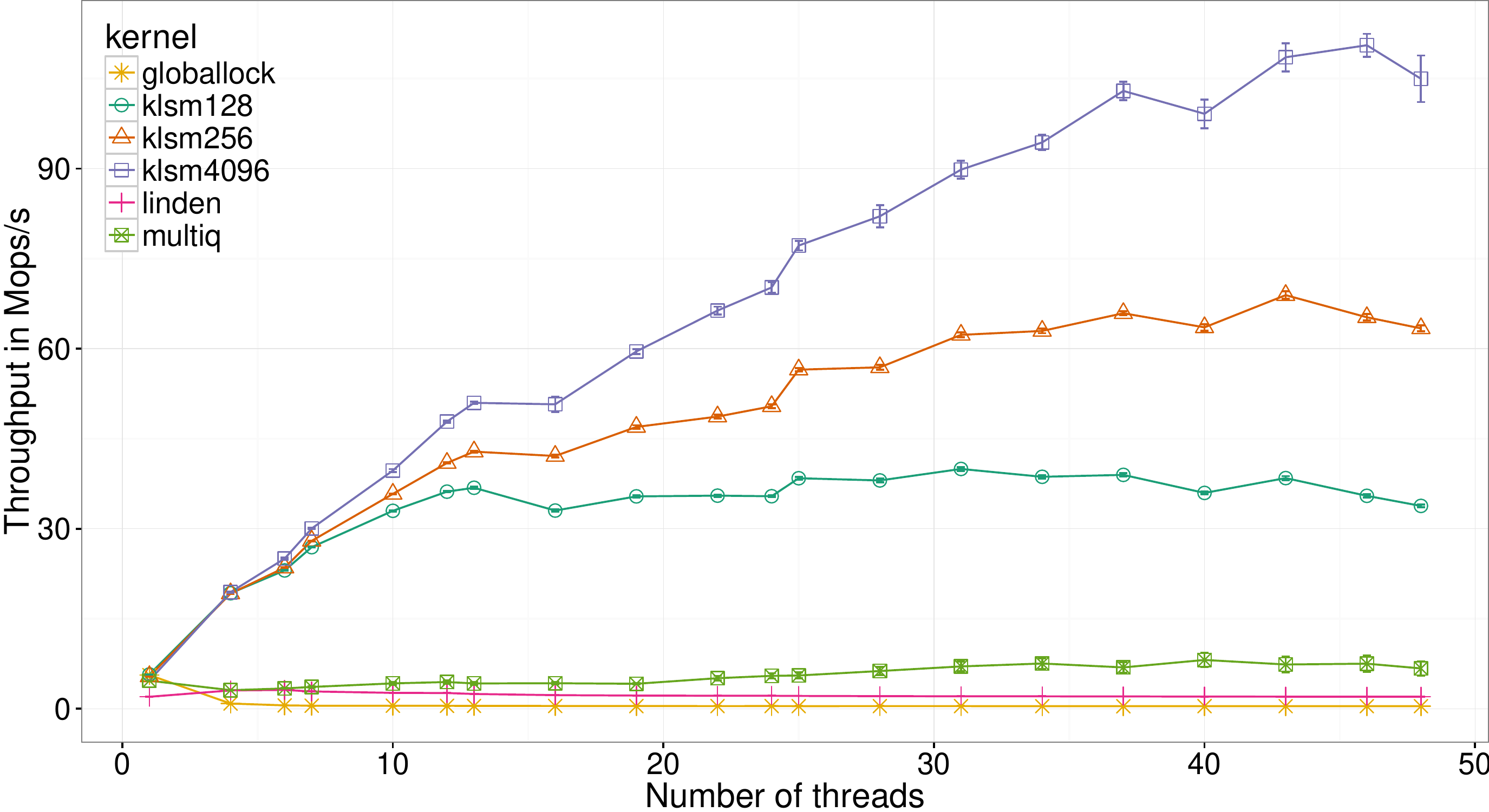}
\subcaption{Uniform workload, uniform keys (32 bits).}
\end{minipage}~%
\begin{minipage}{\columnwidth}
\centering
\includegraphics{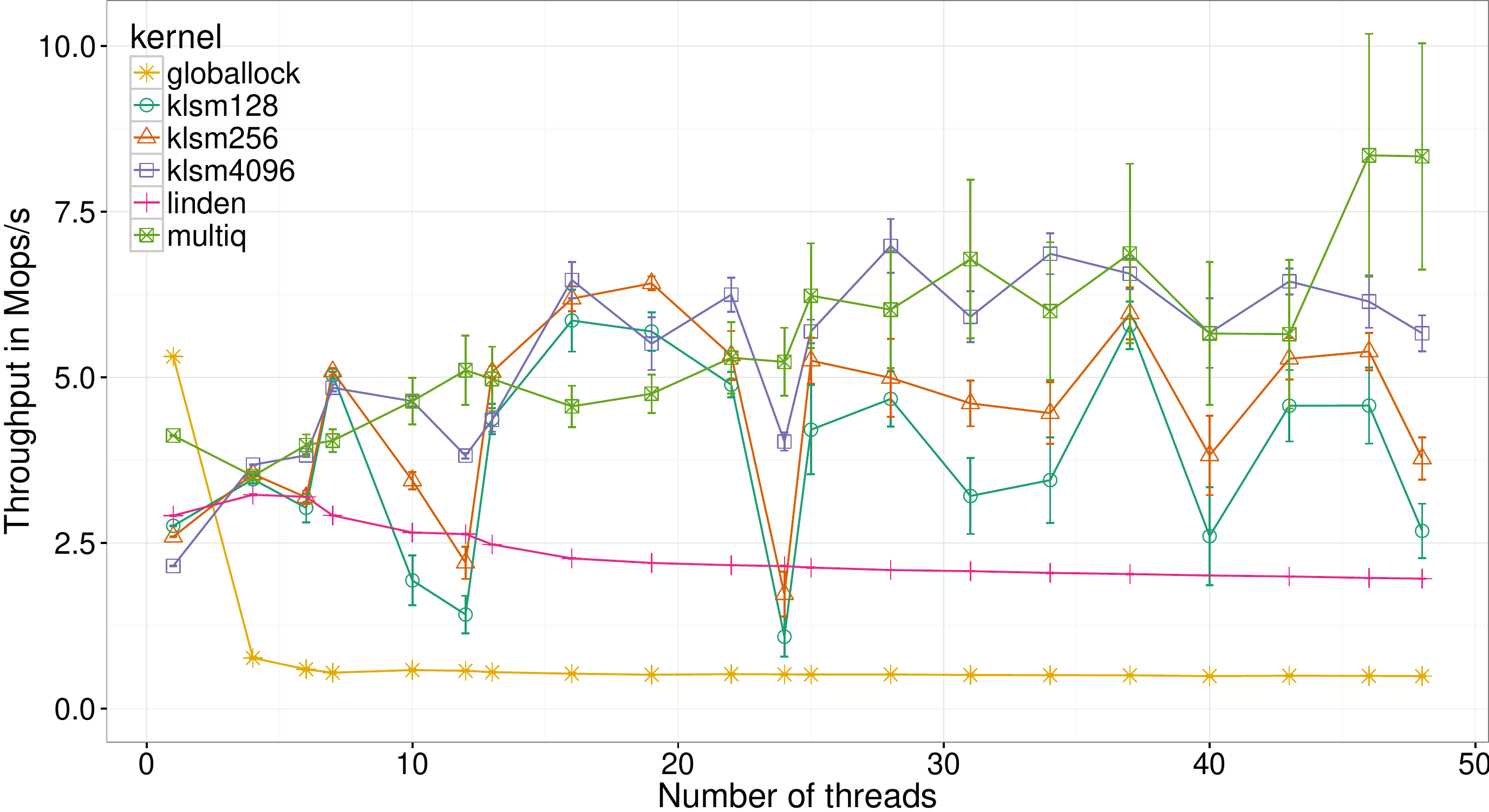}
\subcaption{Uniform workload, ascending keys.}
\end{minipage}

\begin{minipage}{\columnwidth}
\includegraphics{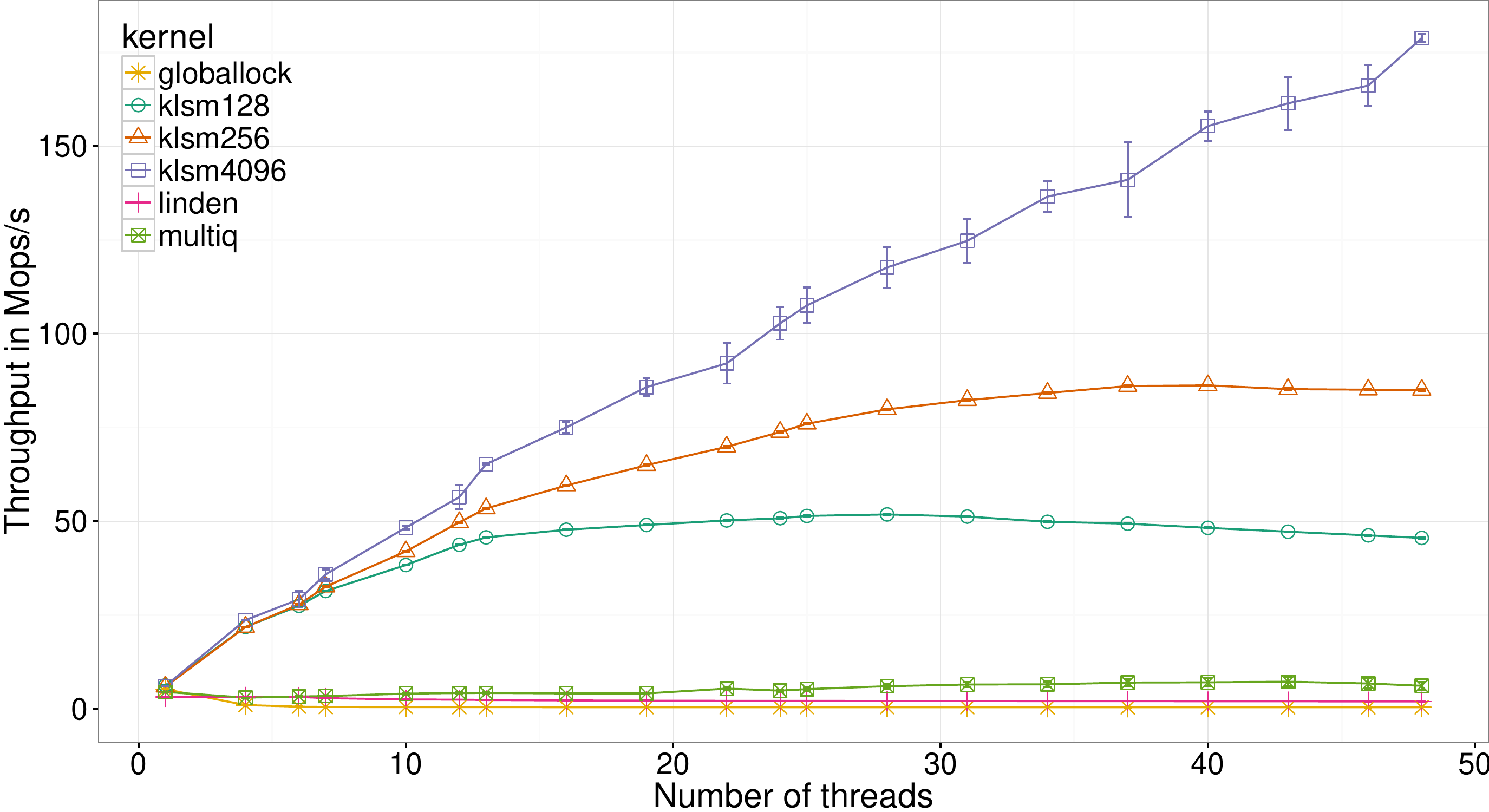}
\subcaption{Uniform workload, descending keys.}
\end{minipage}~%
\begin{minipage}{\columnwidth}
\centering
\includegraphics{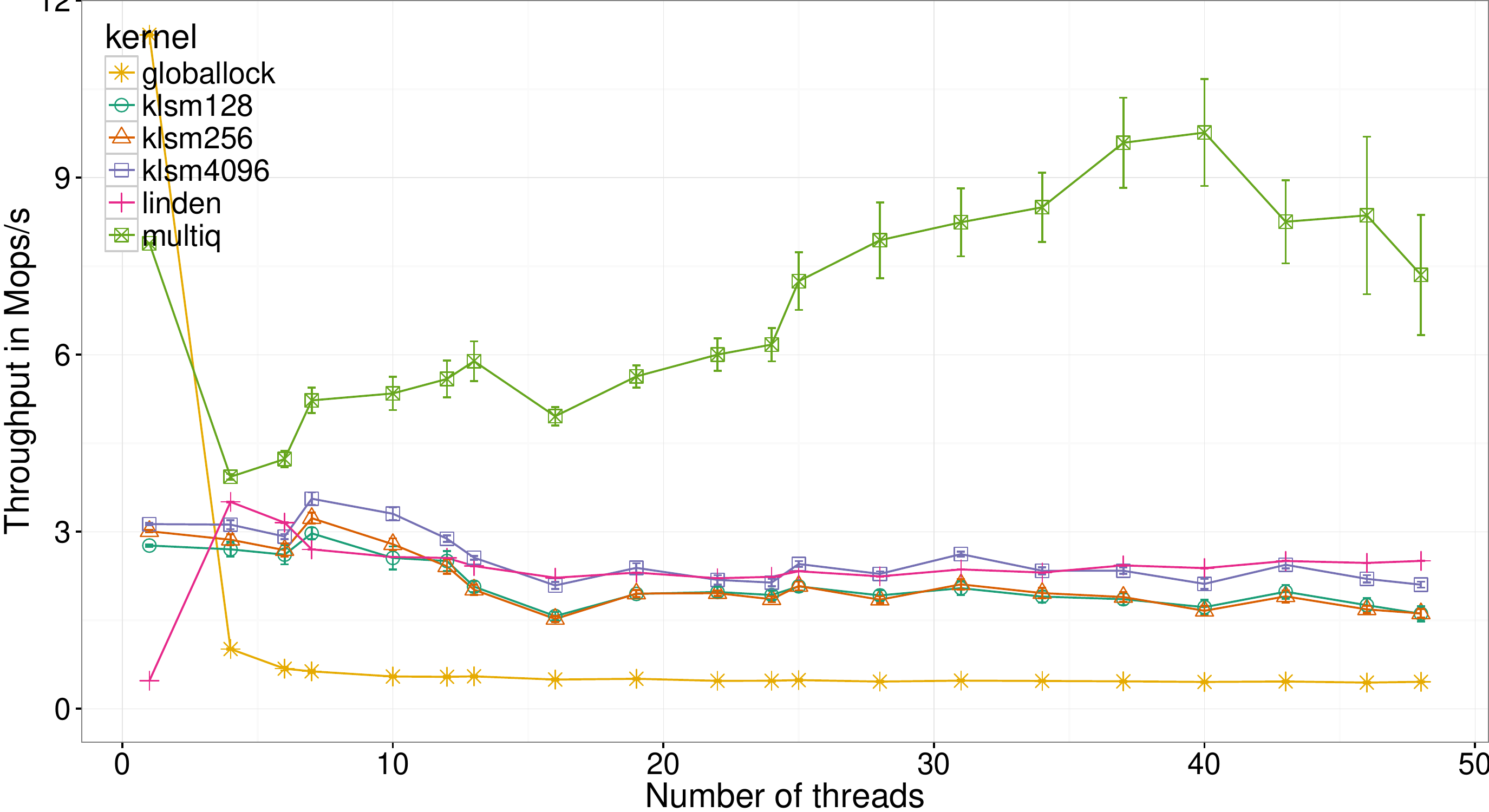}
\subcaption{Split workload, uniform keys (32 bits).}
\end{minipage}

\begin{minipage}{\columnwidth}
\centering
\includegraphics{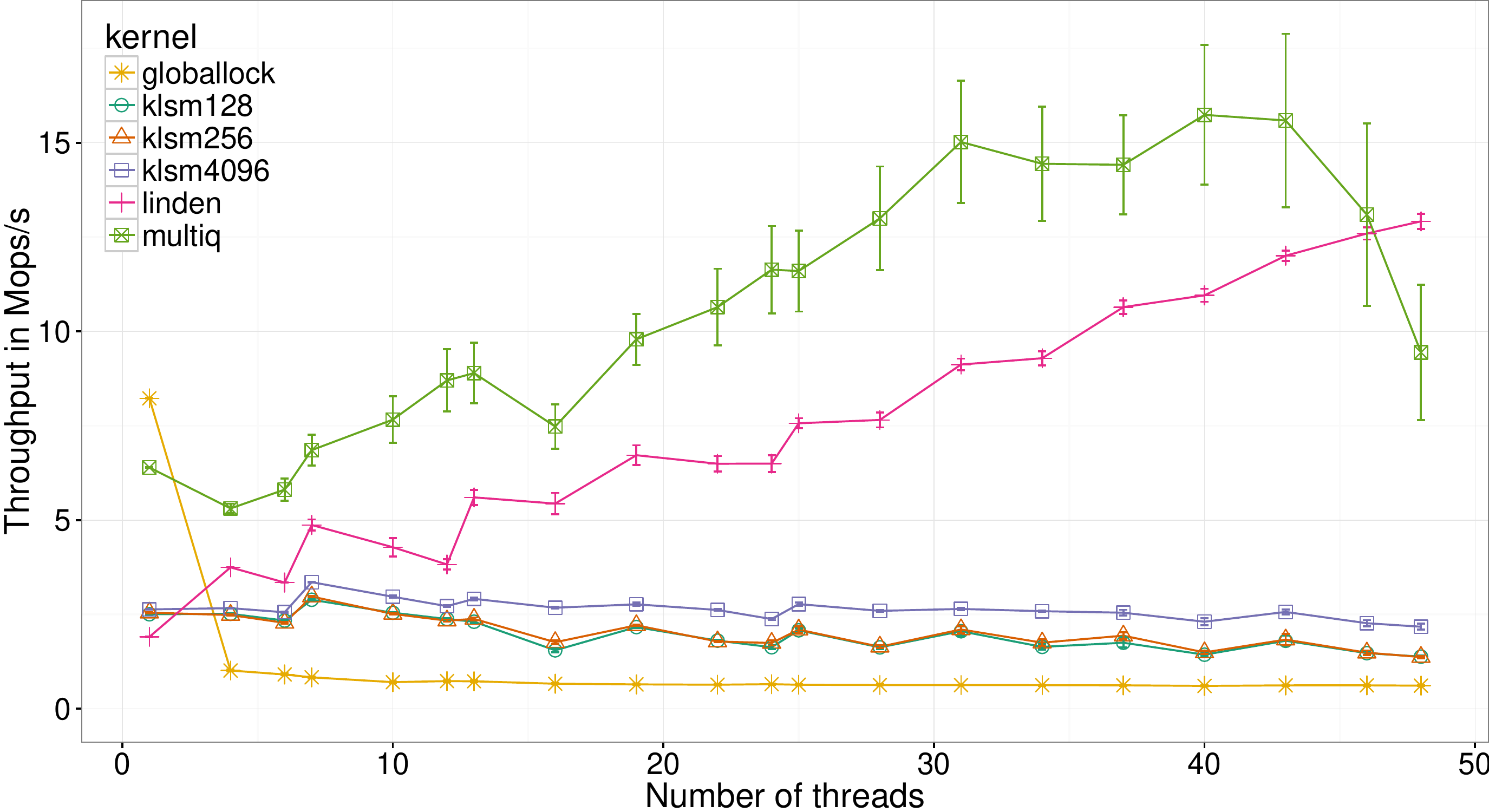}
\subcaption{Split workload, ascending keys.}
\label{fig:saturn_spl_asc}
\end{minipage}~%
\begin{minipage}{\columnwidth}
\centering
\includegraphics{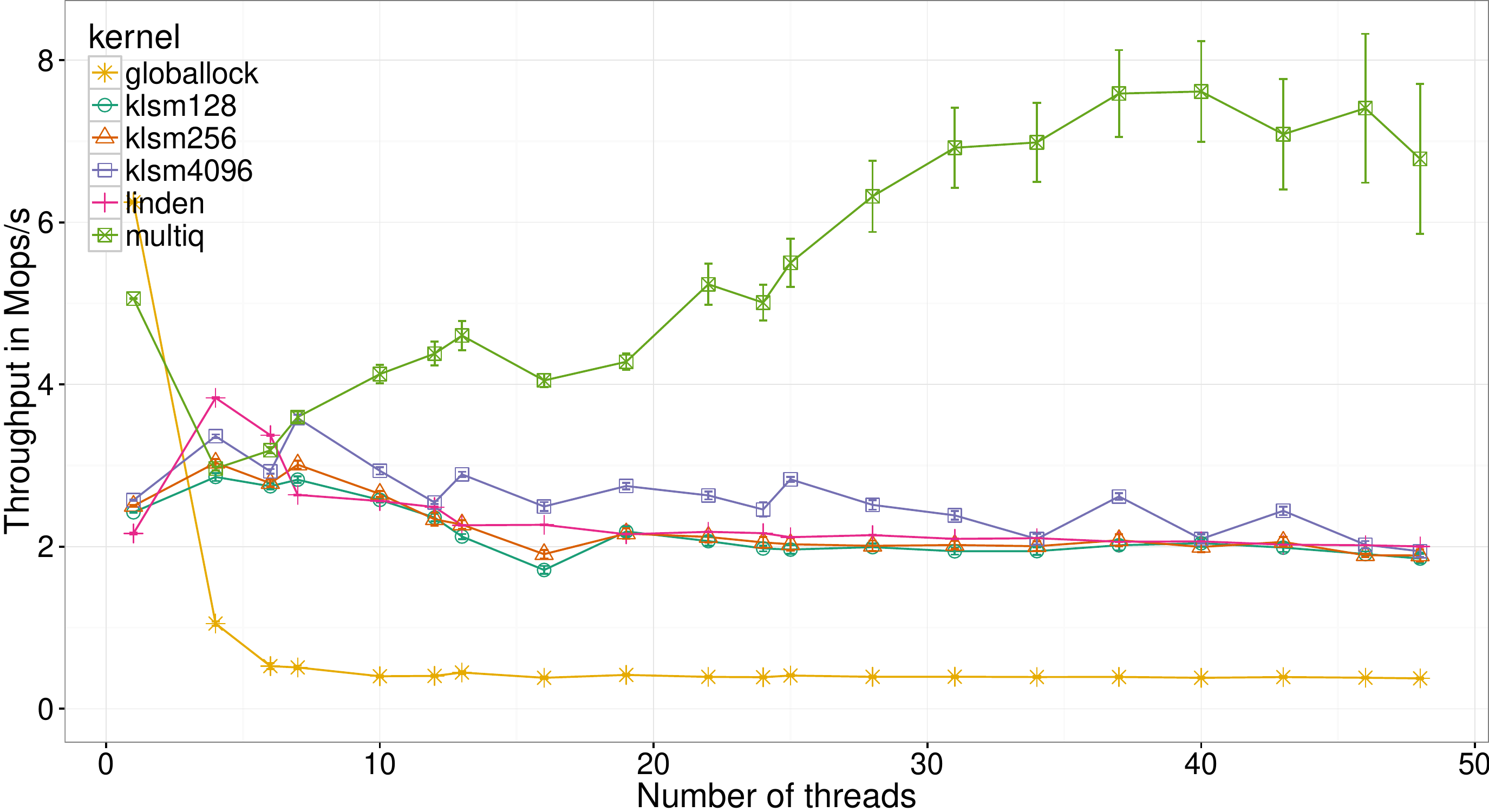}
\subcaption{Split workload, descending keys.}
\end{minipage}

\begin{minipage}{\columnwidth}
\centering
\includegraphics{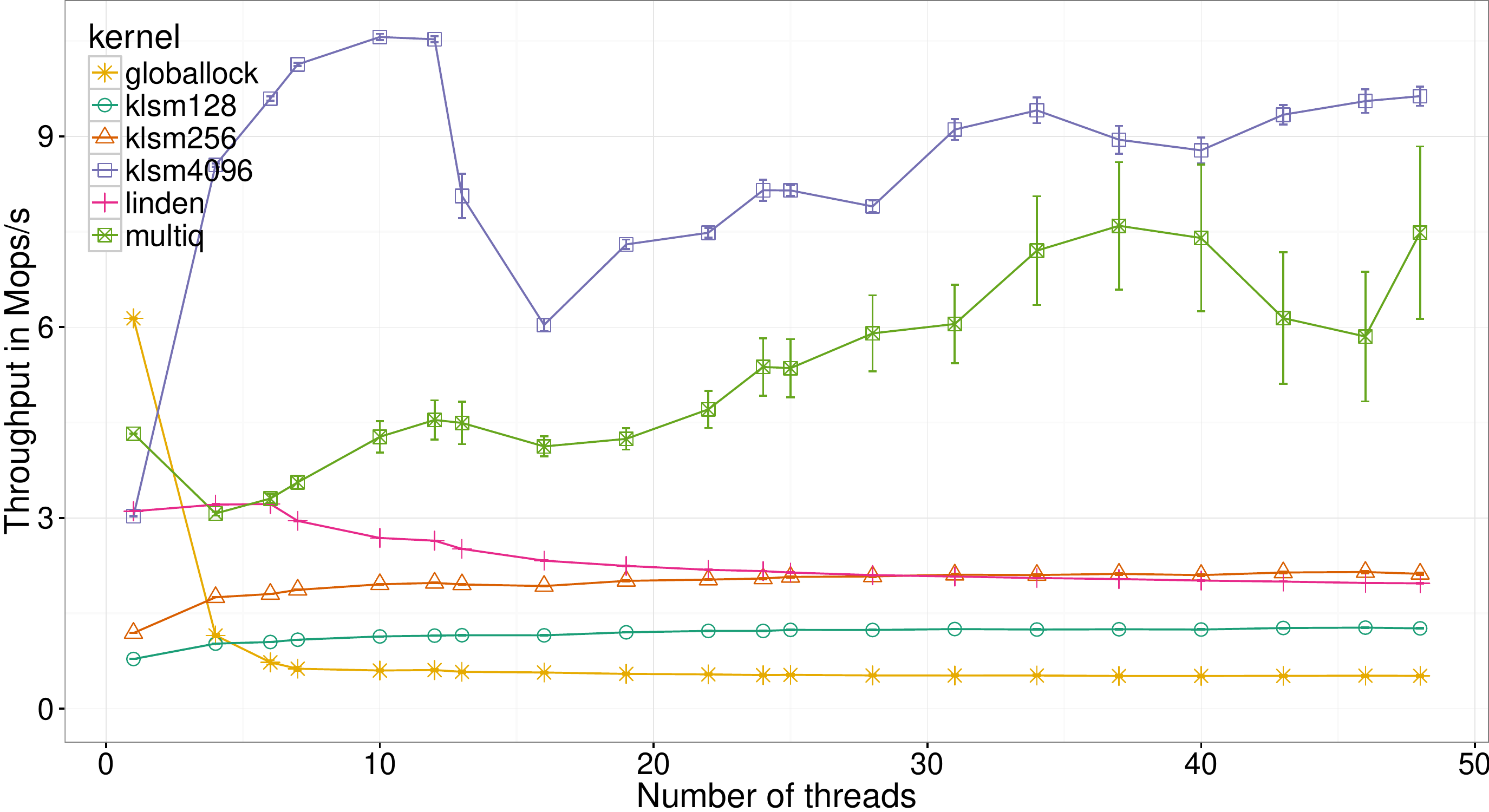}
\subcaption{Uniform workload, uniform keys (8 bits).}
\label{fig:saturn_uni_rst8}
\end{minipage}~%
\begin{minipage}{\columnwidth}
\centering
\includegraphics{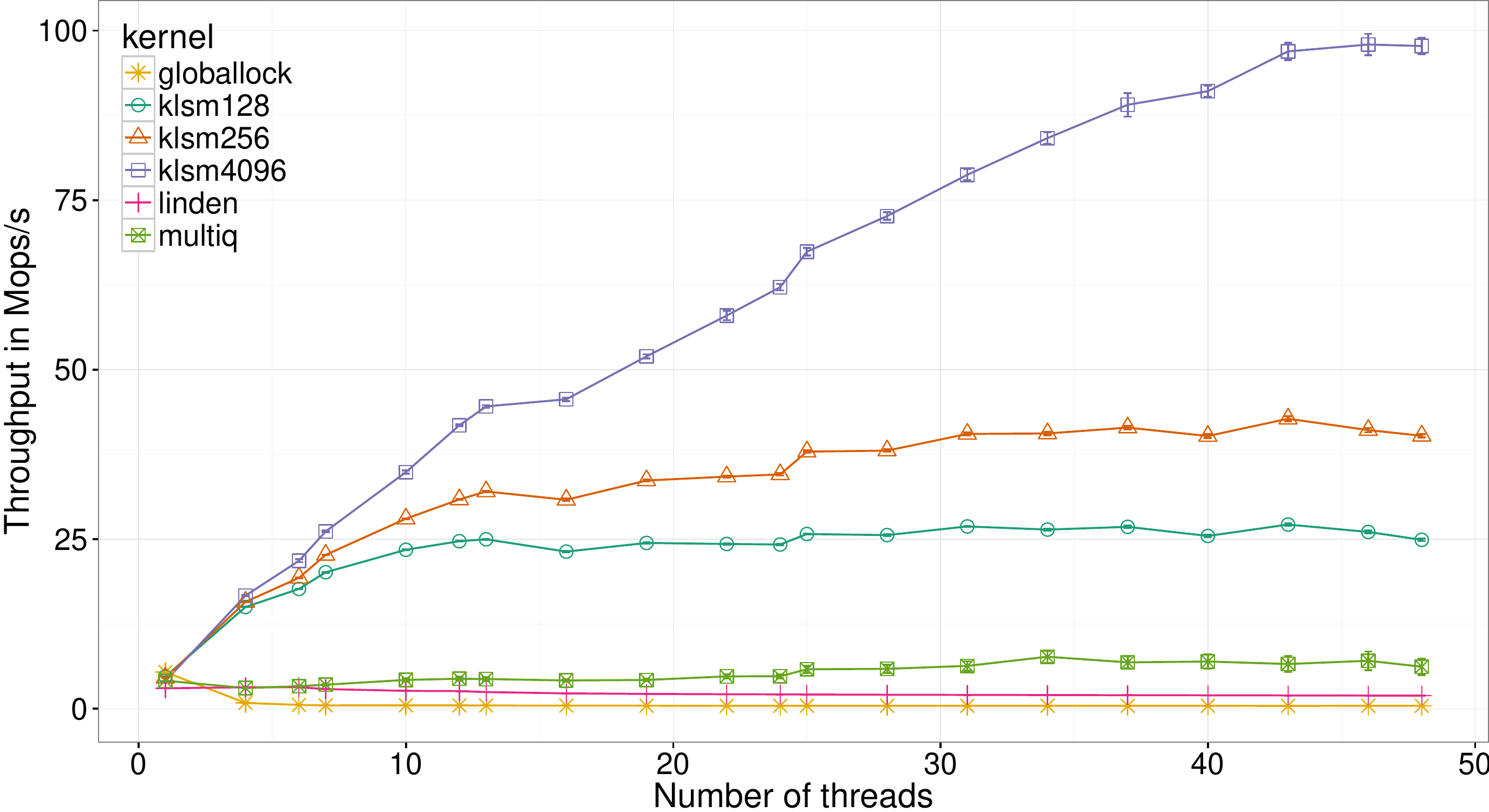}
\subcaption{Uniform workload, uniform keys (16 bits).}
\end{minipage}

\par \vspace{\belowdisplayskip} \vspace{\abovedisplayskip}
\caption{Throughput on \texttt{saturn}.}
\label{fig:saturn}
\end{figure*}

Figure~\ref{fig:saturn} shows results for our AMD Opteron machine
called \texttt{saturn}. Here, MultiQueue has fairly disappointing
throughput, barely achieving the sequential performance of
\texttt{globallock}, and only substantially exceeding it under split
workload and ascending key distribution
(Figure~\ref{fig:saturn_spl_asc}). Surprisingly, with keys restricted
to the 8-bit range (Figure~\ref{fig:saturn_uni_rst8}), the
\texttt{linden} queue has a higher throughput than all other data
structures.  Quality trends in Table~\ref{tbl:saturn} are
consistent with those on \texttt{mars}.

\begin{figure*}[ht]
\centering
\begin{minipage}{\columnwidth}
\includegraphics{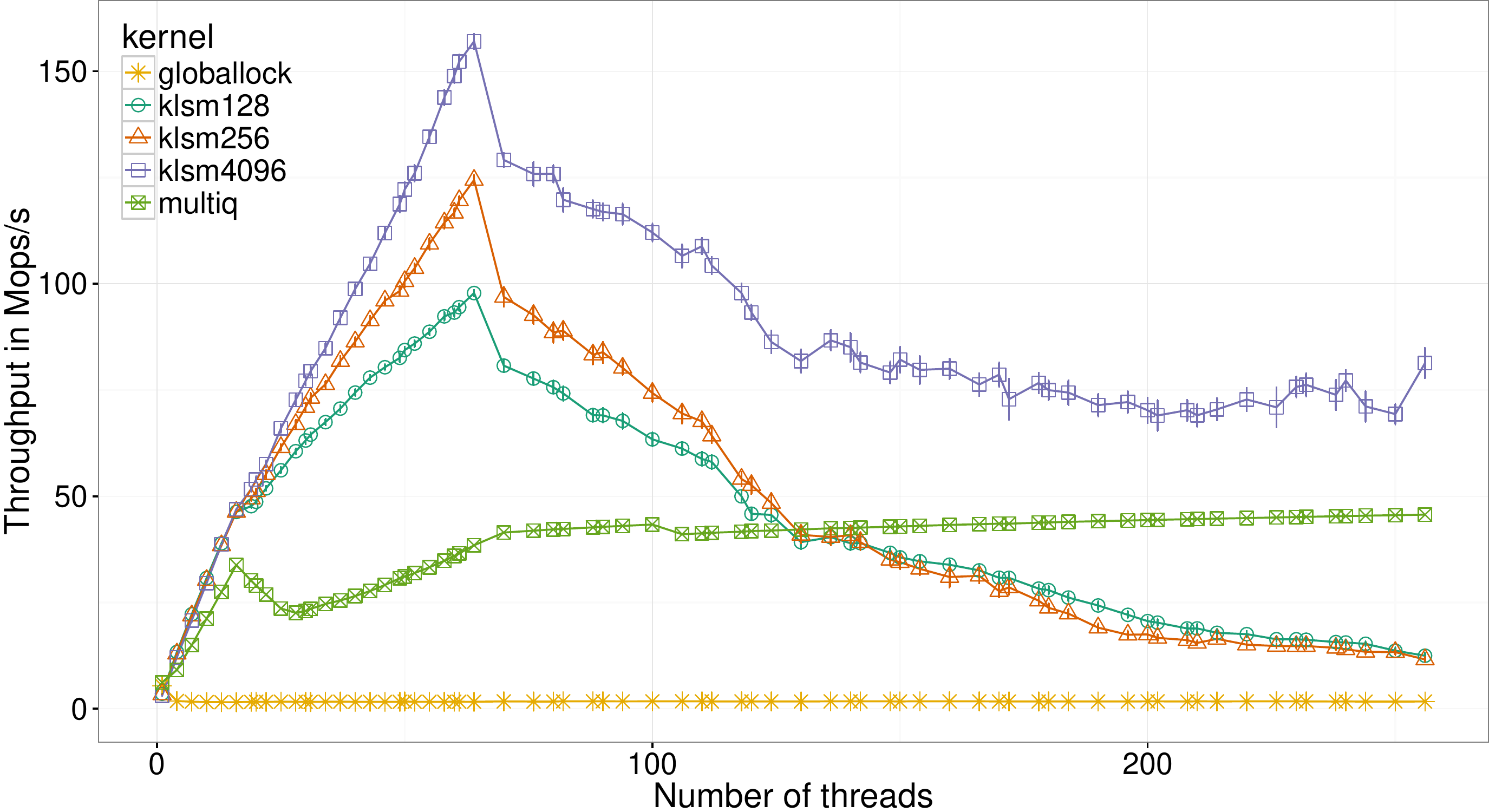}
\subcaption{Uniform workload, uniform keys (32 bits).}
\end{minipage}~%
\begin{minipage}{\columnwidth}
\centering
\includegraphics{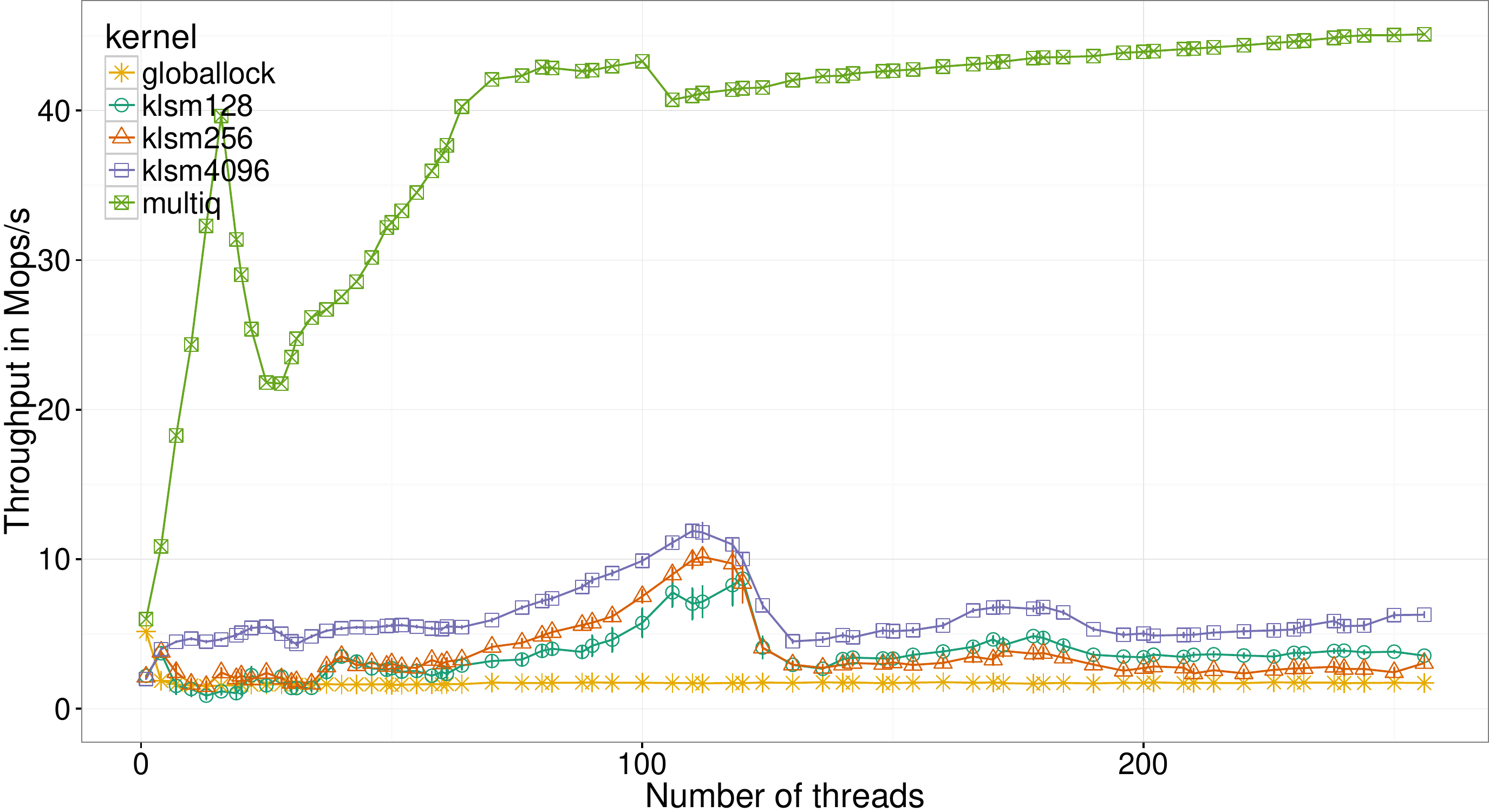}
\subcaption{Uniform workload, ascending keys.}
\end{minipage}

\begin{minipage}{\columnwidth}
\centering
\includegraphics{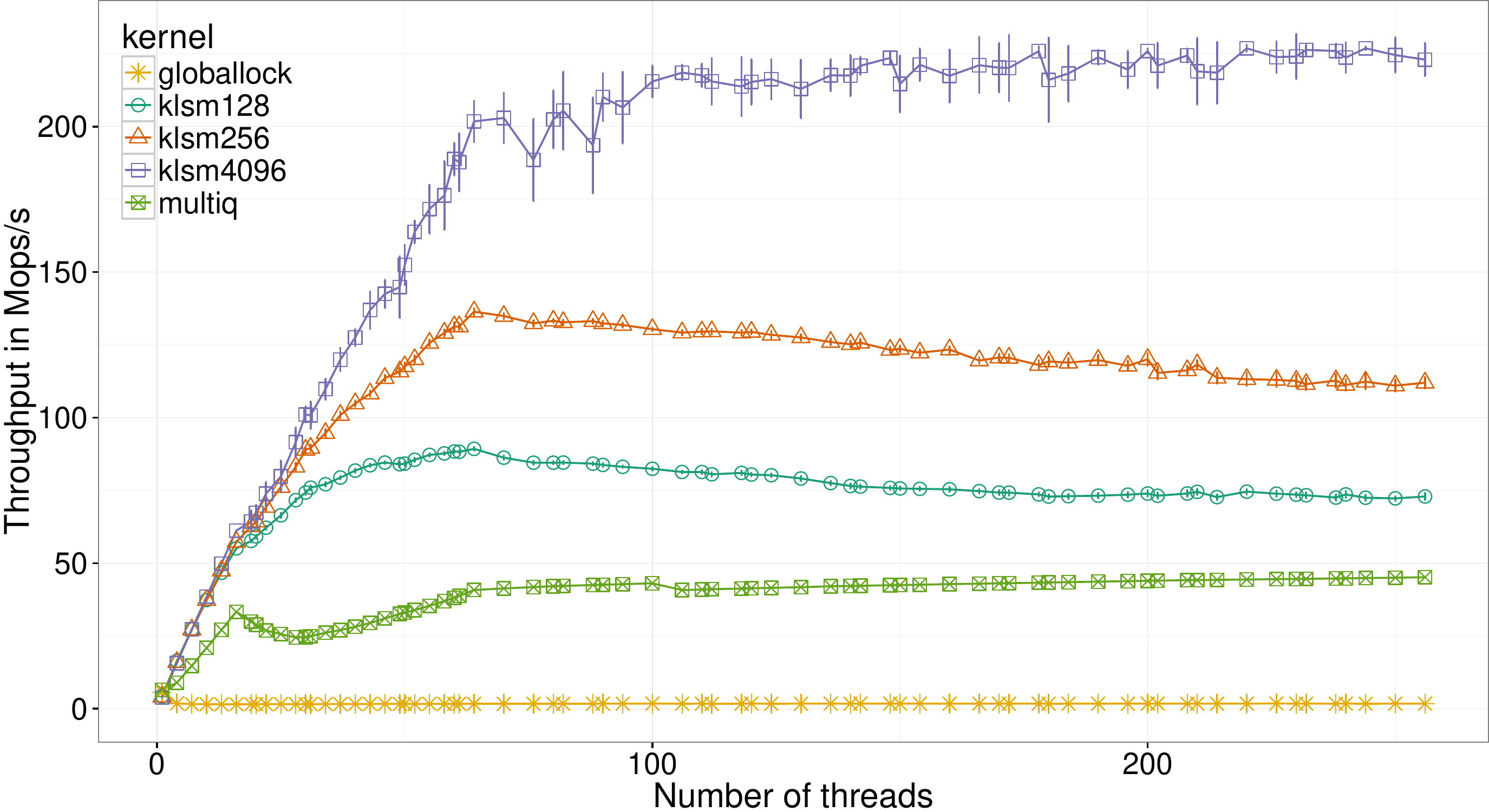}
\subcaption{Uniform workload, descending keys.}
\end{minipage}~%
\begin{minipage}{\columnwidth}
\centering
\includegraphics{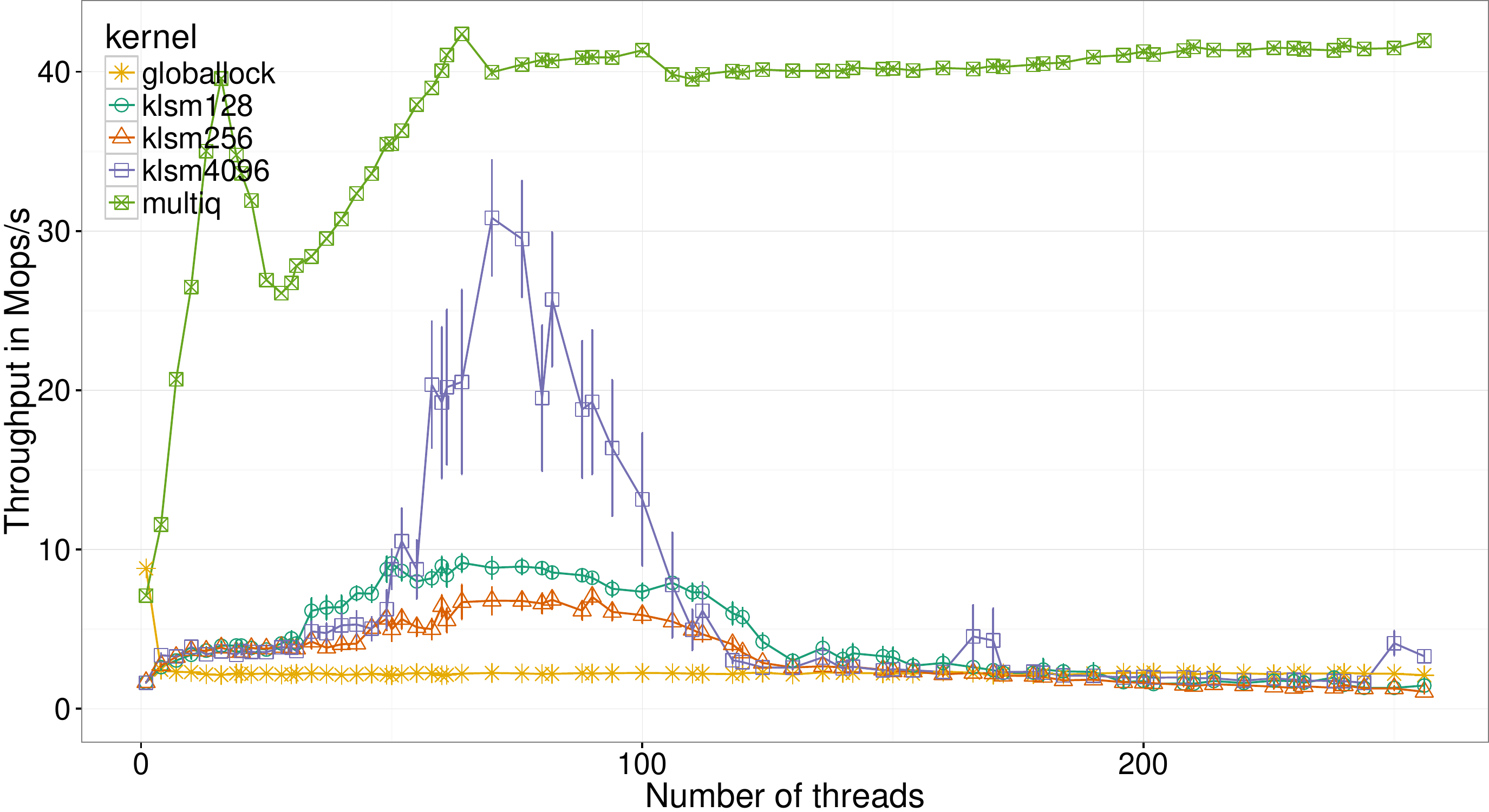}
\subcaption{Split workload, uniform keys (32 bits).}
\label{fig:ceres_spl_uni}
\end{minipage}

\begin{minipage}{\columnwidth}
\centering
\includegraphics{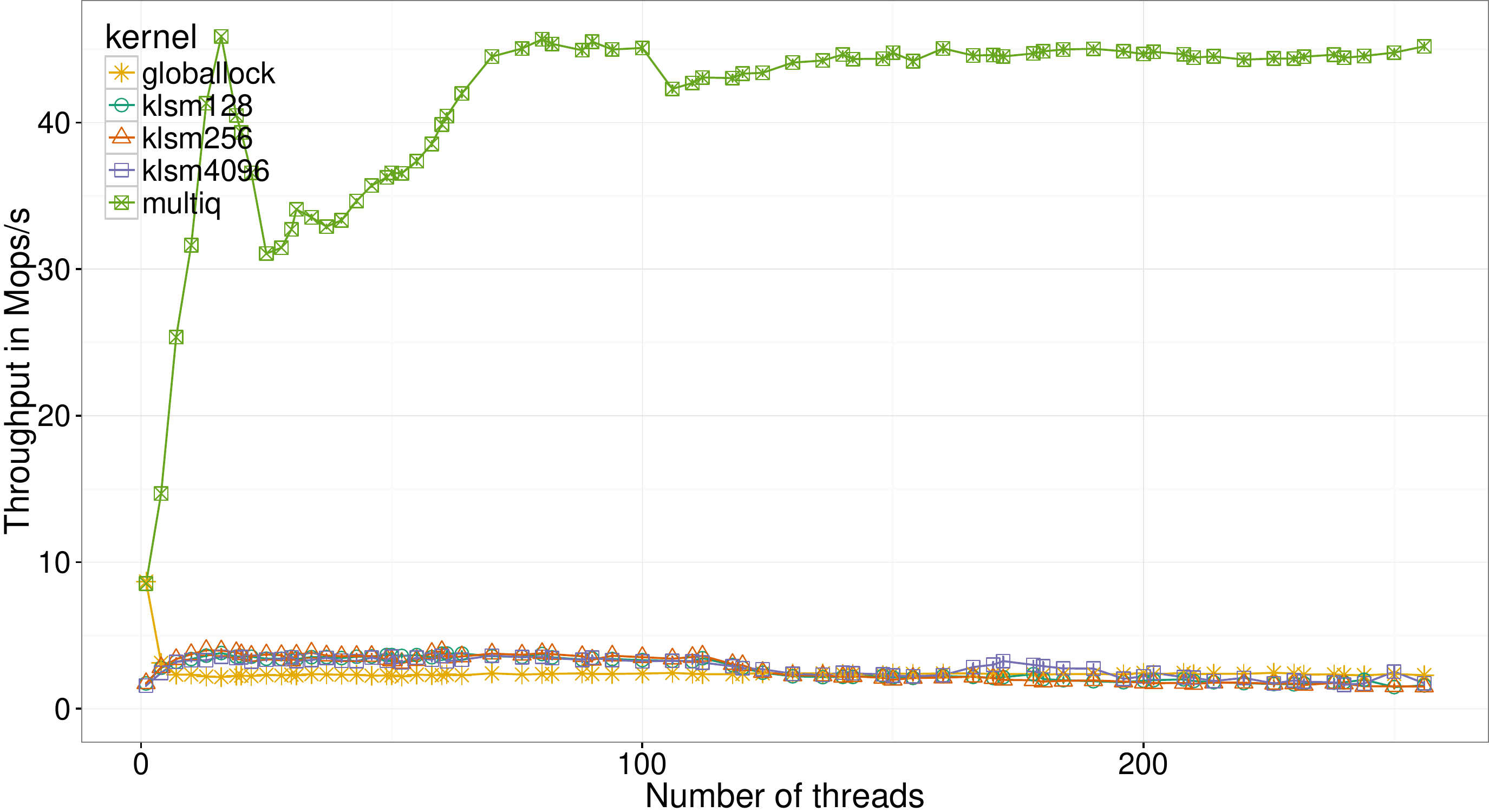}
\subcaption{Split workload, ascending keys.}
\end{minipage}~%
\begin{minipage}{\columnwidth}
\centering
\includegraphics{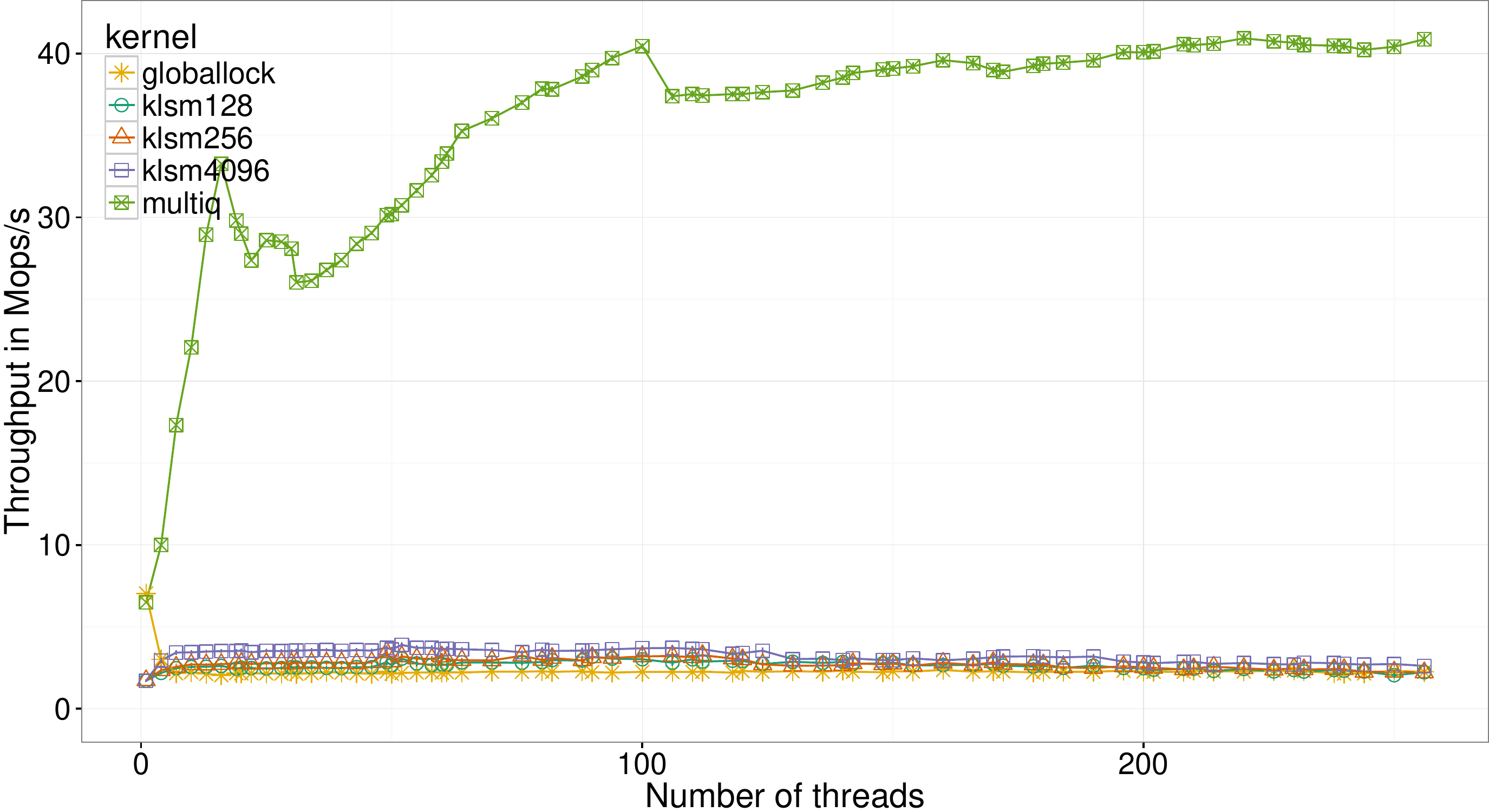}
\subcaption{Split workload, descending keys.}
\end{minipage}

\begin{minipage}{\columnwidth}
\centering
\includegraphics{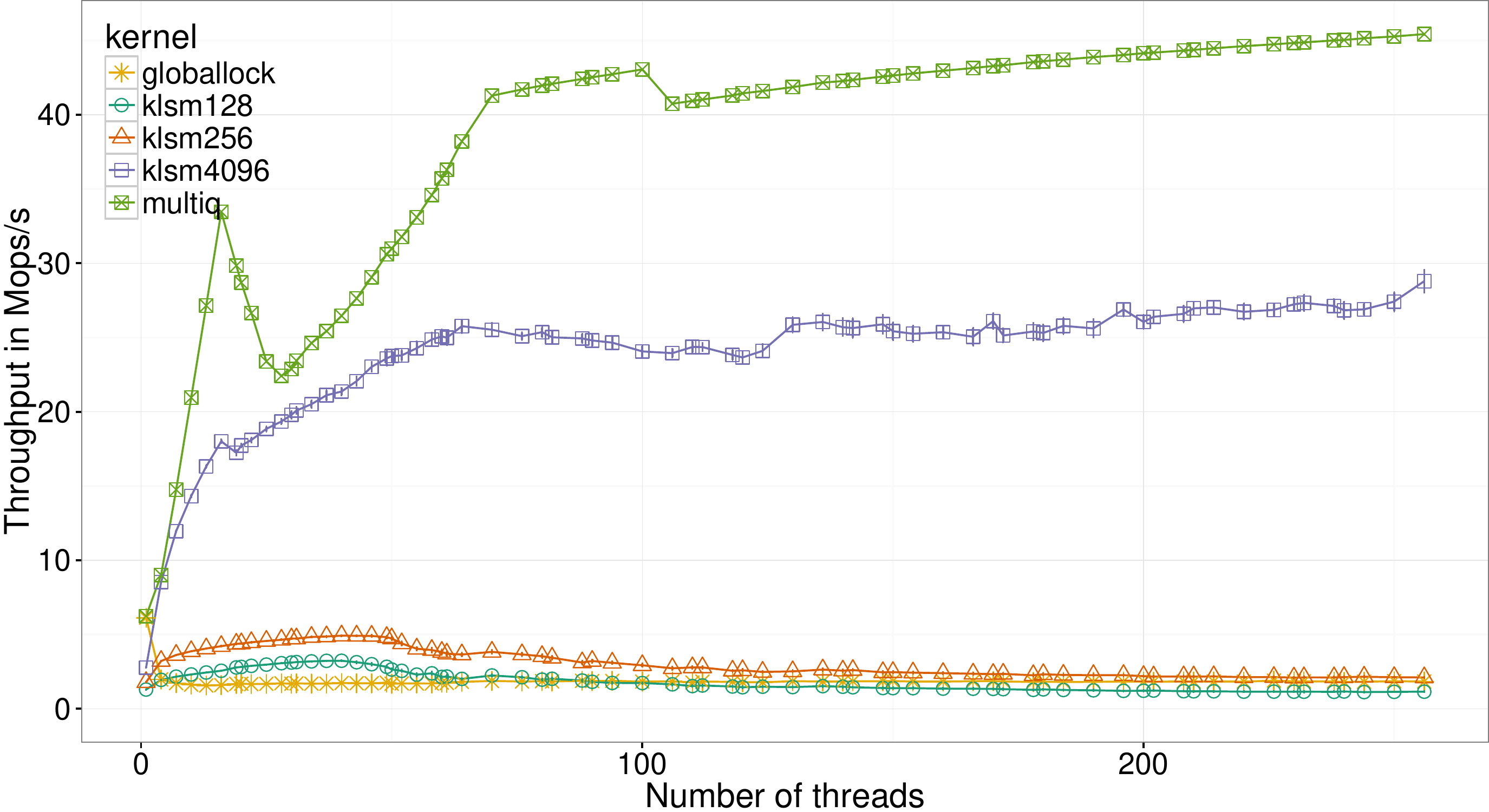}
\subcaption{Uniform workload, uniform keys (8 bits).}
\end{minipage}~%
\begin{minipage}{\columnwidth}
\centering
\includegraphics{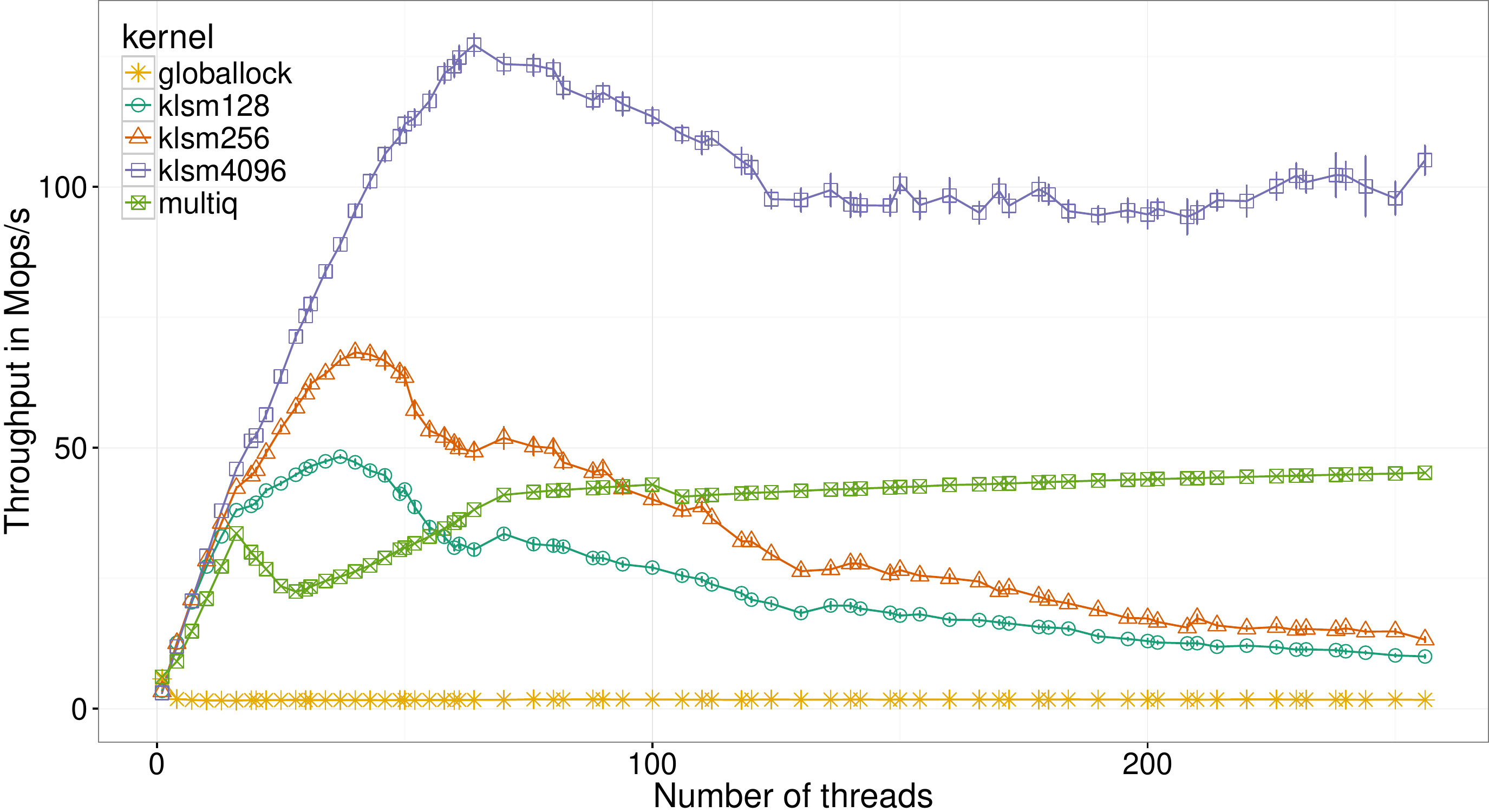}
\subcaption{Uniform workload, uniform keys (16 bits).}
\end{minipage}

\par \vspace{\belowdisplayskip} \vspace{\abovedisplayskip}
\caption{Throughput on \texttt{ceres}.}
\label{fig:ceres}
\end{figure*}

Figure~\ref{fig:ceres} and Table~\ref{tbl:ceres} display throughput
and quality results for \texttt{ceres}. On this machine, we display
results for up to 4-way hyperthreading. As previously on
\texttt{mars}, throughput of tested algorithms does not benefit from
hyperthreading in general. Only \texttt{multiq} appears to
consistently gain further (small) increases at thread counts over 64.
Split workload combined with uniform key distribution
(Figure~\ref{fig:ceres_spl_uni}) causes a local maximum of around 30
\ac{MOps/s} for the \texttt{klsm4096}.

\begin{figure*}[ht]
\centering
\begin{minipage}{\columnwidth}
\includegraphics{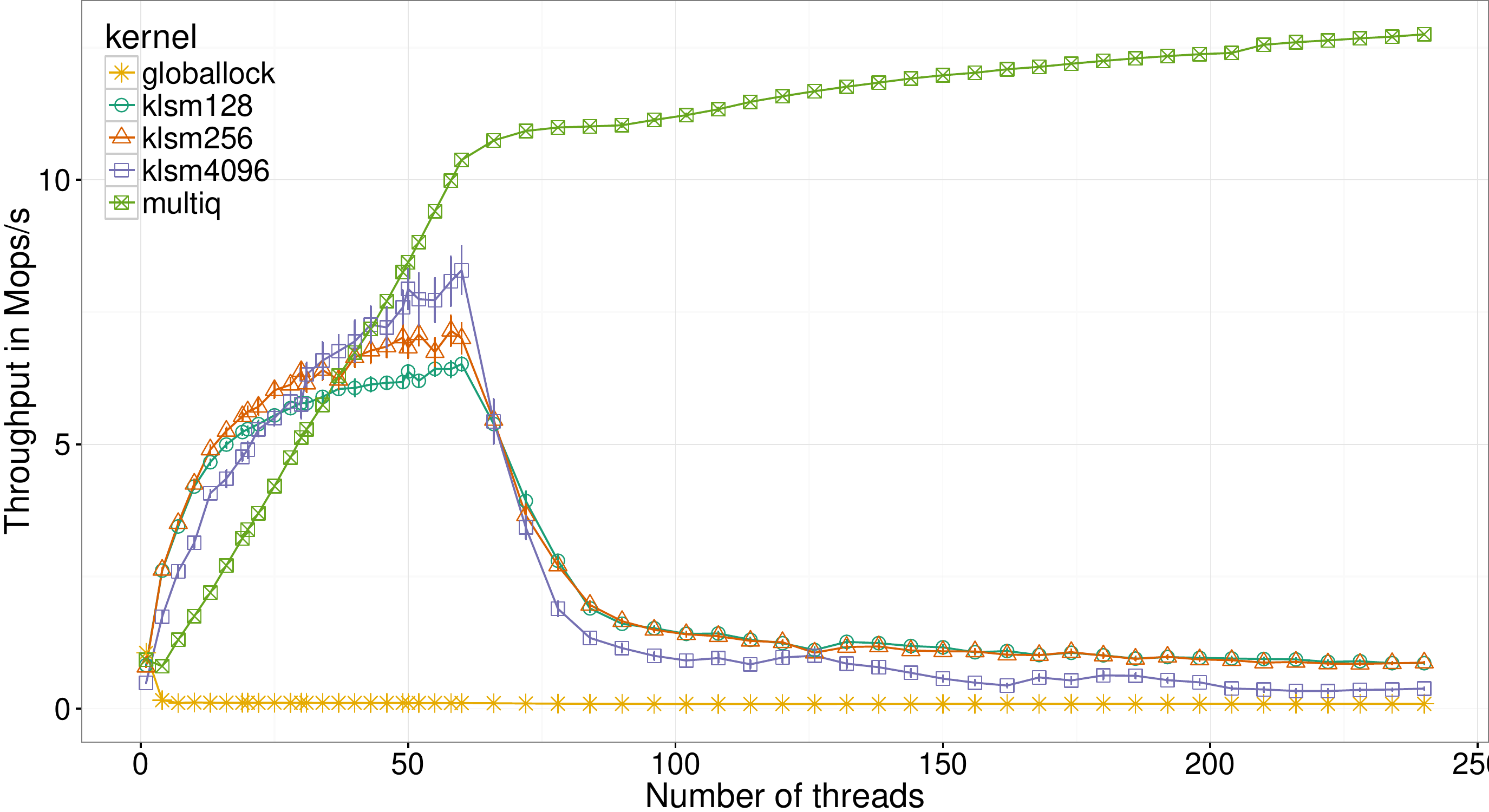}
\subcaption{Uniform workload, uniform keys (32 bits).}
\label{fig:pluto_uni_uni}
\end{minipage}~%
\begin{minipage}{\columnwidth}
\centering
\includegraphics{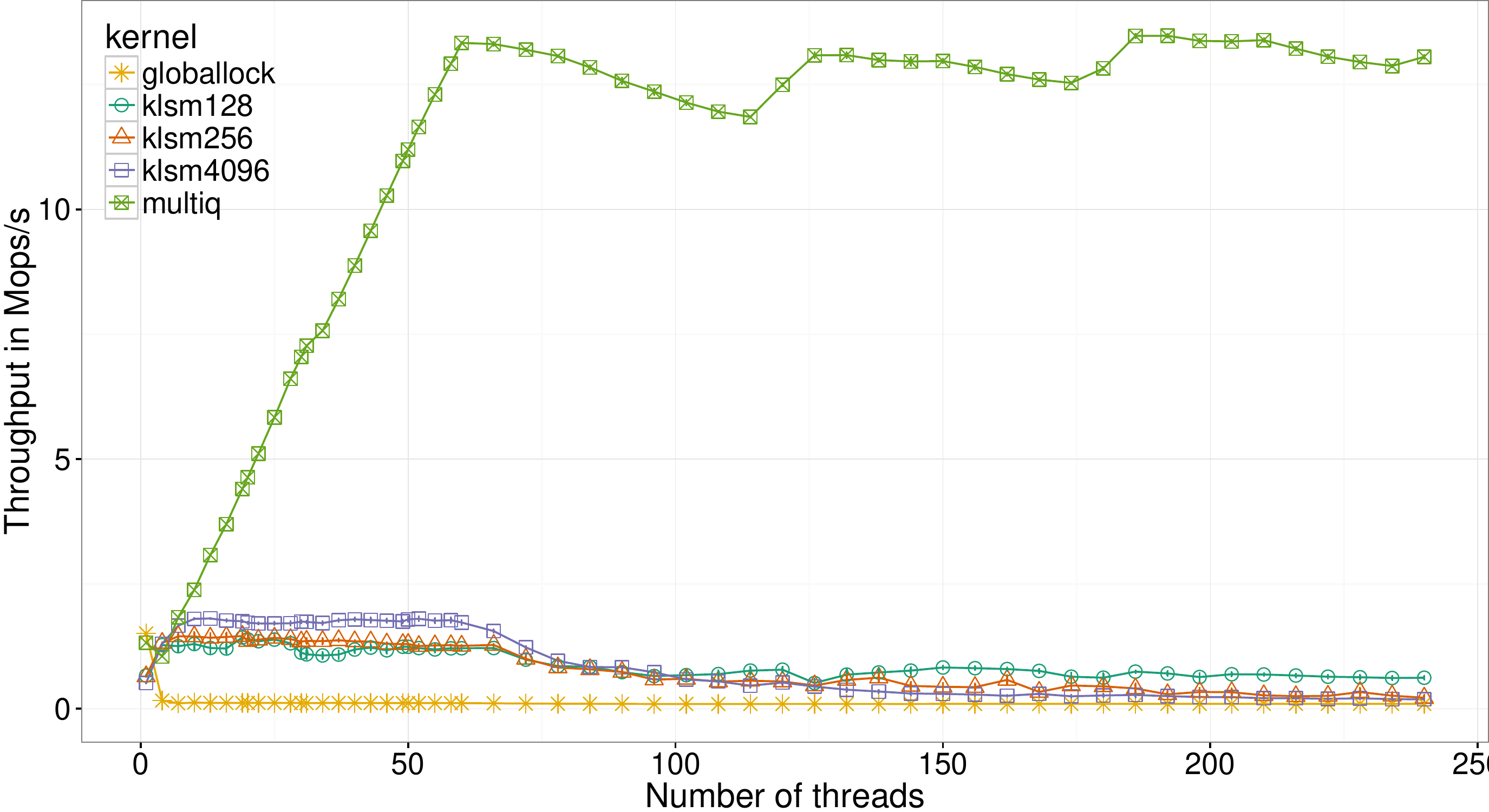}
\subcaption{Uniform workload, ascending keys.}
\end{minipage}

\begin{minipage}{\columnwidth}
\centering
\includegraphics{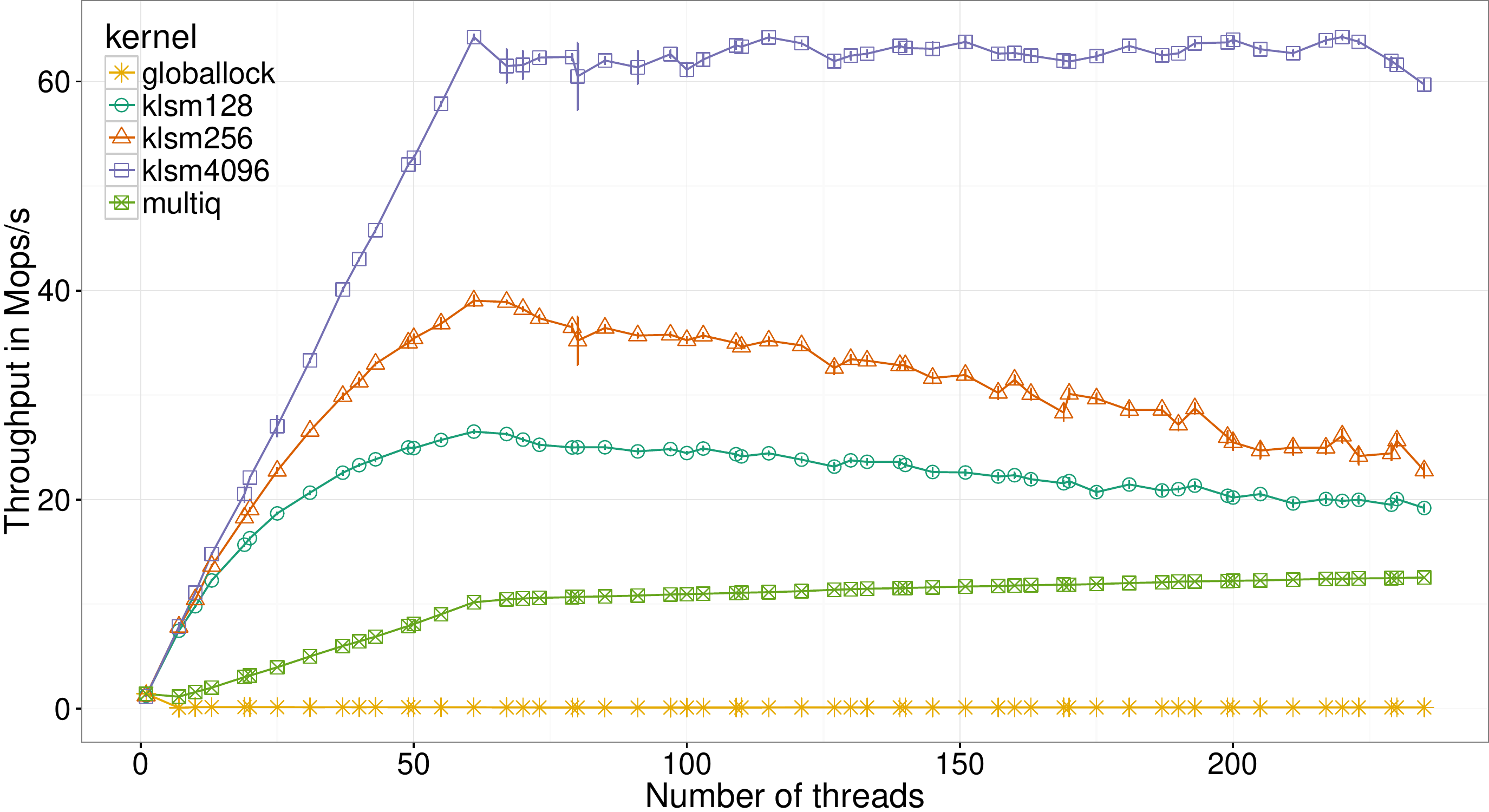}
\subcaption{Uniform workload, descending keys.}
\label{fig:pluto_uni_desc}
\end{minipage}~%
\begin{minipage}{\columnwidth}
\centering
\includegraphics{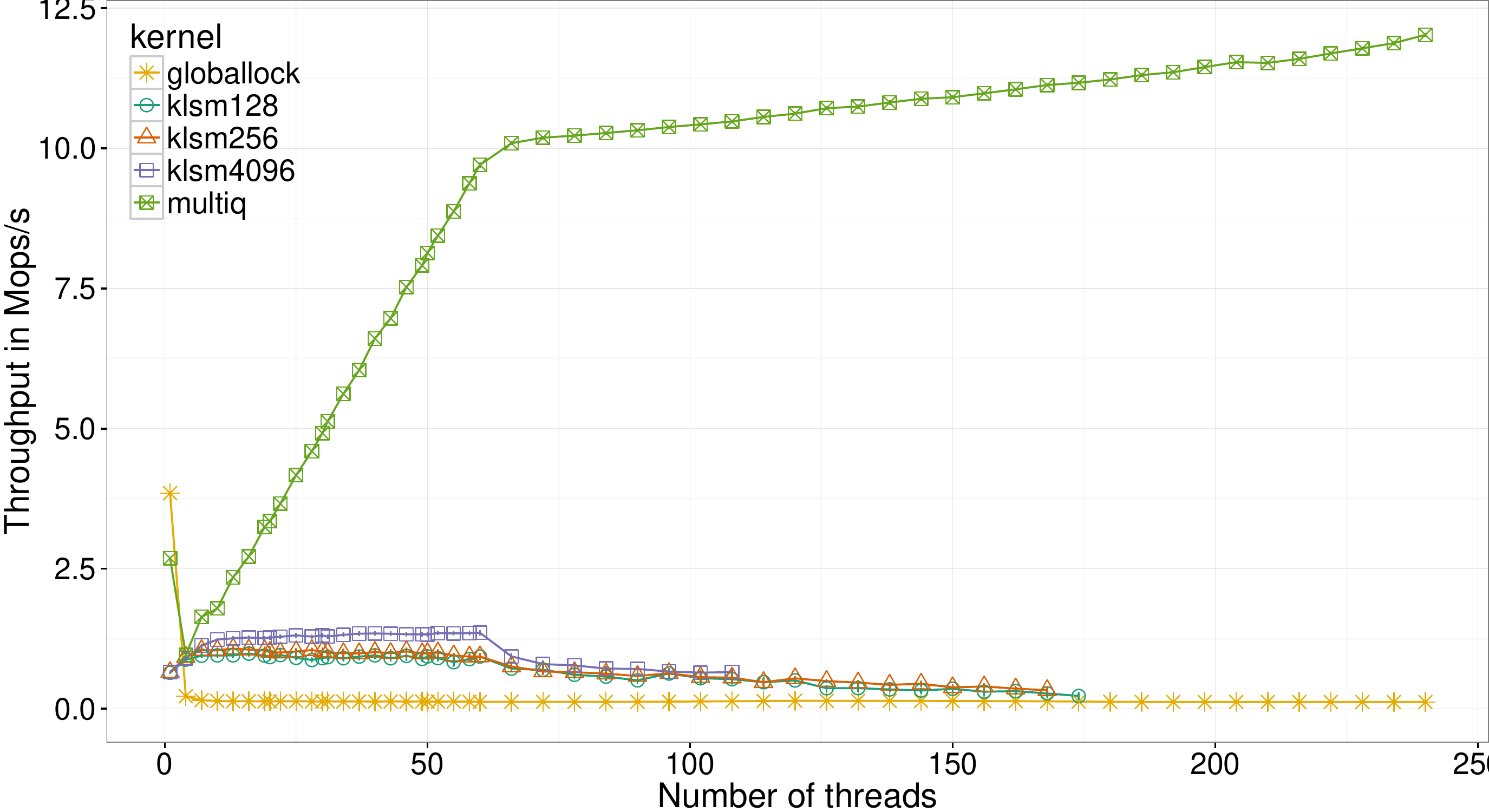}
\subcaption{Split workload, uniform keys (32 bits).}
\end{minipage}

\begin{minipage}{\columnwidth}
\centering
\includegraphics{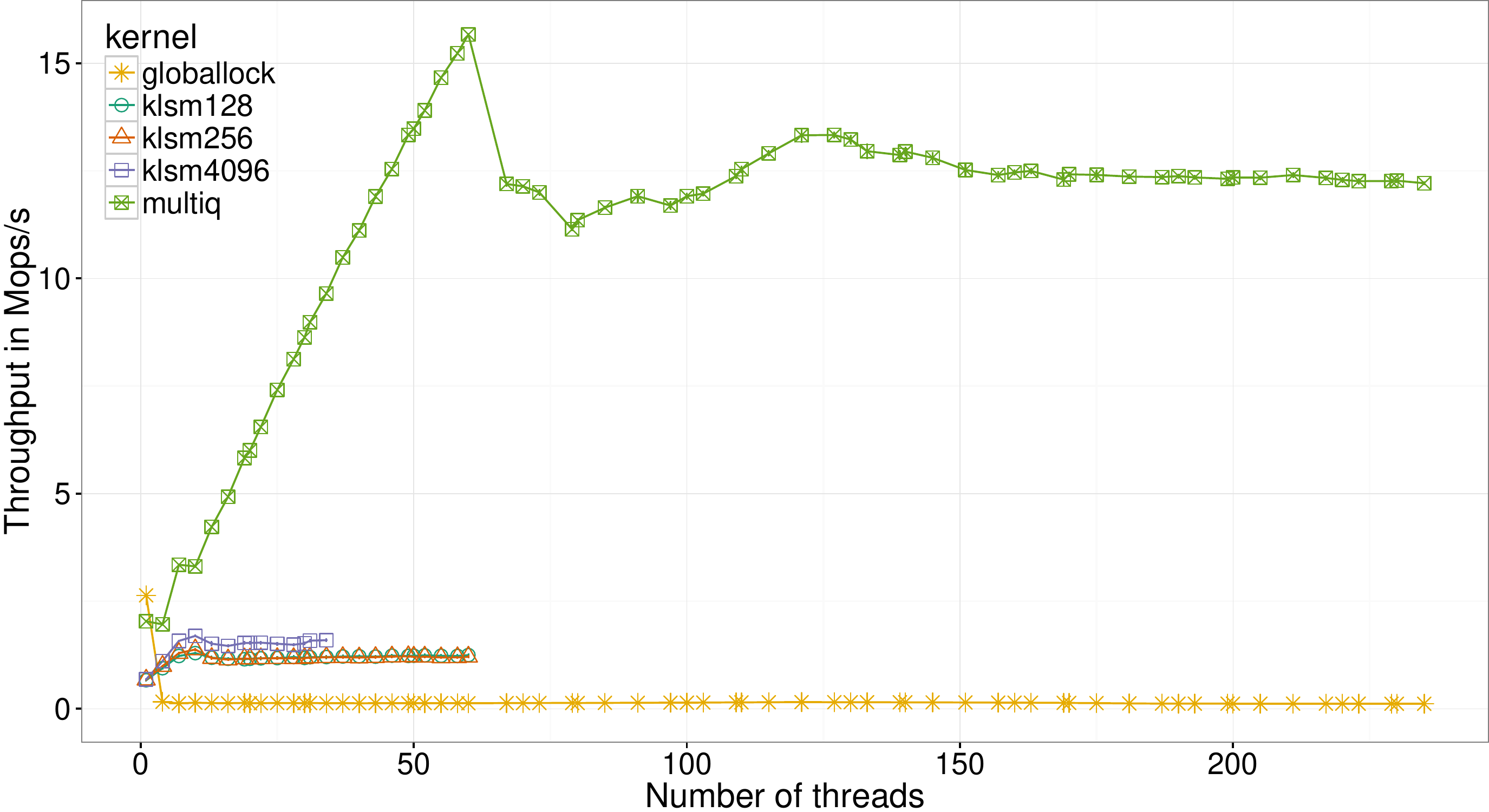}
\subcaption{Split workload, ascending keys.}
\end{minipage}~%
\begin{minipage}{\columnwidth}
\centering
\includegraphics{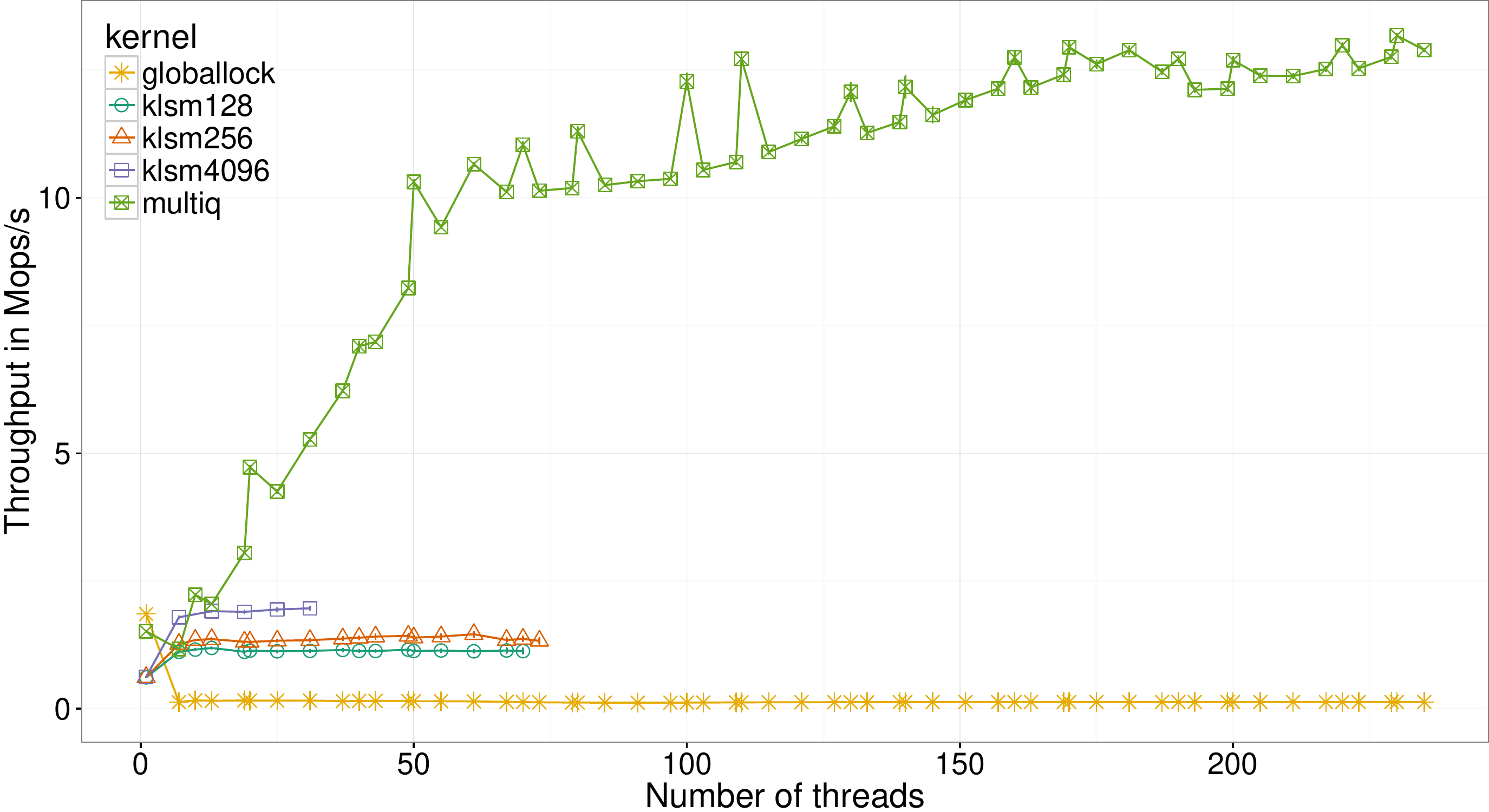}
\subcaption{Split workload, descending keys.}
\end{minipage}

\begin{minipage}{\columnwidth}
\centering
\includegraphics{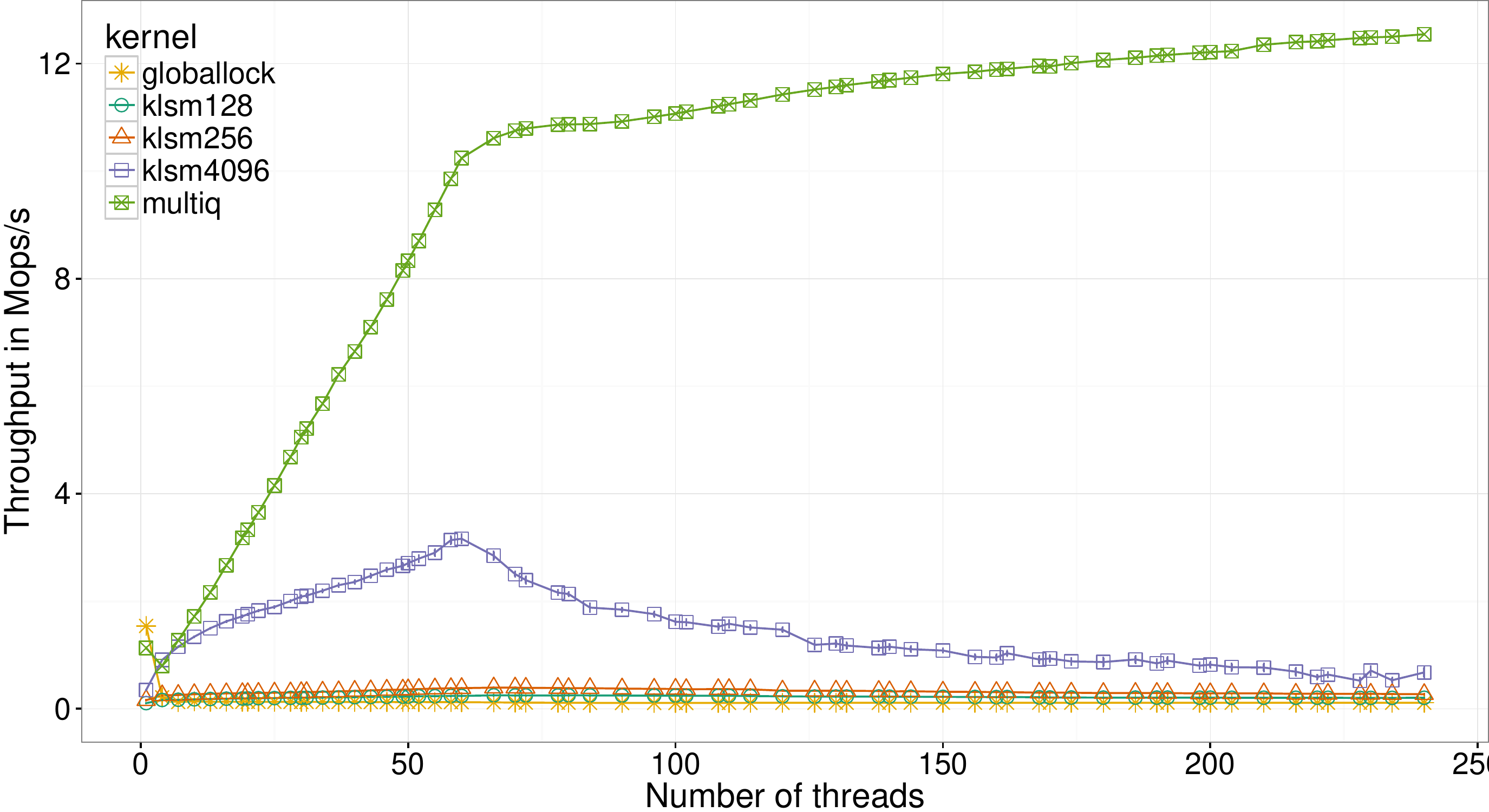}
\subcaption{Uniform workload, uniform keys (8 bits).}
\end{minipage}~%
\begin{minipage}{\columnwidth}
\centering
\includegraphics{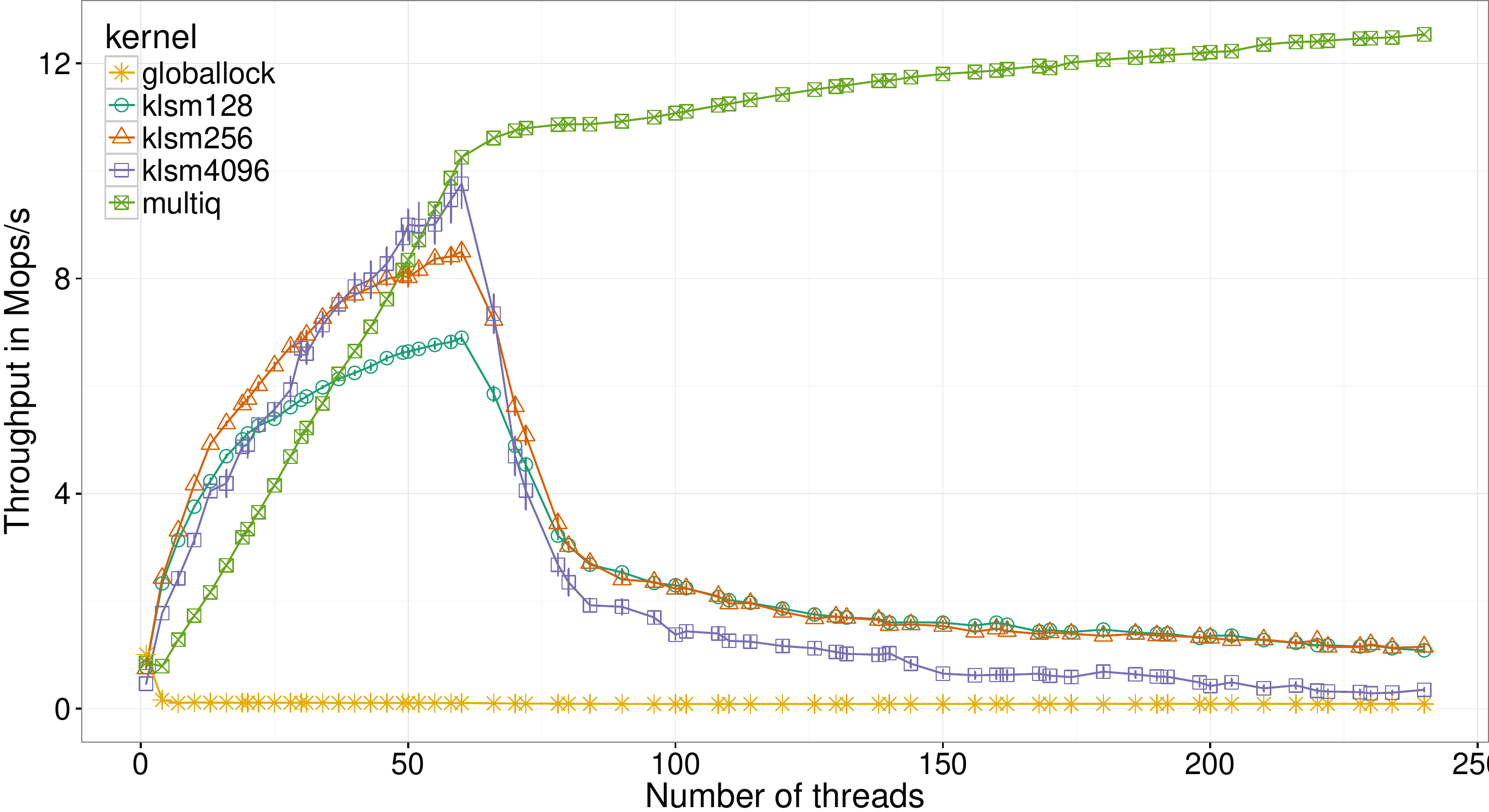}
\subcaption{Uniform workload, uniform keys (16 bits).}
\end{minipage}

\par \vspace{\belowdisplayskip} \vspace{\abovedisplayskip}
\caption{Throughput on \texttt{pluto}.}
\label{fig:pluto}
\end{figure*}

Figure~\ref{fig:pluto} shows throughput results for our Xeon
Phi machine \texttt{pluto} (there is no corresponding quality table,
since we did not run our quality benchmark on this machine).  On
\texttt{pluto}, the \klsm does not match the trend for very high
throughput as previously exhibited for the uniform workload and
uniform key benchmark
(Figure~\ref{fig:pluto_uni_uni}). While scalability is decent up to
the physical core count of 61, its absolute performance is exceeded by
the \texttt{multiq} at higher thread counts. Only in descending key
distribution (Figure~\ref{fig:pluto_uni_desc}) does the \klsm reach
previous heights. Note that this is also the only benchmark in which
throughput on \texttt{pluto} exceeds roughly 15 \ac{MOps/s}. Unfortunately,
the \klsm was unstable at higher thread counts under split workloads and
we omit all data where it is not completely reliable. Again,
hyperthreading results in modest gains for \texttt{multiq}, and in
stagnant performance for all other data structures in the best case.

\begin{figure*}[ht]
\centering
\begin{minipage}{\columnwidth}
\includegraphics{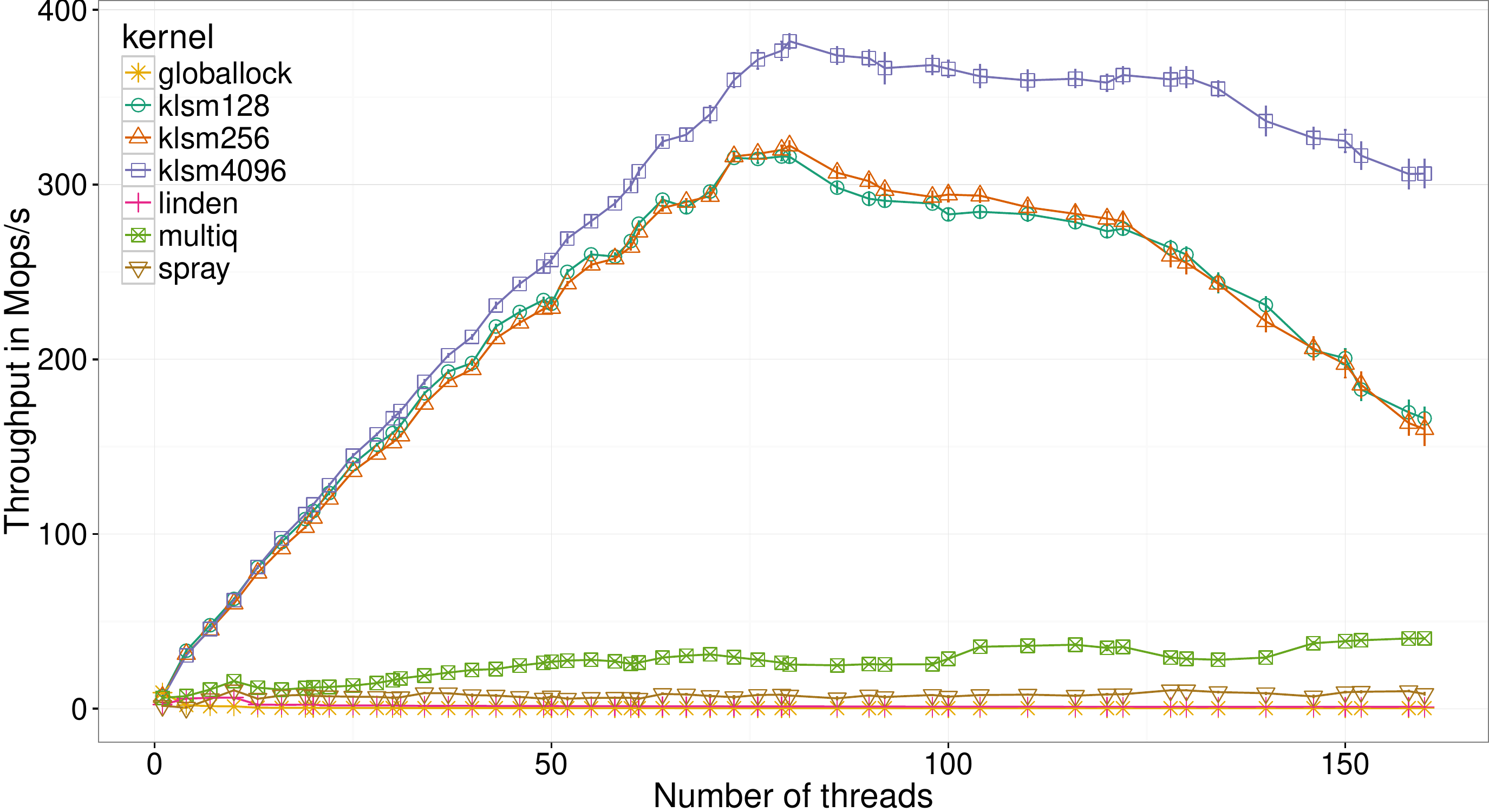}
\subcaption{\texttt{mars}, uniform keys (32 bits).}
\label{fig:mars_alt_uni}
\end{minipage}~%
\begin{minipage}{\columnwidth}
\centering
\includegraphics{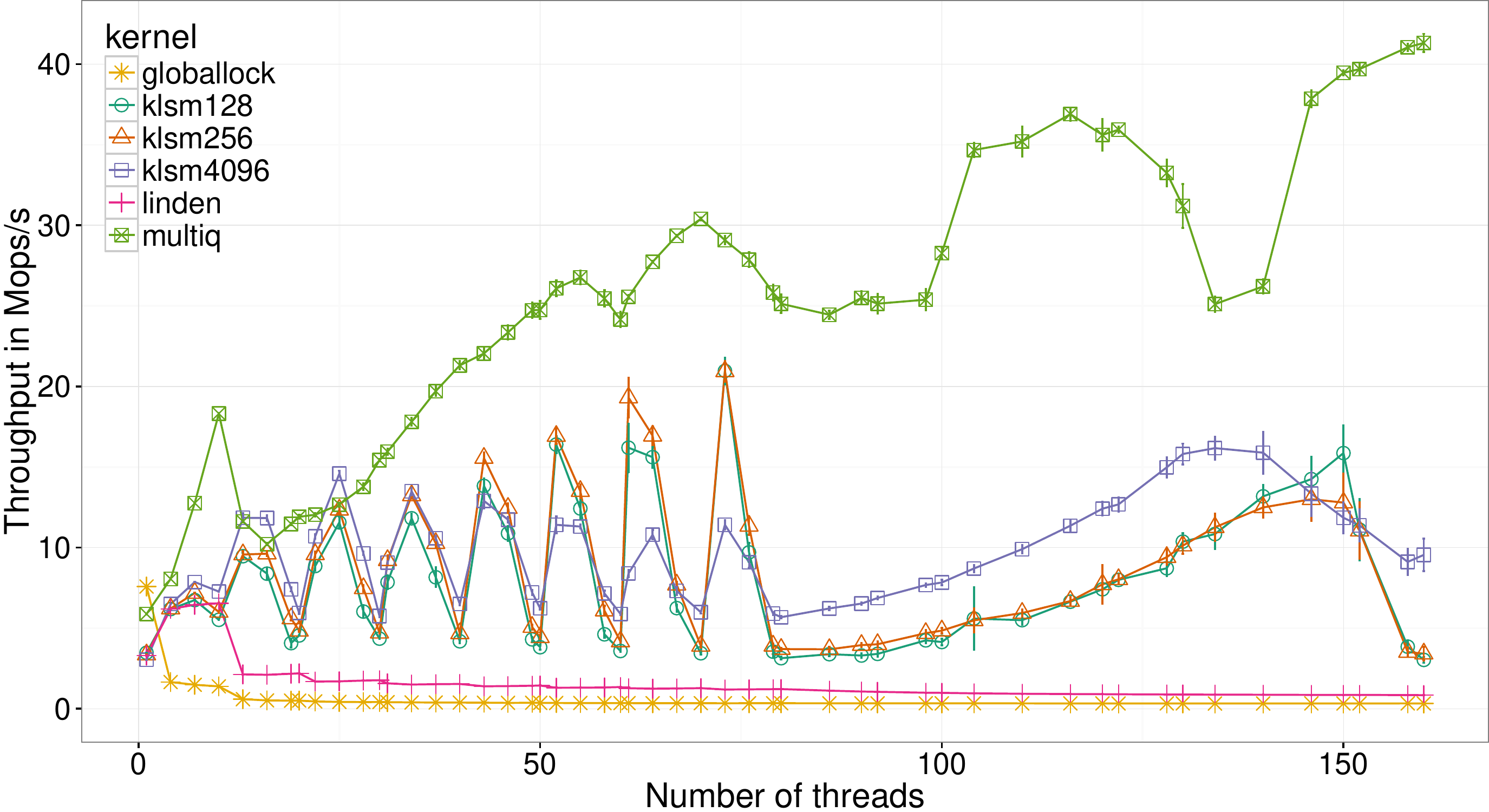}
\subcaption{\texttt{mars}, ascending keys.}
\end{minipage}

\begin{minipage}{\columnwidth}
\centering
\includegraphics{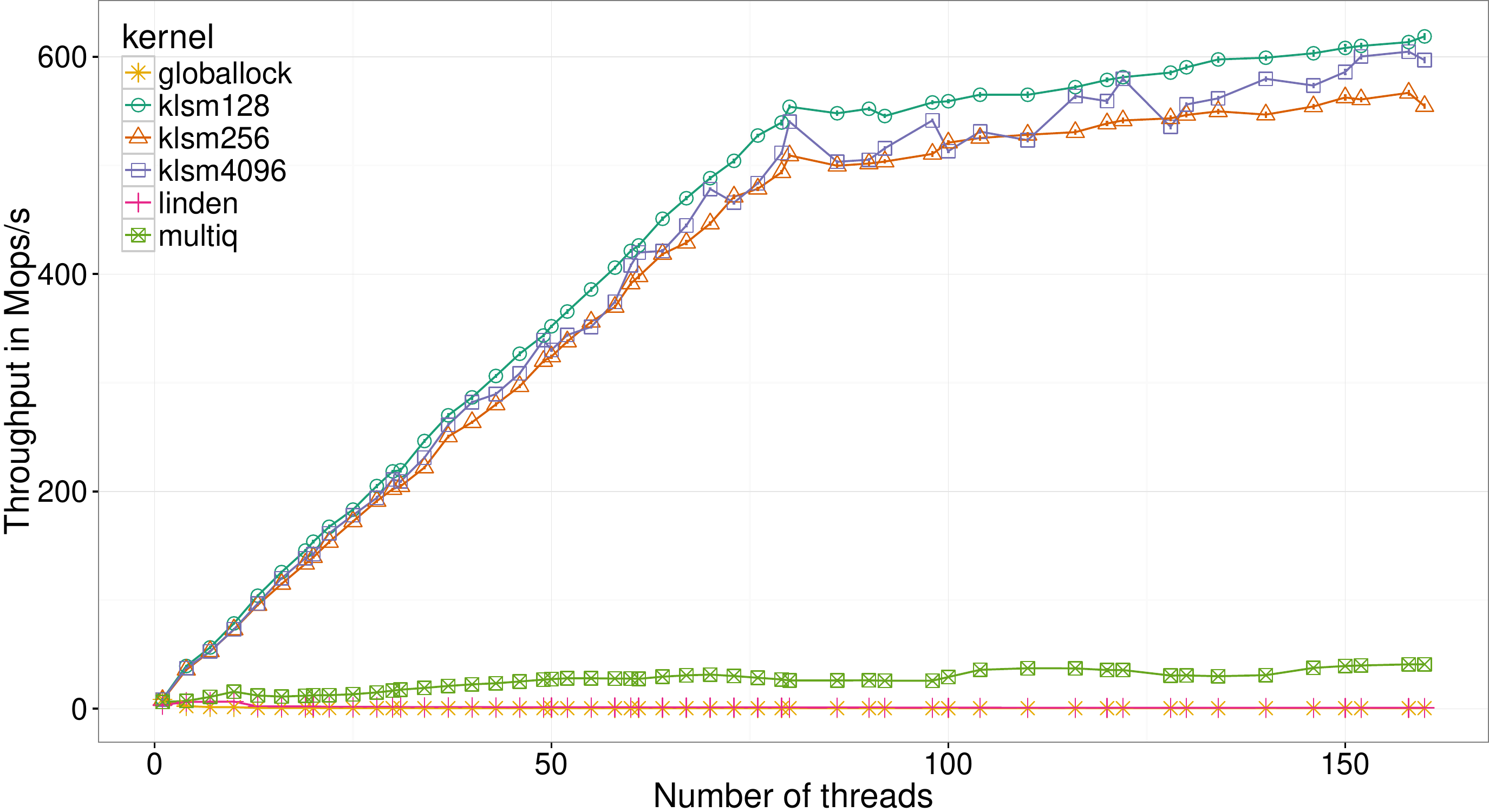}
\subcaption{\texttt{mars}, descending keys.}
\label{fig:mars_alt_desc}
\end{minipage}~%
\begin{minipage}{\columnwidth}
\centering
\includegraphics{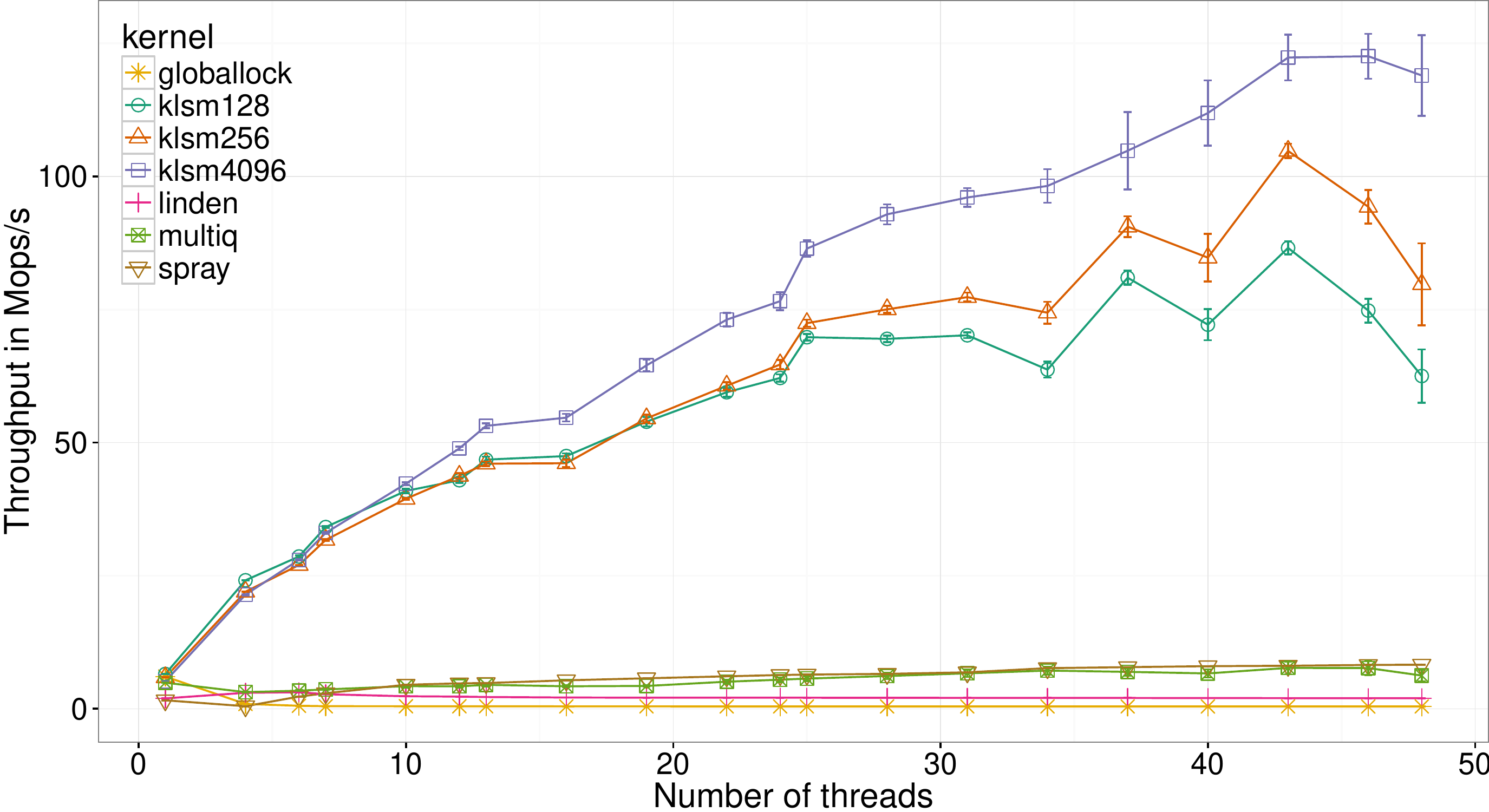}
\subcaption{\texttt{saturn}, uniform keys (32 bits).}
\end{minipage}

\begin{minipage}{\columnwidth}
\centering
\includegraphics{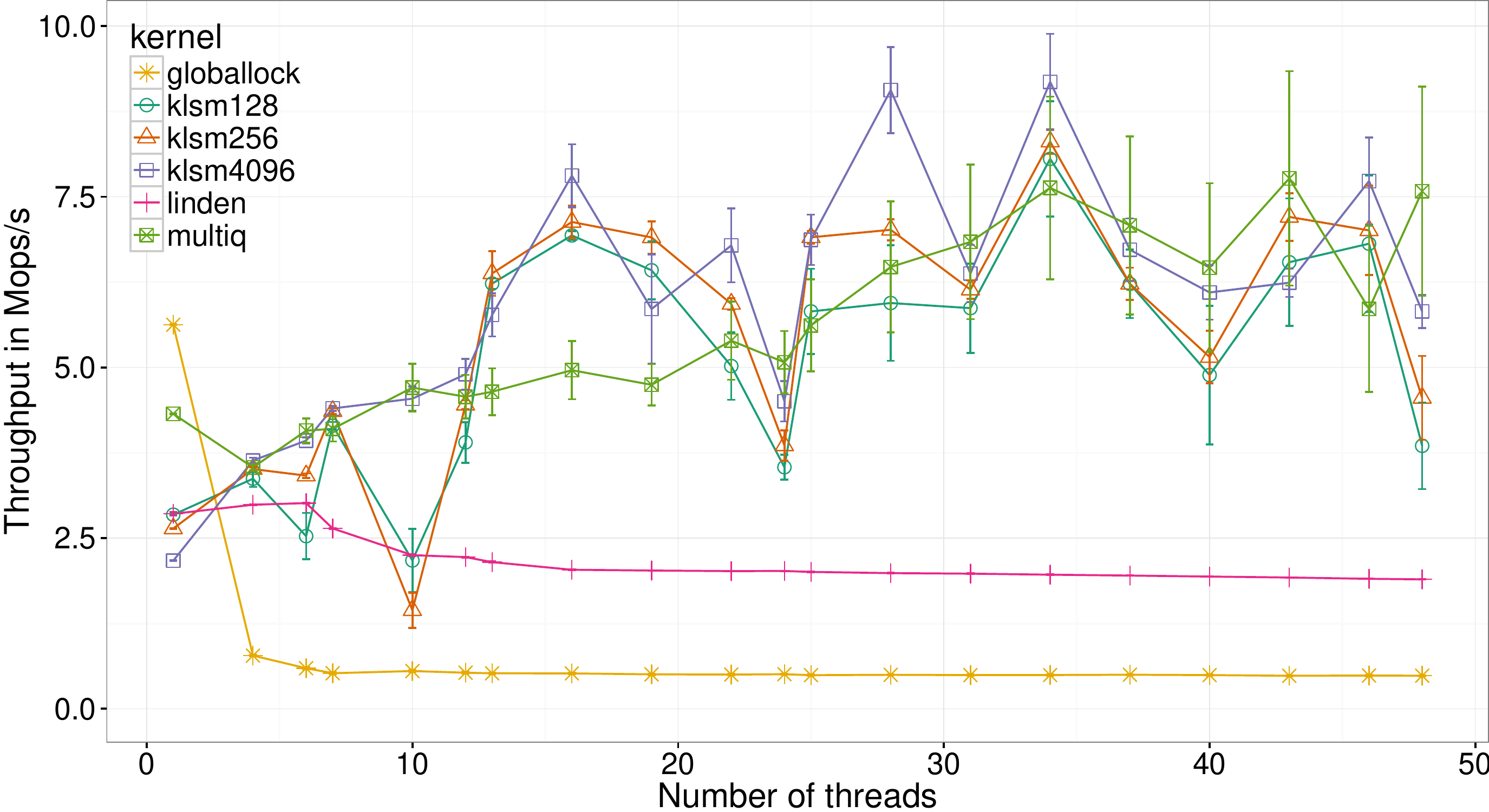}
\subcaption{\texttt{saturn}, ascending keys.}
\end{minipage}~%
\begin{minipage}{\columnwidth}
\centering
\includegraphics{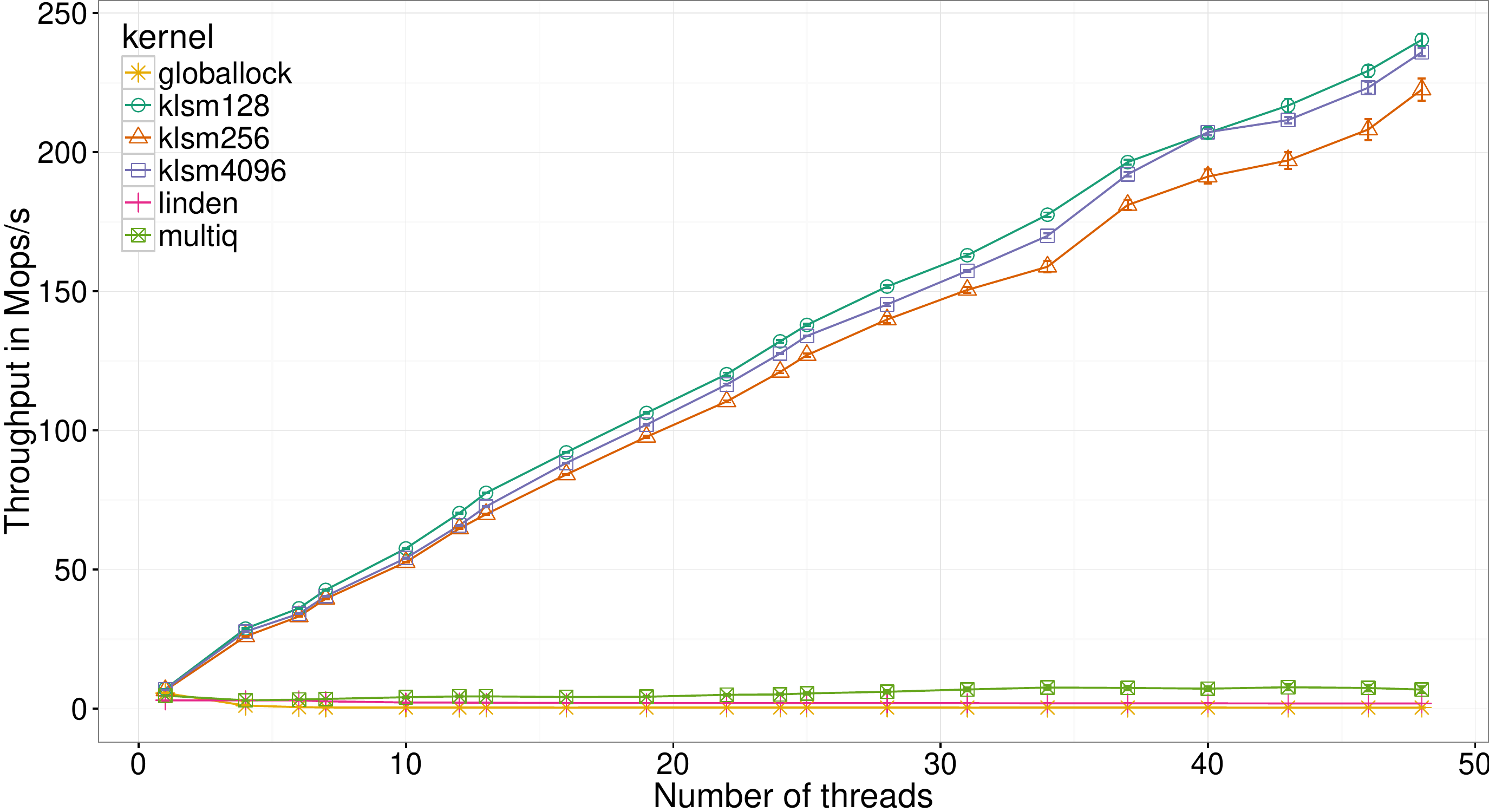}
\subcaption{\texttt{saturn}, descending keys.}
\end{minipage}

\par \vspace{\belowdisplayskip} \vspace{\abovedisplayskip}
\caption{Throughput with alternating workload.}
\label{fig:alt_mars_saturn}
\end{figure*}

\begin{figure*}[ht]
\centering
\begin{minipage}{\columnwidth}
\includegraphics{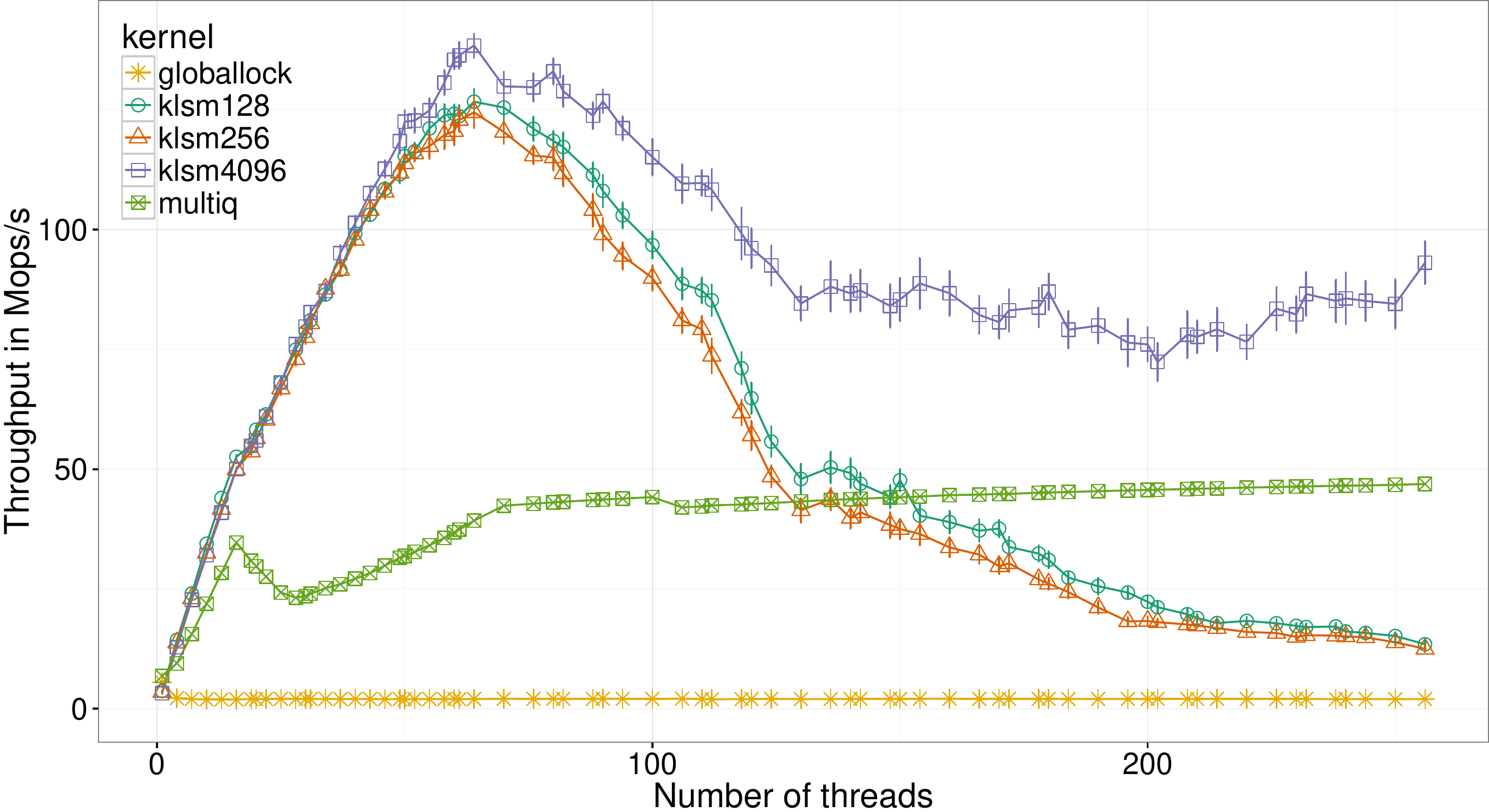}
\subcaption{\texttt{ceres}, uniform keys (32 bits).}
\label{fig:ceres_alt_uni}
\end{minipage}~%
\begin{minipage}{\columnwidth}
\centering
\includegraphics{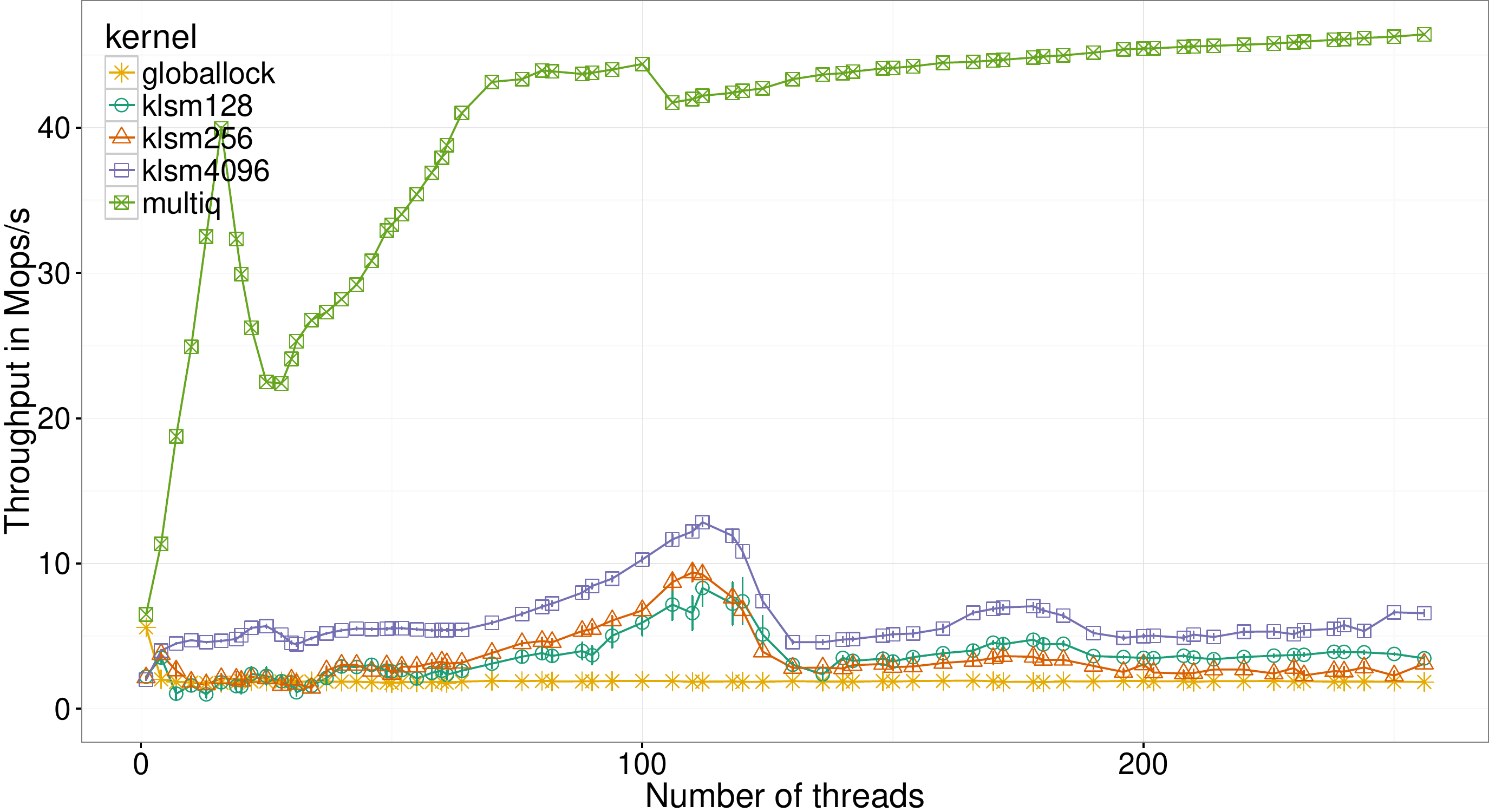}
\subcaption{\texttt{ceres}, ascending keys.}
\end{minipage}

\begin{minipage}{\columnwidth}
\centering
\includegraphics{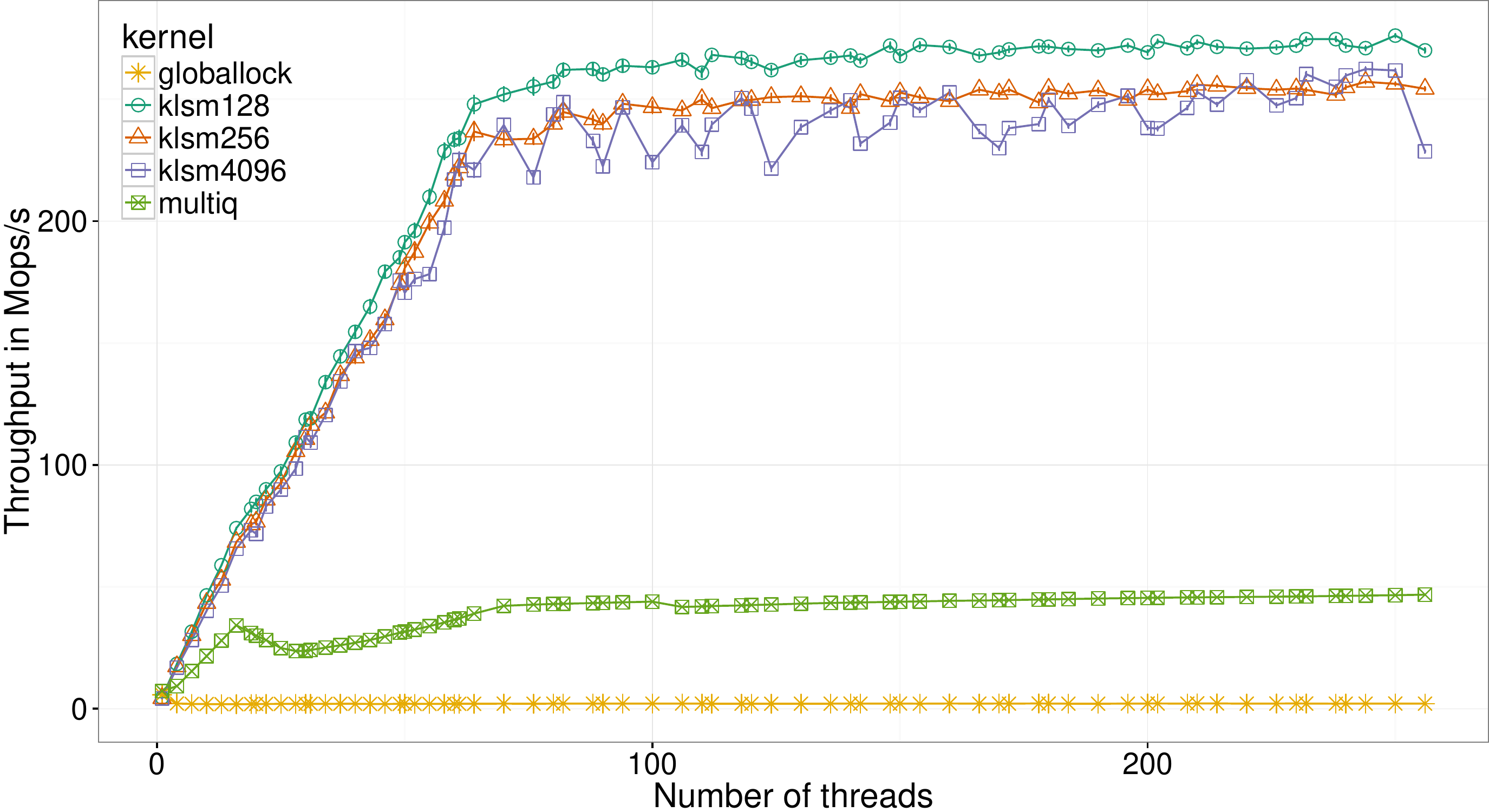}
\subcaption{\texttt{ceres}, descending keys.}
\label{fig:ceres_alt_desc}
\end{minipage}~%
\begin{minipage}{\columnwidth}
\centering
\includegraphics{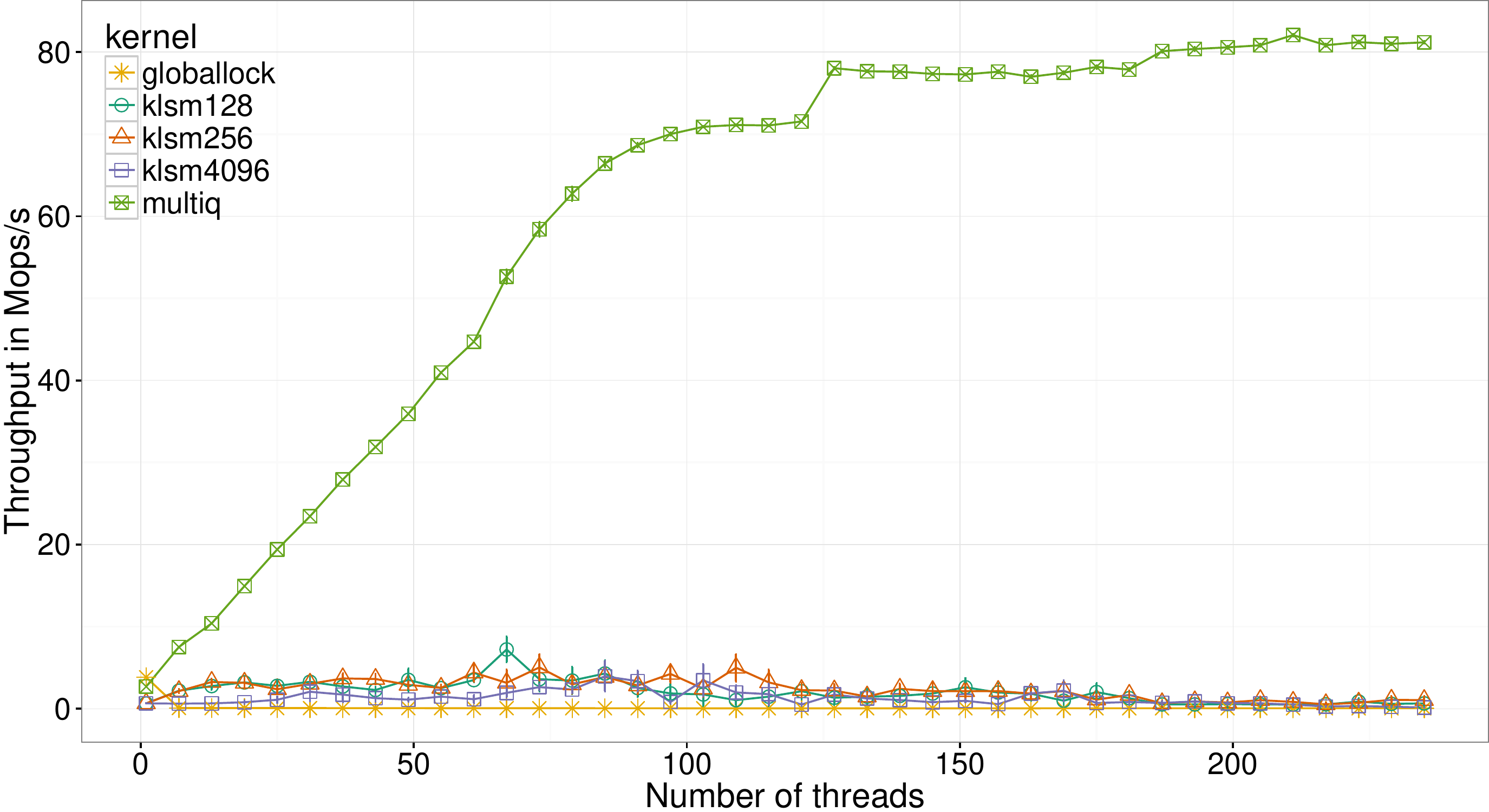}
\subcaption{\texttt{pluto}, uniform keys (32 bits).}
\end{minipage}

\begin{minipage}{\columnwidth}
\centering
\includegraphics{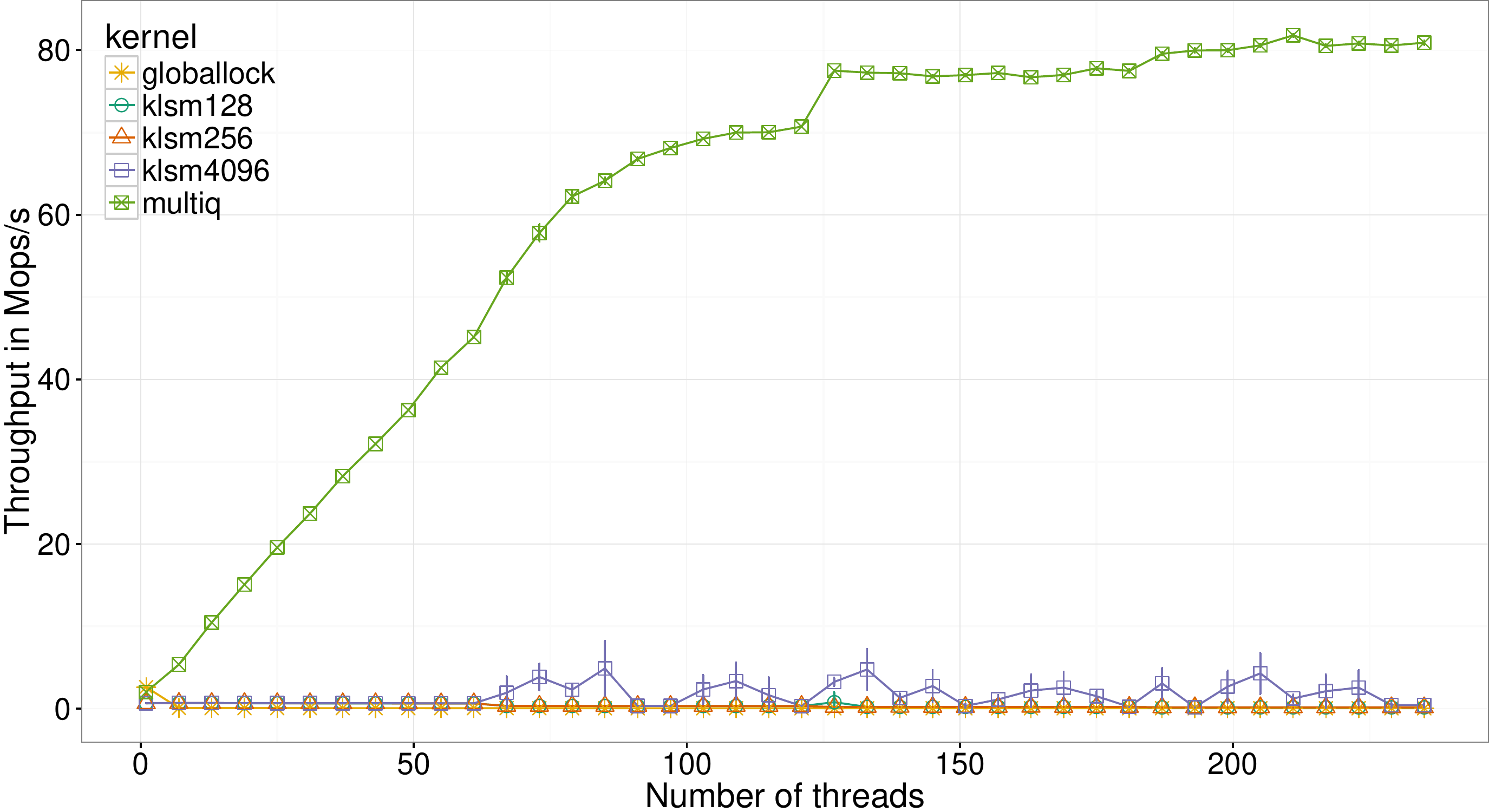}
\subcaption{\texttt{pluto}, ascending keys.}
\end{minipage}~%
\begin{minipage}{\columnwidth}
\centering
\includegraphics{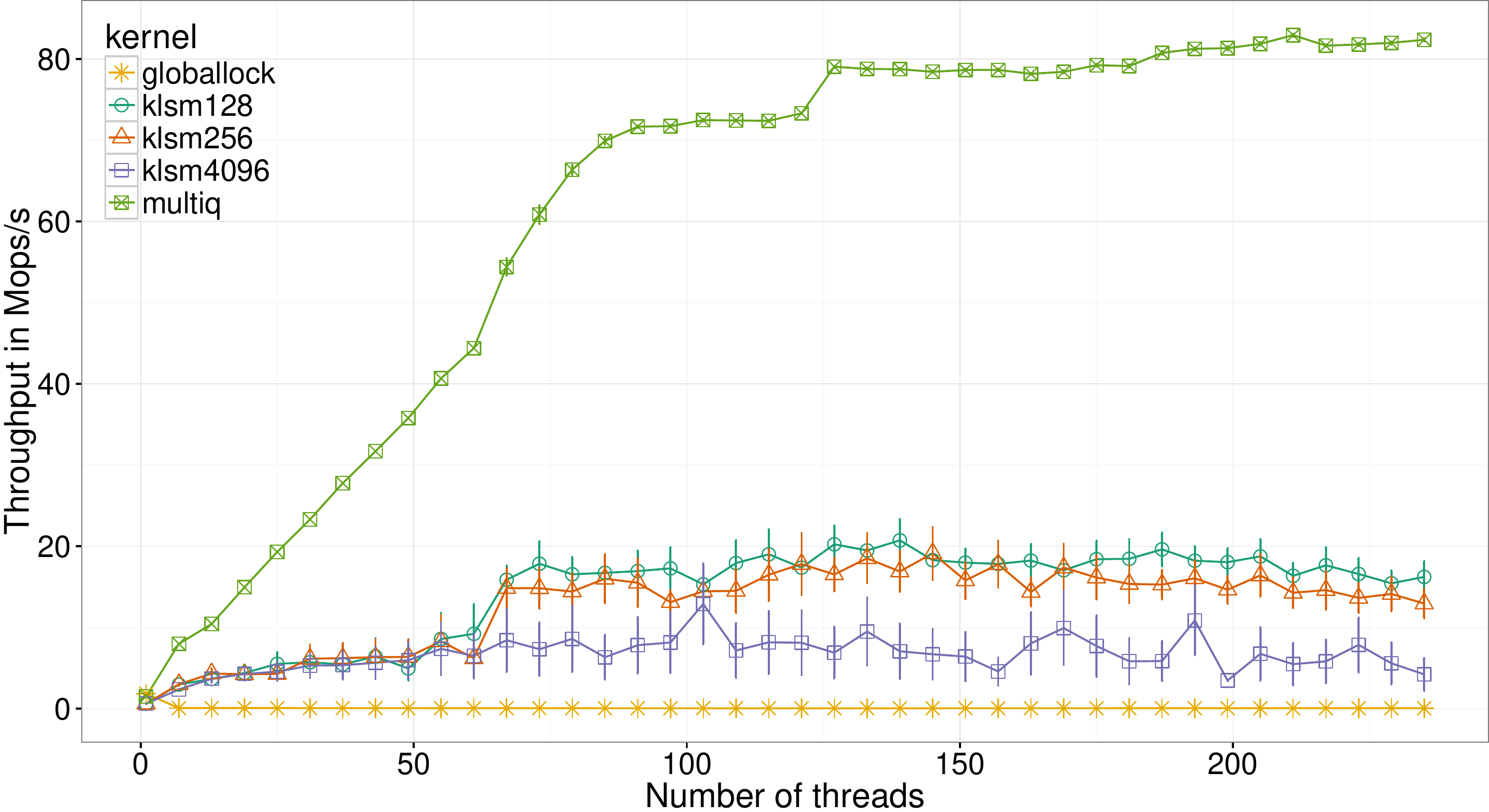}
\subcaption{\texttt{pluto}, descending keys.}
\end{minipage}

\par \vspace{\belowdisplayskip} \vspace{\abovedisplayskip}
\caption{Throughput with alternating workload.}
\label{fig:alt_ceres_pluto}
\end{figure*}

Finally, Figures~\ref{fig:alt_mars_saturn} and
\ref{fig:alt_ceres_pluto} show results for alternating workloads on
all of our machines. Although the alternating workload appears to be
similar to uniform workloads (both perform $50\%$ insertions and
$50\%$ deletions, and are distinguished only by the fact that
operations are strictly alternating in the alternating workload),
there are significant differences in the resulting throughput.
Uniform keys on \texttt{mars} (Figure~\ref{fig:mars_alt_uni}) show
increases for the \klsm in both throughput (to almost 400 \ac{MOps/s})
and scalability, with all \klsm variants ($k \in \{128, 256, 4096\}$)
scaling almost equally well until 80 threads. Likewise, descending
keys (Figure~\ref{fig:mars_alt_desc}) sees all \klsm variants reaching
a new throughput peak of around 600 \ac{MOps/s}.  Behavior on
\texttt{saturn} is similar, in which uniform and descending keys show
improved throughput and scalability for the \klsm, while results for
ascending keys remain unchanged from the uniform workload benchmark.
On \texttt{ceres}, only scalability seems to improve while throughput
is again roughly unchanged compared to uniform workload. Finally, on
\texttt{pluto}, the \klsm surprisingly does not perform well in any
case, not even under descending keys (which led to good results when
using uniform workload).  However, \texttt{multiq} throughput
increases by almost a factor of $8$, reaching over 80 \ac{MOps/s} in
all cases.

In general, the \klsm priority queue seems superior to the other
priority queues in specific scenarios: in uniform workload combined
with uniformly chosen 32- or 16-bit keys, and with descending key distribution,
throughput is almost 10 times that of other priority queues.
However, in most other benchmarks its performance is
disappointing. This appears to be due to the differing loads placed on
its component data structures: whenever the extremely scalable
\ac{DLSM} is highly utilized, throughput increases; and when the load
shifts towards the \ac{SLSM}, throughput drops. The fact that the
\klsm is composed of two priority queue designs seems to cause it to
be highly sensitive towards changes in its environment.

The MultiQueue does not seem to have the same potential for raw
performance as the \klsm at its peak.  However, in the majority of
cases it still outperforms all other tested priority queues by a good
margin. And most significantly, its behavior is extremely stable
across all of our benchmark types. Quality results show that
relaxation of the MultiQueue is fairly high, but it appears to grow
linearly with the thread count.

The \texttt{linden} queue generally only scales as long as
participating processors are located on the same physical socket;
however, a split workload combined with ascending key generation is
the exception to this rule, in which the \texttt{linden} queue is
often able to scale well until the maximal thread count.

\begin{sloppypar}
\printbibliography
\end{sloppypar}


\clearpage

\FloatBarrier
\clearpage
\FloatBarrier

\begin{table*}
\begin{minipage}{\columnwidth}
\small
\centering
\begin{tabular}{lrrrrrr}
  \toprule
& \multicolumn{2}{c}{20 threads} & \multicolumn{2}{c}{40 threads} & \multicolumn{2}{c}{80 threads} \\ \cmidrule(r){2-3}\cmidrule(r){4-5}\cmidrule(r){6-7}  & Mean & St.D. & Mean & St.D. & Mean & St.D. \\ 
  \midrule
klsm128 & 32 & 29 & 57 & 48 & 298 & 288 \\ 
  klsm256 & 42 & 42 & 68 & 57 & 635 & 464 \\ 
  klsm4096 & 422 & 729 & 1124 & 1287 & 13469 & 13980 \\ 
  multiq & 1163 & 3607 & 2296 & 7881 & 3753 & 12856 \\ 
   \bottomrule
\end{tabular}\subcaption{Uniform workload, uniform key.}
\label{tbl:mars_uni_uni}
\end{minipage}~%
\begin{minipage}{\columnwidth}
\small
\centering
\begin{tabular}{lrrrrrr}
  \toprule
& \multicolumn{2}{c}{20 threads} & \multicolumn{2}{c}{40 threads} & \multicolumn{2}{c}{80 threads} \\ \cmidrule(r){2-3}\cmidrule(r){4-5}\cmidrule(r){6-7}  & Mean & St.D. & Mean & St.D. & Mean & St.D. \\ 
  \midrule
klsm128 & 21 & 18 & 22 & 19 & 26 & 22 \\ 
  klsm256 & 38 & 33 & 38 & 33 & 42 & 37 \\ 
  klsm4096 & 499 & 469 & 479 & 451 & 496 & 467 \\ 
  multiq & 101 & 120 & 202 & 239 & 451 & 566 \\ 
   \bottomrule
\end{tabular}\subcaption{Uniform workload, ascending keys.}
\label{tbl:mars_uni_asc}
\end{minipage}

\par \vspace{\belowdisplayskip} \vspace{\abovedisplayskip}

\begin{minipage}{\columnwidth}
\small
\centering
\begin{tabular}{lrrrrrr}
  \toprule
& \multicolumn{2}{c}{20 threads} & \multicolumn{2}{c}{40 threads} & \multicolumn{2}{c}{80 threads} \\ \cmidrule(r){2-3}\cmidrule(r){4-5}\cmidrule(r){6-7}  & Mean & St.D. & Mean & St.D. & Mean & St.D. \\ 
  \midrule
klsm128 & 241 & 175 & 490 & 340 & 880 & 649 \\ 
  klsm256 & 472 & 341 & 942 & 667 & 1765 & 1234 \\ 
  klsm4096 & 2261 & 2601 & 2954 & 2728 & 3913 & 3709 \\ 
  multiq & 329 & 1708 & 674 & 3641 & 1277 & 5985 \\ 
   \bottomrule
\end{tabular}\subcaption{Uniform workload, descending keys.}
\label{tbl:mars_uni_desc}
\end{minipage}~%
\begin{minipage}{\columnwidth}
\small
\centering
\begin{tabular}{lrrrrrr}
  \toprule
& \multicolumn{2}{c}{20 threads} & \multicolumn{2}{c}{40 threads} & \multicolumn{2}{c}{80 threads} \\ \cmidrule(r){2-3}\cmidrule(r){4-5}\cmidrule(r){6-7}  & Mean & St.D. & Mean & St.D. & Mean & St.D. \\ 
  \midrule
klsm128 & 129 & 70 & 294 & 176 & 916 & 520 \\ 
  klsm256 & 217 & 133 & 557 & 340 & 1762 & 1174 \\ 
  klsm4096 & 4497 & 2495 & 11999 & 9176 & 41466 & 25387 \\ 
  multiq & 198 & 617 & 506 & 1530 & 2528 & 7492 \\ 
   \bottomrule
\end{tabular}\subcaption{Split workload, uniform keys (32 bits).}
\label{tbl:mars_spl_uni}
\end{minipage}

\par \vspace{\belowdisplayskip} \vspace{\abovedisplayskip}

\begin{minipage}{\columnwidth}
\small
\centering
\begin{tabular}{lrrrrrr}
  \toprule
& \multicolumn{2}{c}{20 threads} & \multicolumn{2}{c}{40 threads} & \multicolumn{2}{c}{80 threads} \\ \cmidrule(r){2-3}\cmidrule(r){4-5}\cmidrule(r){6-7}  & Mean & St.D. & Mean & St.D. & Mean & St.D. \\ 
  \midrule
klsm128 & 20 & 16 & 27 & 62 & 44 & 152 \\ 
  klsm256 & 34 & 29 & 44 & 97 & 78 & 290 \\ 
  klsm4096 & 439 & 400 & 894 & 2772 & 1530 & 4027 \\ 
  multiq & 163 & 634 & 514 & 1308 & 1031 & 2107 \\ 
   \bottomrule
\end{tabular}\subcaption{Split workload, ascending keys.}
\label{tbl:mars_spl_asc}
\end{minipage}~%
\begin{minipage}{\columnwidth}
\small
\centering
\begin{tabular}{lrrrrrr}
  \toprule
& \multicolumn{2}{c}{20 threads} & \multicolumn{2}{c}{40 threads} & \multicolumn{2}{c}{80 threads} \\ \cmidrule(r){2-3}\cmidrule(r){4-5}\cmidrule(r){6-7}  & Mean & St.D. & Mean & St.D. & Mean & St.D. \\ 
  \midrule
klsm128 & 561 & 203 & 1138 & 340 & 2251 & 702 \\ 
  klsm256 & 1098 & 426 & 2385 & 541 & 4555 & 1286 \\ 
  klsm4096 & 13226 & 7884 & 34502 & 12221 & 56529 & 29092 \\ 
  multiq & 1252 & 6376 & 6504 & 24567 & 1453 & 5442 \\ 
   \bottomrule
\end{tabular}\subcaption{Split workload, descending keys.}
\label{tbl:mars_spl_desc}
\end{minipage}

\par \vspace{\belowdisplayskip} \vspace{\abovedisplayskip}

\begin{minipage}{\columnwidth}
\small
\centering
\begin{tabular}{lrrrrrr}
  \toprule
& \multicolumn{2}{c}{20 threads} & \multicolumn{2}{c}{40 threads} & \multicolumn{2}{c}{80 threads} \\ \cmidrule(r){2-3}\cmidrule(r){4-5}\cmidrule(r){6-7}  & Mean & St.D. & Mean & St.D. & Mean & St.D. \\ 
  \midrule
klsm128 & 992 & 1252 & 1006 & 1192 & 1059 & 1111 \\ 
  klsm256 & 1001 & 1384 & 1022 & 1399 & 1174 & 1306 \\ 
  klsm4096 & 1091 & 2274 & 1320 & 2480 & 12654 & 13036 \\ 
  multiq & 1675 & 4263 & 2620 & 8243 & 3812 & 12166 \\ 
   \bottomrule
\end{tabular}\subcaption{Uniform workload, uniform keys (8 bits).}
\label{tbl:mars_uni_rst8}
\end{minipage}~%
\begin{minipage}{\columnwidth}
\small
\centering
\begin{tabular}{lrrrrrr}
  \toprule
& \multicolumn{2}{c}{20 threads} & \multicolumn{2}{c}{40 threads} & \multicolumn{2}{c}{80 threads} \\ \cmidrule(r){2-3}\cmidrule(r){4-5}\cmidrule(r){6-7}  & Mean & St.D. & Mean & St.D. & Mean & St.D. \\ 
  \midrule
klsm128 & 36 & 34 & 58 & 49 & 267 & 165 \\ 
  klsm256 & 43 & 44 & 64 & 56 & 551 & 392 \\ 
  klsm4096 & 268 & 470 & 481 & 854 & 14060 & 14217 \\ 
  multiq & 1173 & 3579 & 2398 & 8092 & 3744 & 12367 \\ 
   \bottomrule
\end{tabular}\subcaption{Uniform workload, uniform keys (16 bits).}
\label{tbl:mars_uni_rst16}
\end{minipage}

\par \vspace{\belowdisplayskip} \vspace{\abovedisplayskip}

\caption{Rank error on \texttt{mars}.}
\label{tbl:mars}
\end{table*}

\FloatBarrier


\clearpage

\FloatBarrier
\clearpage
\FloatBarrier

\begin{table*}
\begin{minipage}{\columnwidth}
\small
\centering
\begin{tabular}{lrrrrrr}
  \toprule
& \multicolumn{2}{c}{12 threads} & \multicolumn{2}{c}{24 threads} & \multicolumn{2}{c}{48 threads} \\ \cmidrule(r){2-3}\cmidrule(r){4-5}\cmidrule(r){6-7}  & Mean & St.D. & Mean & St.D. & Mean & St.D. \\ 
  \midrule
klsm128 & 21 & 19 & 32 & 27 & 74 & 77 \\ 
  klsm256 & 31 & 32 & 43 & 39 & 120 & 136 \\ 
  klsm4096 & 310 & 452 & 2412 & 2096 & 3319 & 3006 \\ 
  multiq & 277 & 625 & 899 & 2499 & 2298 & 6544 \\ 
   \bottomrule
\end{tabular}\subcaption{Uniform workload, uniform keys (32 bits).}
\label{tbl:saturn_uni_uni}
\end{minipage}~%
\begin{minipage}{\columnwidth}
\small
\centering
\begin{tabular}{lrrrrrr}
  \toprule
& \multicolumn{2}{c}{12 threads} & \multicolumn{2}{c}{24 threads} & \multicolumn{2}{c}{48 threads} \\ \cmidrule(r){2-3}\cmidrule(r){4-5}\cmidrule(r){6-7}  & Mean & St.D. & Mean & St.D. & Mean & St.D. \\ 
  \midrule
klsm128 & 21 & 17 & 21 & 18 & 23 & 20 \\ 
  klsm256 & 38 & 33 & 38 & 33 & 39 & 35 \\ 
  klsm4096 & 515 & 484 & 486 & 466 & 748 & 1130 \\ 
  multiq & 60 & 71 & 121 & 143 & 243 & 287 \\ 
   \bottomrule
\end{tabular}\subcaption{Uniform workload, ascending keys.}
\label{tbl:saturn_uni_asc}
\end{minipage}

\par \vspace{\belowdisplayskip} \vspace{\abovedisplayskip}

\begin{minipage}{\columnwidth}
\small
\centering
\begin{tabular}{lrrrrrr}
  \toprule
& \multicolumn{2}{c}{12 threads} & \multicolumn{2}{c}{24 threads} & \multicolumn{2}{c}{48 threads} \\ \cmidrule(r){2-3}\cmidrule(r){4-5}\cmidrule(r){6-7}  & Mean & St.D. & Mean & St.D. & Mean & St.D. \\ 
  \midrule
klsm128 & 143 & 113 & 284 & 213 & 582 & 447 \\ 
  klsm256 & 262 & 215 & 532 & 419 & 1092 & 797 \\ 
  klsm4096 & 1406 & 1697 & 2720 & 3141 & 3948 & 4034 \\ 
  multiq & 193 & 964 & 381 & 1853 & 754 & 3418 \\ 
   \bottomrule
\end{tabular}\subcaption{Uniform workload, descending keys.}
\label{tbl:saturn_uni_desc}
\end{minipage}~%
\begin{minipage}{\columnwidth}
\small
\centering
\begin{tabular}{lrrrrrr}
  \toprule
& \multicolumn{2}{c}{12 threads} & \multicolumn{2}{c}{24 threads} & \multicolumn{2}{c}{48 threads} \\ \cmidrule(r){2-3}\cmidrule(r){4-5}\cmidrule(r){6-7}  & Mean & St.D. & Mean & St.D. & Mean & St.D. \\ 
  \midrule
klsm128 & 61 & 31 & 219 & 118 & 257 & 140 \\ 
  klsm256 & 198 & 114 & 220 & 113 & 950 & 524 \\ 
  klsm4096 & 1982 & 1218 & 2832 & 1523 & 5482 & 3825 \\ 
  multiq & 81 & 140 & 172 & 332 & 548 & 1476 \\ 
   \bottomrule
\end{tabular}\subcaption{Split workload, uniform keys (32 bits).}
\label{tbl:saturn_spl_uni}
\end{minipage}

\par \vspace{\belowdisplayskip} \vspace{\abovedisplayskip}

\begin{minipage}{\columnwidth}
\small
\centering
\begin{tabular}{lrrrrrr}
  \toprule
& \multicolumn{2}{c}{12 threads} & \multicolumn{2}{c}{24 threads} & \multicolumn{2}{c}{48 threads} \\ \cmidrule(r){2-3}\cmidrule(r){4-5}\cmidrule(r){6-7}  & Mean & St.D. & Mean & St.D. & Mean & St.D. \\ 
  \midrule
klsm128 & 19 & 15 & 19 & 16 & 20 & 17 \\ 
  klsm256 & 34 & 29 & 34 & 29 & 35 & 30 \\ 
  klsm4096 & 446 & 411 & 422 & 403 & 405 & 376 \\ 
  multiq & 62 & 93 & 118 & 150 & 219 & 267 \\ 
   \bottomrule
\end{tabular}\subcaption{Split workload, ascending keys.}
\label{tbl:saturn_spl_asc}
\end{minipage}~%
\begin{minipage}{\columnwidth}
\small
\centering
\begin{tabular}{lrrrrrr}
  \toprule
& \multicolumn{2}{c}{12 threads} & \multicolumn{2}{c}{24 threads} & \multicolumn{2}{c}{48 threads} \\ \cmidrule(r){2-3}\cmidrule(r){4-5}\cmidrule(r){6-7}  & Mean & St.D. & Mean & St.D. & Mean & St.D. \\ 
  \midrule
klsm128 & 192 & 159 & 431 & 284 & 835 & 786 \\ 
  klsm256 & 362 & 290 & 759 & 620 & 1559 & 1564 \\ 
  klsm4096 & 6471 & 4024 & 12923 & 8545 & 24879 & 17970 \\ 
  multiq & 337 & 1780 & 329 & 1557 & 372 & 1115 \\ 
   \bottomrule
\end{tabular}\subcaption{Split workload, descending keys.}
\label{tbl:saturn_spl_desc}
\end{minipage}

\par \vspace{\belowdisplayskip} \vspace{\abovedisplayskip}


\begin{minipage}{\columnwidth}
\small
\centering
\begin{tabular}{lrrrrrr}
  \toprule
& \multicolumn{2}{c}{12 threads} & \multicolumn{2}{c}{24 threads} & \multicolumn{2}{c}{48 threads} \\ \cmidrule(r){2-3}\cmidrule(r){4-5}\cmidrule(r){6-7}  & Mean & St.D. & Mean & St.D. & Mean & St.D. \\ 
  \midrule
klsm128 & 986 & 1129 & 992 & 1119 & 1015 & 1108 \\ 
  klsm256 & 989 & 1253 & 1002 & 1241 & 1039 & 1234 \\ 
  klsm4096 & 1233 & 1858 & 1398 & 1973 & 2537 & 3540 \\ 
  multiq & 1160 & 2039 & 1478 & 3094 & 2669 & 7097 \\ 
   \bottomrule
\end{tabular}\subcaption{Uniform workload, uniform keys (8 bits).}
\label{tbl:saturn_uni_rst8}
\end{minipage}~%
\begin{minipage}{\columnwidth}
\small
\centering
\begin{tabular}{lrrrrrr}
  \toprule
& \multicolumn{2}{c}{12 threads} & \multicolumn{2}{c}{24 threads} & \multicolumn{2}{c}{48 threads} \\ \cmidrule(r){2-3}\cmidrule(r){4-5}\cmidrule(r){6-7}  & Mean & St.D. & Mean & St.D. & Mean & St.D. \\ 
  \midrule
klsm128 & 26 & 24 & 37 & 32 & 82 & 76 \\ 
  klsm256 & 39 & 43 & 52 & 54 & 156 & 194 \\ 
  klsm4096 & 355 & 553 & 698 & 1097 & 2793 & 3934 \\ 
  multiq & 342 & 843 & 898 & 2478 & 2314 & 6898 \\ 
   \bottomrule
\end{tabular}\subcaption{Uniform workload, uniform keys (16 bits).}
\label{tbl:saturn_uni_rst16}
\end{minipage}

\par \vspace{\belowdisplayskip} \vspace{\abovedisplayskip}

\caption{Rank error on \texttt{saturn}.}
\label{tbl:saturn}
\end{table*}

\FloatBarrier


\clearpage

\FloatBarrier
\clearpage
\FloatBarrier

\begin{table*}
\begin{minipage}{\columnwidth}
\small
\centering
\begin{tabular}{lrrrrrr}
  \toprule
& \multicolumn{2}{c}{16 threads} & \multicolumn{2}{c}{32 threads} & \multicolumn{2}{c}{64 threads} \\ \cmidrule(r){2-3}\cmidrule(r){4-5}\cmidrule(r){6-7}  & Mean & St.D. & Mean & St.D. & Mean & St.D. \\ 
  \midrule
klsm128 & 30 & 26 & 58 & 49 & 147 & 130 \\ 
  klsm256 & 43 & 44 & 79 & 72 & 231 & 231 \\ 
  klsm4096 & 289 & 488 & 594 & 1032 & 5831 & 7604 \\ 
  multiq & 1258 & 4657 & 1995 & 6603 & 3315 & 11720 \\ 
   \bottomrule
\end{tabular}\subcaption{Uniform workload, uniform keys (32 bits).}
\label{tbl:ceres_uni_uni}
\end{minipage}~%
\begin{minipage}{\columnwidth}
\small
\centering
\begin{tabular}{lrrrrrr}
  \toprule
& \multicolumn{2}{c}{16 threads} & \multicolumn{2}{c}{32 threads} & \multicolumn{2}{c}{64 threads} \\ \cmidrule(r){2-3}\cmidrule(r){4-5}\cmidrule(r){6-7}  & Mean & St.D. & Mean & St.D. & Mean & St.D. \\ 
  \midrule
klsm128 & 20 & 17 & 23 & 19 & 25 & 22 \\ 
  klsm256 & 37 & 32 & 39 & 33 & 43 & 37 \\ 
  klsm4096 & 513 & 477 & 493 & 460 & 528 & 493 \\ 
  multiq & 81 & 95 & 163 & 192 & 500 & 669 \\ 
   \bottomrule
\end{tabular}\subcaption{Uniform workload, ascending keys.}
\label{tbl:ceres_uni_asc}
\end{minipage}

\par \vspace{\belowdisplayskip} \vspace{\abovedisplayskip}

\begin{minipage}{\columnwidth}
\small
\centering
\begin{tabular}{lrrrrrr}
  \toprule
& \multicolumn{2}{c}{16 threads} & \multicolumn{2}{c}{32 threads} & \multicolumn{2}{c}{64 threads} \\ \cmidrule(r){2-3}\cmidrule(r){4-5}\cmidrule(r){6-7}  & Mean & St.D. & Mean & St.D. & Mean & St.D. \\ 
  \midrule
klsm128 & 215 & 153 & 427 & 310 & 1064 & 848 \\ 
  klsm256 & 400 & 292 & 796 & 556 & 2057 & 1562 \\ 
  klsm4096 & 1097 & 961 & 4080 & 3970 & 7742 & 6931 \\ 
  multiq & 295 & 1912 & 535 & 2766 & 1085 & 5348 \\ 
   \bottomrule
\end{tabular}\subcaption{Uniform workload, descending keys.}
\label{tbl:ceres_uni_desc}
\end{minipage}~%
\begin{minipage}{\columnwidth}
\small
\centering
\begin{tabular}{lrrrrrr}
  \toprule
& \multicolumn{2}{c}{16 threads} & \multicolumn{2}{c}{32 threads} & \multicolumn{2}{c}{64 threads} \\ \cmidrule(r){2-3}\cmidrule(r){4-5}\cmidrule(r){6-7}  & Mean & St.D. & Mean & St.D. & Mean & St.D. \\ 
  \midrule
klsm128 & 126 & 76 & 250 & 144 & 871 & 548 \\ 
  klsm256 & 218 & 127 & 486 & 287 & 1723 & 1027 \\ 
  klsm4096 & 4231 & 2614 & 13006 & 7627 & 16781 & 13269 \\ 
  multiq & 376 & 1195 & 491 & 1290 & 969 & 2417 \\ 
   \bottomrule
\end{tabular}\subcaption{Split workload, uniform keys (32 bits).}
\label{tbl:ceres_spl_uni}
\end{minipage}

\par \vspace{\belowdisplayskip} \vspace{\abovedisplayskip}

\begin{minipage}{\columnwidth}
\small
\centering
\begin{tabular}{lrrrrrr}
  \toprule
& \multicolumn{2}{c}{16 threads} & \multicolumn{2}{c}{32 threads} & \multicolumn{2}{c}{64 threads} \\ \cmidrule(r){2-3}\cmidrule(r){4-5}\cmidrule(r){6-7}  & Mean & St.D. & Mean & St.D. & Mean & St.D. \\ 
  \midrule
klsm128 & 19 & 15 & 21 & 18 & 25 & 22 \\ 
  klsm256 & 34 & 28 & 36 & 30 & 41 & 35 \\ 
  klsm4096 & 435 & 400 & 453 & 411 & 465 & 417 \\ 
  multiq & 2342 & 10594 & 479 & 1160 & 5077 & 8644 \\ 
   \bottomrule
\end{tabular}\subcaption{Split workload, ascending keys.}
\label{tbl:ceres_spl_asc}
\end{minipage}~%
\begin{minipage}{\columnwidth}
\small
\centering
\begin{tabular}{lrrrrrr}
  \toprule
& \multicolumn{2}{c}{16 threads} & \multicolumn{2}{c}{32 threads} & \multicolumn{2}{c}{64 threads} \\ \cmidrule(r){2-3}\cmidrule(r){4-5}\cmidrule(r){6-7}  & Mean & St.D. & Mean & St.D. & Mean & St.D. \\ 
  \midrule
klsm128 & 381 & 195 & 696 & 376 & 1115 & 626 \\ 
  klsm256 & 779 & 389 & 1009 & 919 & 1584 & 1211 \\ 
  klsm4096 & 10584 & 6209 & 14480 & 11416 & 27109 & 22042 \\ 
  multiq & 3690 & 18381 & 257 & 702 & 1363 & 4570 \\ 
   \bottomrule
\end{tabular}\subcaption{Split workload, descending keys.}
\label{tbl:ceres_spl_desc}
\end{minipage}

\par \vspace{\belowdisplayskip} \vspace{\abovedisplayskip}

\begin{minipage}{\columnwidth}
\small
\centering
\begin{tabular}{lrrrrrr}
  \toprule
& \multicolumn{2}{c}{16 threads} & \multicolumn{2}{c}{32 threads} & \multicolumn{2}{c}{64 threads} \\ \cmidrule(r){2-3}\cmidrule(r){4-5}\cmidrule(r){6-7}  & Mean & St.D. & Mean & St.D. & Mean & St.D. \\ 
  \midrule
klsm128 & 997 & 1404 & 1006 & 1454 & 1108 & 1552 \\ 
  klsm256 & 1007 & 1579 & 1028 & 1610 & 1212 & 1721 \\ 
  klsm4096 & 1102 & 2241 & 1491 & 2693 & 8001 & 9726 \\ 
  multiq & 1838 & 6381 & 2249 & 7262 & 3582 & 12676 \\ 
   \bottomrule
\end{tabular}\subcaption{Uniform workload, uniform keys (8 bits).}
\label{tbl:ceres_uni_rst8}
\end{minipage}~%
\begin{minipage}{\columnwidth}
\small
\centering
\begin{tabular}{lrrrrrr}
  \toprule
& \multicolumn{2}{c}{16 threads} & \multicolumn{2}{c}{32 threads} & \multicolumn{2}{c}{64 threads} \\ \cmidrule(r){2-3}\cmidrule(r){4-5}\cmidrule(r){6-7}  & Mean & St.D. & Mean & St.D. & Mean & St.D. \\ 
  \midrule
klsm128 & 33 & 32 & 63 & 55 & 142 & 119 \\ 
  klsm256 & 45 & 47 & 77 & 73 & 242 & 275 \\ 
  klsm4096 & 306 & 515 & 638 & 1201 & 4137 & 5347 \\ 
  multiq & 1280 & 5041 & 1768 & 5602 & 3421 & 12338 \\ 
   \bottomrule
\end{tabular}\subcaption{Uniform workload, uniform keys (16 bits).}
\label{tbl:ceres_uni_rst16}
\end{minipage}

\par \vspace{\belowdisplayskip} \vspace{\abovedisplayskip}

\caption{Rank error on \texttt{ceres}.}
\label{tbl:ceres}
\end{table*}

\FloatBarrier


\clearpage

\FloatBarrier


\clearpage

\FloatBarrier
\clearpage

\FloatBarrier
\clearpage

\begin{table*}
\begin{minipage}{\columnwidth}
\small
\centering
\begin{tabular}{lrrrrrr}
  \toprule
& \multicolumn{2}{c}{20 threads} & \multicolumn{2}{c}{40 threads} & \multicolumn{2}{c}{80 threads} \\ \cmidrule(r){2-3}\cmidrule(r){4-5}\cmidrule(r){6-7}  & Mean & St.D. & Mean & St.D. & Mean & St.D. \\ 
  \midrule
klsm128 & 41 & 39 & 26 & 21 & 443 & 277 \\ 
  klsm256 & 25 & 30 & 43 & 50 & 1377 & 885 \\ 
  klsm4096 & 457 & 726 & 1628 & 1615 & 9486 & 13502 \\ 
  multiq & 1084 & 3243 & 2422 & 8278 & 2890 & 7927 \\ 
   \bottomrule
\end{tabular}\subcaption{\texttt{mars}, uniform keys (32 bits).}
\label{tbl:mars_alt_uni}
\end{minipage}~%
\begin{minipage}{\columnwidth}
\small
\centering
\begin{tabular}{lrrrrrr}
  \toprule
& \multicolumn{2}{c}{20 threads} & \multicolumn{2}{c}{40 threads} & \multicolumn{2}{c}{80 threads} \\ \cmidrule(r){2-3}\cmidrule(r){4-5}\cmidrule(r){6-7}  & Mean & St.D. & Mean & St.D. & Mean & St.D. \\ 
  \midrule
klsm128 & 21 & 18 & 22 & 19 & 26 & 22 \\ 
  klsm256 & 38 & 33 & 39 & 34 & 42 & 37 \\ 
  klsm4096 & 495 & 463 & 490 & 464 & 484 & 449 \\ 
  multiq & 100 & 118 & 202 & 238 & 413 & 490 \\ 
   \bottomrule
\end{tabular}\subcaption{\texttt{mars}, ascending keys.}
\label{tbl:mars_alt_asc}
\end{minipage}

\par \vspace{\belowdisplayskip} \vspace{\abovedisplayskip}

\begin{minipage}{\columnwidth}
\small
\centering
\begin{tabular}{lrrrrrr}
  \toprule
& \multicolumn{2}{c}{20 threads} & \multicolumn{2}{c}{40 threads} & \multicolumn{2}{c}{80 threads} \\ \cmidrule(r){2-3}\cmidrule(r){4-5}\cmidrule(r){6-7}  & Mean & St.D. & Mean & St.D. & Mean & St.D. \\ 
  \midrule
klsm128 & 10 & 7 & 24 & 13 & 40 & 30 \\ 
  klsm256 & 11 & 7 & 23 & 13 & 41 & 28 \\ 
  klsm4096 & 8 & 6 & 20 & 15 & 41 & 36 \\ 
  multiq & 334 & 1634 & 674 & 3640 & 1194 & 4935 \\ 
   \bottomrule
\end{tabular}\subcaption{\texttt{mars}, descending keys.}
\label{tbl:mars_alt_desc}
\end{minipage}~%
\begin{minipage}{\columnwidth}
\small
\centering
\begin{tabular}{lrrrrrr}
  \toprule
& \multicolumn{2}{c}{12 threads} & \multicolumn{2}{c}{24 threads} & \multicolumn{2}{c}{48 threads} \\ \cmidrule(r){2-3}\cmidrule(r){4-5}\cmidrule(r){6-7}  & Mean & St.D. & Mean & St.D. & Mean & St.D. \\ 
  \midrule
klsm128 & 17 & 20 & 22 & 23 & 65 & 72 \\ 
  klsm256 & 31 & 39 & 34 & 39 & 115 & 165 \\ 
  klsm4096 & 379 & 599 & 396 & 591 & 915 & 900 \\ 
  multiq & 340 & 839 & 815 & 2178 & 2216 & 6763 \\ 
   \bottomrule
\end{tabular}\subcaption{\texttt{saturn}, uniform keys (32 bits).}
\label{tbl:saturn_alt_uni}
\end{minipage}

\par \vspace{\belowdisplayskip} \vspace{\abovedisplayskip}

\begin{minipage}{\columnwidth}
\small
\centering
\begin{tabular}{lrrrrrr}
  \toprule
& \multicolumn{2}{c}{12 threads} & \multicolumn{2}{c}{24 threads} & \multicolumn{2}{c}{48 threads} \\ \cmidrule(r){2-3}\cmidrule(r){4-5}\cmidrule(r){6-7}  & Mean & St.D. & Mean & St.D. & Mean & St.D. \\ 
  \midrule
klsm128 & 20 & 17 & 21 & 18 & 23 & 20 \\ 
  klsm256 & 37 & 32 & 38 & 33 & 40 & 35 \\ 
  klsm4096 & 509 & 478 & 492 & 467 & 849 & 1402 \\ 
  multiq & 60 & 70 & 120 & 142 & 244 & 287 \\ 
   \bottomrule
\end{tabular}\subcaption{\texttt{saturn}, ascending keys.}
\label{tbl:saturn_alt_asc}
\end{minipage}~%
\begin{minipage}{\columnwidth}
\small
\centering
\begin{tabular}{lrrrrrr}
  \toprule
& \multicolumn{2}{c}{12 threads} & \multicolumn{2}{c}{24 threads} & \multicolumn{2}{c}{48 threads} \\ \cmidrule(r){2-3}\cmidrule(r){4-5}\cmidrule(r){6-7}  & Mean & St.D. & Mean & St.D. & Mean & St.D. \\ 
  \midrule
klsm128 & 6 & 5 & 11 & 8 & 24 & 16 \\ 
  klsm256 & 6 & 4 & 11 & 7 & 24 & 17 \\ 
  klsm4096 & 5 & 4 & 14 & 10 & 22 & 19 \\ 
  multiq & 182 & 858 & 364 & 1612 & 766 & 3619 \\ 
   \bottomrule
\end{tabular}\subcaption{\texttt{saturn}, descending keys.}
\label{tbl:saturn_alt_desc}
\end{minipage}

\par \vspace{\belowdisplayskip} \vspace{\abovedisplayskip}

\begin{minipage}{\columnwidth}
\small
\centering
\begin{tabular}{lrrrrrr}
  \toprule
& \multicolumn{2}{c}{16 threads} & \multicolumn{2}{c}{32 threads} & \multicolumn{2}{c}{64 threads} \\ \cmidrule(r){2-3}\cmidrule(r){4-5}\cmidrule(r){6-7}  & Mean & St.D. & Mean & St.D. & Mean & St.D. \\ 
  \midrule
klsm128 & 19 & 21 & 30 & 32 & 99 & 129 \\ 
  klsm256 & 30 & 36 & 50 & 71 & 151 & 195 \\ 
  klsm4096 & 355 & 576 & 958 & 1317 & 3602 & 6421 \\ 
  multiq & 1268 & 4727 & 1944 & 6454 & 3433 & 12385 \\ 
   \bottomrule
\end{tabular}\subcaption{\texttt{ceres}, uniform keys (32 bits).}
\label{tbl:ceres_alt_uni}
\end{minipage}~%
\begin{minipage}{\columnwidth}
\small
\centering
\begin{tabular}{lrrrrrr}
  \toprule
& \multicolumn{2}{c}{16 threads} & \multicolumn{2}{c}{32 threads} & \multicolumn{2}{c}{64 threads} \\ \cmidrule(r){2-3}\cmidrule(r){4-5}\cmidrule(r){6-7}  & Mean & St.D. & Mean & St.D. & Mean & St.D. \\ 
  \midrule
klsm128 & 20 & 17 & 22 & 19 & 25 & 22 \\ 
  klsm256 & 36 & 31 & 39 & 33 & 42 & 37 \\ 
  klsm4096 & 521 & 485 & 491 & 461 & 532 & 491 \\ 
  multiq & 80 & 95 & 163 & 192 & 1049 & 1993 \\ 
   \bottomrule
\end{tabular}\subcaption{\texttt{ceres}, ascending keys.}
\label{tbl:ceres_alt_asc}
\end{minipage}

\par \vspace{\belowdisplayskip} \vspace{\abovedisplayskip}

\begin{minipage}{\columnwidth}
\small
\centering
\begin{tabular}{lrrrrrr}
  \toprule
& \multicolumn{2}{c}{16 threads} & \multicolumn{2}{c}{32 threads} & \multicolumn{2}{c}{64 threads} \\ \cmidrule(r){2-3}\cmidrule(r){4-5}\cmidrule(r){6-7}  & Mean & St.D. & Mean & St.D. & Mean & St.D. \\ 
  \midrule
klsm128 & 10 & 6 & 16 & 11 & 33 & 23 \\ 
  klsm256 & 9 & 6 & 17 & 11 & 35 & 25 \\ 
  klsm4096 & 9 & 7 & 19 & 15 & 37 & 29 \\ 
  multiq & 280 & 1706 & 551 & 3118 & 1071 & 5497 \\ 
   \bottomrule
\end{tabular}\subcaption{\texttt{ceres}, descending keys.}
\label{tbl:ceres_alt_desc}
\end{minipage}

\par \vspace{\belowdisplayskip} \vspace{\abovedisplayskip}

\caption{Rank error with alternating workload.}
\label{tbl:alt}
\end{table*}

\end{refsection}

\end{document}

%% file: acronyms.tex
\acrodef{BST}{Binary Search Tree}
\acrodef{CAS}{Compare-And-Swap}
\acrodef{CBPQ}{Chunk-Based Priority Queue}
\acrodef{DCAS}{Double-Compare-And-Swap}
\acrodef{DCSS}{Double-Compare-Single-Swap}
\acrodef{DLSM}{Dist\-rib\-uted LSM}
\acrodef{DS}{Data Structure}
\acrodef{FAA}{Fetch-And-Add}
\acrodef{FAO}{Fetch-And-Or}
\acrodef{FIFO}{First In First Out}
\acrodef{GSL}{GNU Scientific Library}
\acrodef{LIFO}{Last In First Out}
\acrodef{LSM}{Log-Structured Merge-Tree}
\acrodef{MOps}{million operations}
\acrodef{MOps/s}{million operations per second}
\acrodef{NUMA}{non-uniform memory access}
\acrodef{PQ}{Priority Queue}
\acrodef{pthreads}{POSIX Threads}
\acrodef{RAM}{Random Access Memory}
\acrodef{SL}{SkipList}
\acrodef{SLSM}{Shared LSM}
\acrodef{STL}{C++ Standard Library}
\acrodef{TAS}{Test-And-Set}